\documentstyle[12pt]{report}
\begin{document}
\pagenumbering{roman} \pagestyle{myheadings}
\baselineskip = 16pt
\newcounter{appen}
\setcounter{appen}{1}
\def\beqar {\begin{eqnarray}}
\def\eeqar {\end{eqnarray}}
\def\beq {\begin{equation}}
\def\eeq {\end{equation}}
\def\A{{\cal A}}
\def\B{{\cal B}}
\def\C{{\cal C}}
\def\F{{\cal F}}
\def\G{{\cal G}}
\def\H{{\cal H}}
\def\J{{\cal J}}
\def\K{{\cal K}}
\def\S{{\cal S}}
\def\O{{\cal O}}
\def\P{{\cal P}}
\def\L{{\cal L}}
\def\D{{\cal D}}
\def\N{{\cal N}}
\def\M{{\cal M}}
\def\Z{{\cal Z}}
\def\al{\alpha}
\def\bt{\beta}
\def\del{\delta}
\def\Del{\Delta}
\def\ga{\gamma}
\def\Ga{\Gamma}
\def\ka{\kappa}
\def\ep{\epsilon}
\def\la{\lambda}
\def\La{\Lambda}
\def\om{\omega}
\def\Om{\Omega}
\def\th{\theta}
\def\vf{{\varphi}}
\def\et{\eta}
\def\si{\sigma}
\def\Si{\Sigma}
\def\zt{\zeta}
\def\p{\phi}
\def\d{\partial}
\def\bd{{\bar \partial}}
\def\nb{\nabla}
\def\Ad{{\dot A}}
\def\Bd{{\dot B}}
\def\Cd{{\dot C}}
\def\bz{{\bar z}}
\def\bZ{{\bar Z}}
\def\bu{{\bar u}}
\def\bpi{{\bar \pi}}
\def\bom{{\bar \omega}}
\def\bxi{{\bar \xi}}
\def\bzt{{\bar \zeta}}
\def\bth{{\bar \theta}}
\def\qu{\frac{1}{4}}
\def\hf{\frac{1}{2}}
\def\half{\frac{1}{2}}
\def\shf{\frac{1}{\sqrt{2}}}
\def\<{\langle}\def\bra{\langle}
\def\>{\rangle}\def\ket{\rangle}
\def\Tr{{\rm Tr}}
\def\STr{{\rm STr}}
\def\tr{{\rm tr}}
\def\sgn{{\rm sgn}}
\def\diag{{\rm diag}}\def\rank{{\rm rank}}
\def\det{{\rm det}}\def\dim{{\rm dim}}
\def\cp{{\bf CP}}
\def \slashint {{\begin{array}{ c }\hskip
  .015in\lceil\\\rfloor\\\end{array}}}
\def\adot{{\dot a}}
\def\bdot {{\dot b}}

\begin{titlepage}
\vspace*{60pt}
 \centerline{\Large CONSTRUCTION OF FUZZY SPACES}
\vspace*{20pt} \centerline{\Large AND}
 \vspace*{20pt}
 \centerline{\Large THEIR APPLICATIONS TO MATRIX MODELS}
 \vspace*{40pt}
  \centerline{\Large by}
\vspace*{40pt} \centerline{\large YASUHIRO ABE}

\vspace*{180pt}

\centerline{A dissertation submitted to the Graduate Faculty in
Physics in partial fulfillment} \vspace{0pt} \centerline{of the
requirements for the degree of Doctor of Philosophy,}\vspace{0pt}
\centerline{The City University of New York}

\vfill \centerline{2006}
\end{titlepage}

\setcounter{page}{2}

\vspace*{380pt} \vfill
\newpage


\vspace{10pt} \centerline{\large Abstract}

\vspace{20pt} \centerline{CONSTRUCTION OF FUZZY SPACES}
 \vspace{10pt}
 \centerline{AND}
 \vspace{10pt}
  \centerline{THEIR APPLICATIONS TO MATRIX MODELS}
\vspace{10pt}
 \centerline{by}
 \vspace{10pt}
\centerline{Yasuhiro Abe}
 \vspace{10pt}
Advisor: $~~$ Professor V. Parameswaran Nair
 \vspace{10pt}

\noindent Quantization of spacetime by means of finite dimensional
matrices is the basic idea of fuzzy spaces. There remains an issue
of quantizing time, however, the idea is simple and it provides an
interesting interplay of various ideas in mathematics and physics.
Shedding some light on such an interplay is the main theme of this
dissertation. The dissertation roughly separates into two parts.
In the first part, we consider rather mathematical aspects of
fuzzy spaces, namely, their construction. We begin with a review
of construction of fuzzy complex projective spaces $\cp^k$
($k=1,2,\cdots$) in relation to geometric quantization. This
construction facilitates defining symbols and star products on
fuzzy $\cp^k$. Algebraic construction of fuzzy $\cp^k$ is also
discussed. We then present construction of fuzzy $S^4$, utilizing
the fact that $\cp^3$ is an $S^2$ bundle over $S^4$. Fuzzy $S^4$
is obtained by imposing an additional algebraic constraint on
fuzzy $\cp^3$. Consequently it is proposed that coordinates on
fuzzy $S^4$ are described by certain block-diagonal matrices. It
is also found that fuzzy $S^8$ can analogously be constructed.

In the second part of this dissertation, we consider applications of
fuzzy spaces to physics. We first consider theories of gravity on
fuzzy spaces, anticipating that they may offer a novel way of
regularizing spacetime dynamics. We obtain actions for gravity on
fuzzy $S^2$ and on fuzzy $\cp^2$ in terms of finite dimensional
matrices. Application to M(atrix) theory is also discussed. With an
introduction of extra potentials to the theory, we show that it also
has new brane solutions whose transverse directions are described by
fuzzy $S^4$ and fuzzy $\cp^3$. The extra potentials can be
considered as fuzzy versions of differential forms or fluxes, which
enable us to discuss compactification models of M(atrix) theory. In
particular, compactification down to fuzzy $S^4$ is discussed and a
realistic matrix model of M-theory in four-dimensions is proposed.

\tableofcontents


\chapter{Introduction}
\pagenumbering{arabic}

Studies of fuzzy spaces cross over a variety of concepts in
mathematics and physics. The basic idea of fuzzy spaces is to
describe compact spaces in terms of finite dimensional $(N\times
N)$-matrices such that they give a concrete realization of
noncommutative spaces \cite{Connes1,Madore1,Landi1,FG1}. Use
of fuzzy spaces in physics was suggested by Madore around 1992
\cite{mad1}. Since then, fuzzy spaces have been an active area of
research. Some of the earlier developments can be found in
\cite{Ydri1,BalRev,Stein1}. For recent reviews on fuzzy spaces,
one may refer to \cite{NairRev1,NairRev2,BKVrev}.

\section{Matrix realization of noncommutative geometry}

Definition of fuzzy spaces can be made from a framework of
noncommutative geometry initiated by Connes
\cite{Connes1,Connes2}, where it has been shown that the usual
differential calculus on a Riemannian manifold $\M$ can be
constructed by the so-called spectral triple $(\A , \H , \D)$;
$\A$ is the algebra of smooth bounded functions on $\M$, $\H$ is
the Hilbert space of square-integrable spinor functions on $\M$
(or sections of the irreducible spinor bundle) and $\D$ is the
Dirac operator on $\M$, carrying the information of metric and
Levi-Civita spin connection. With a slight modification of Connes'
idea, Fr\"{o}hlich and Gaw\c{e}dzki have also indicated that the
Riemannian geometry can be constructed by the abstract triplet
$(\A, \H, \Delta)$, where $\Delta$ is the Laplace-Beltrami
operator on $\M$ \cite{FG1}. Fuzzy spaces are then defined by a
sequence of triples \beq  ( Mat_{N}, \H_N , \Delta_N )
\label{1-1}\eeq where $Mat_{N}$ is a matrix algebra of $(N\times
N)$-matrices which act on the $N$-dimensional Hilbert space $\H_N$
and $\Delta_N$ is a matrix analog of the Laplacian. The inner
product of matrix algebra is defined by $\< A, B \> = \frac{1}{N}
\Tr (A^\dag B)$. The Laplacian $\Delta_N$ contains information of
metrical and other geometrical properties of $\M$. For example,
the dimension of the manifold $\M$ relates to the $N$-dependence
of the number of eigenvalues in $\Delta_N$.

Since fuzzy spaces are described by finite dimensional matrices,
due to the Cayley-Hamilton theorem, there is a natural cut-off on
the number of modes for matrix functions on fuzzy spaces. So one
can use fuzzy spaces to construct regularized field theories in
much the same way that lattice gauge theories are built. Various
interesting features of field theories on fuzzy spaces have been
reported; for example, existence of topological solutions such as
monopoles and instantons, appearance of the so-called UV-IR
mixing, and evasion of the fermion doubling problem which appears
in the lattice regularization. For these and other aspects of
fuzzy spaces, one may refer to \cite{GKP1}-\cite{Sasa} and, in
particular, to \cite{BKVrev} for a review.

\section{Relation to geometric quantization}

Construction of fuzzy spaces is closely related to quantization
programs in the construction of quantum Hilbert spaces from
classical phase spaces. It is known that there exist different
quantization schemes such as canonical quantization and functional
integral (or path integral) quantization. In either case, the
quantum theory is described by a unitary irreducible
representation (UIR) of the algebra of symmetry on a Hilbert
space.\footnote{The Hilbert space structure may not be apparent in
the path integral approach, where one is interested in computation
of correlation functions or S-matrices, however, it is in general
possible to define the quantum theory in terms of a UIR of the
operator algebra on a Hilbert space.} Physical observables are
given by hermitian operators which generate unitary
transformations on the Hilbert space. In the classical theory, the
operators correspond to functions on a phase space which generate
canonical transformations. The basic idea of quantization is to
have a correspondence between the algebra of Poisson brackets
represented by functions on a phase space or a symplectic manifold
$\M_s$ and the algebra of commutation rules represented
irreducibly by operators on a Hilbert space $\H$. The hermitian
operators can be represented by ($N\times N$)-matrices where $N$
is the dimension of the Hilbert space $\H$. From this point of
view, the quantization programs are essentially equivalent to the
construction of fuzzy spaces. The matrix version of Laplacian
$\Delta_N$ in (\ref{1-1}) can be obtained as a double commutator.
This is observed as follows; consider Heisenberg commutation rules
of quantum mechanics $[\hat{x}, \hat{x}]=[\hat{p} , \hat{p}]=0$,
$[\hat{x}, \hat{p}]=i$ with $\hat{x}\psi = x \psi$, $\hat{p} \psi
= -i \frac{\d}{\d x} \psi$ where $\psi (x)$ is a wavefunction,
then the Laplacian is expressed as $\Delta f(\hat{x}) = -
[\hat{p}, [\hat{p}, f(\hat{x})]]$ where $f(\hat{x})$ is a function
of $\hat{x}$. Note that the wavefunction depends only on $x$
instead of $(x,p)$. This is necessary to have an irreducible
representation of the operator algebra. It is also related to the
notion of polarization or holomorphic condition in a framework of
geometric quantization.

Geometric quantization would be a mathematically more rigorous
quantization scheme \cite{Snia,Wood,Pere,NairBook}. It turns out
to be very useful in quantizing many systems, including the
Chern-Simons theory \cite{QCS,Hu}. In geometric quantization, one
considers the so-called prequantum line bundle which is a line
bundle on a phase space. The curvature or the first Chern class of
the line bundle can naturally be chosen as a symplectic two-form
$\Om_{s}$. By use of the line bundle, one can show an explicit
correspondence between the algebra of Poisson brackets and the
algebra of commutators. The upshot of geometric quantization is
that a quantum Hilbert space is given by sections of a `polarized'
line bundle. The above wavefunction $\psi (x)$ corresponds to this
polarized line bundle, while unpolarized one would lead to a
function $\psi (x,p)$. Usually, we impose a complex structure on
the phase space and identify the symplectic two-form as a
K\"{a}hler form or some multiple thereof. In this case, the
easiest polarization condition to use is a holomorphic condition
on a complex line bundle. This gives what is known as the
K\"{a}hler polarization. The idea of forming a Hilbert space as
holomorphic sections of a complex line bundle was in fact
exploited in the studies of representation of compact Lie groups
by Borel, Weil and Bott. They showed that, for any compact Lie
group $G$, all UIR's of $G$ are realized by holomorphic sections
of a complex line bundle on a coset space $G/T$, where $T$ is the
maximal torus of $G$ and $G/T$ is proven to be a K\"{a}hler
manifold. (The group $G$ acts on the space of holomorphic
sections, or a Hilbert space, as right translation.) For detailed
description of this Borel-Weil-Bott theory or theorem, one may
refer to \cite{Snia,Wood,Pere,Hu}.

Utilizing a quantization program, we can obtain a finite
dimensional Hilbert space $\H_N$ in (\ref{1-1}) for any compact
symplectic manifold $\M_{s}$. The matrix algebra $Mat_N$ is given
by the algebra of operators acting on $\H_N$. As mentioned
earlier, the Laplacian $\Delta_N$ in (\ref{1-1}) is naturally
obtained upon the determination of $Mat_N$. Construction of fuzzy
spaces is therefore implemented by quantization of compact
symplectic manifolds. A family of such manifolds is given by the
so-called co-adjoint orbits of a compact semi-simple Lie group
$G$. (For semi-simple Lie groups, there is no difference between
co-adjoint and adjoint orbits.) It is known that the co-adjoint
orbits can be quantized when their symplectic two-forms satisfy a
Dirac-type quantization condition. For quantization of co-adjoint
orbits, one may refer to \cite{Snia,Wood,Pere}. The co-adjoint
orbit of a compact semi-simple Lie group $G$, with its Lie algebra
being $\underline{G}$, is given by $\{g t g^{-1}:  g \in G \}$
where $t \in \underline{G}$. The co-adjoint orbit is then
considered as a coset space $G/H_t$ where $H_t$ is a subset of $G$
defined by $H_t = \{g \in G: [g,t]=0 \}$. When $H_t$ coincides
with the maximal torus of $G$, the co-adjoint orbit becomes the
above mentioned space $G/T$, This space, known as a flag manifold,
has the maximal dimension of the co-adjoint orbits, {\it i.e.},
$\dim G - \rank G$. An example of such a space is
$\frac{SU(3)}{U(1)\times U(1)}$ where $t$ is given by $t \sim
\diag (1,-1,0)$ corresponding to $\la_3$ in terms of the Gell-Mann
matrices $\la_a$ ($a=1,2,\cdots, 8$) for $\underline{SU(3)}$. When
$t$ has degeneracy, the co-adjoint orbits are called degenerate
and their dimensions are given by $\dim G - \dim H_t$. An example
is $SU(3)/U(2)$ with $t \sim \la_8 \sim \diag (1,1,-2)$. This
coset is equivalent to the four-dimensional complex projective
space $\cp^2$. Since we are interested in a finite dimensional UIR
of $G$, the compact group $G$ is to be chosen as $U(n)$ or its
subgroup. In this case, the generator $t$ always includes a
$\underline{U(1)}$ element of $\underline{U(n)}$. Consequently,
the subset $H_t \subset G$, known as the stabilizer of $t$,
contains the $U(1)$ element of $U(n)$. This is a fact of some
significance particularly in considering gauge theories on fuzzy
spaces.

In quantizing the co-adjoint orbit $G/H_t$, the Hilbert space is
given by holomorphic sections of a complex line bundle over
$G/H_t$. The holomorphic sections correspond to a UIR of $G$. (The
holomorphicity allows the extension of the $G$-action to a
$G^{{\bf C}}$-action, where $G^{{\bf C}}$ is the complexification
of $G$. Note that any compact group can be complexified; this is
known as Chevalley's complexification of compact Lie groups.) We
can now make direct use of geometric quantization to construct the
fuzzy version of $G/H_t$. In fact, fuzzy spaces which have been
constructed so far, to be consistent with the definition of
(\ref{1-1}), all fit into this class of coset spaces. Namely, they
are fuzzy $S^2 = SU(2)/U(1)$, fuzzy $\cp^2 = SU(3)/U(2)$ and fuzzy
$\cp^k = SU(k+1)/U(k)$ $(k=1,2,3,\cdots )$ in general
\cite{mad1,bal1,bal2}.

A detailed construction of fuzzy $\cp^k$ in the same spirit as
geometric quantization has been carried out by Karabali and Nair
\cite{KaraNair1,nair4,nair5}, where the complex line bundle over
$\cp^k = SU(k+1)/U(k)$ is expressed in terms of the Wigner
$\D$-functions for $SU(k+1)$ which, by definition, give a UIR of
$SU(k+1)$. Symbols and star products, notion of functions and
their product algebra in commutative space mapped from
noncommutative counterparts, are explicitly defined in terms of
the $\D$-functions. In the next chapter, we shall recapitulate
these results.

For those manifolds that do not have a symplectic structure, there
exist no quantization schemes. This is the main reason for the
difficulty encountered in construction of odd-dimensional fuzzy
spaces and fuzzy spheres of dimension higher than two.
Construction of higher dimensional fuzzy spheres has been proposed
in \cite{ram2,ram1,kim2,oco1,dol1,dop}
\footnote{
There is also another type of construction with extra constraints
that are expressed in terms of Nambu brackets \cite{jabbari}. The purpose
of this construction is for an application to a matrix model of M-theory.
This construction reveals interesting features in M-theory, particularly 
in brane solutions to M-theory, however, strictly speaking, it does not lead 
to pure spheres in a commutative limit due to the Nambu-bracket constraints.
}.
Each proposal starts from
a co-adjoint orbit such as $\frac{SO(2k+1)}{U(k)}$,
$\frac{SO(k+2)}{SO(k)\times SO(2)}$ ($k=1,2,\cdots $). Factors
irrelevant to the sphere in such a co-adjoint orbit are projected
out in a sort of brute-force way. As a result, the resulting fuzzy
spheres break either associativity or closure of the algebra.
These fuzzy spheres are therefore not compatible with the
definition (\ref{1-1}) where fuzzy spaces are defined by the
matrix algebra on a finite dimensional Hilbert space. One way to
avoid this problem is to impose an extra constraint on a Lie
algebra of $G$ so that the co-adjoint orbit $G/H_t$ (or its
multiple) globally defines a sphere under the algebraic extra
constraint. This is a natural prescription for proper construction
of fuzzy spheres because functions on fuzzy spaces are described
by matrix representation of the algebra $\underline{G}$. The fuzzy
spheres are embedded in ${\bf R}^{\dim G}$ and its algebra is a
subset of $\underline{G}$, preserving closure and associativity.
It is by use of this idea of introducing an extra constraint that
fuzzy $S^3/ {\bf Z}_2$ is constructed from fuzzy $S^2 \times S^2$
in \cite{nair1}. The same idea proves to be applicable to
construction of fuzzy $S^4$ from fuzzy $\cp^3$ \cite{Abe1}. This
construction utilizes the fact that $\cp^3$ is an $S^2$-bundle
over $S^4$, or a Hopf fibration of $S^7$ as an $S^3$-bundle over
$S^4$. Utilizing a Hopf fibration of $S^{15}$ as an $S^7$-bundle
over $S^8$, one can similarly construct fuzzy $S^8$ from fuzzy
$\cp^7$ with some algebraic constraint. In chapter 3, we shall
discuss construction of higher dimensional spheres along these
lines, focusing on the case of fuzzy $S^4$.

\section{Applications to physics}

The fact that co-adjoint orbits are given by coset spaces is
important in application of fuzzy spaces to physics. The coset
space $G/H$ naturally gives rise to an interpretation of $G$ as an
$H$-bundle over $G/H$ or more generally a sum of $H^{(i)}$-bundles
over $G/H$, with $H$ being a direct product of $H^{(i)}$'s
($i=1,2, \cdots$). As mentioned earlier, $H$ always contains a
$U(1)$ group, so at least one of the $H^{(i)}$'s can be identified
as $U(1)$. The corresponding $U(1)$-bundle gives a complex line
bundle whose holomorphic sections are, as discussed earlier,
regarded as a Hilbert space $\H_N$. There is an interesting
correspondence between $\H_N$ and the Hilbert space of the lowest
Landau level, which is a restricted energy level for charged
particles in a strong magnetic field. Physical observables in such
a system are projected onto the lowest Landau level. As a result,
they acquire noncommutativity and it is possible to identify them
with the observables on fuzzy spaces. (For further description of
this correspondence, see a recent review \cite{NairRev1}; for the
Landau problem and its relation to fuzzy sphere and more general
Riemann surfaces, see \cite{NairPoly1,KNP,poly}.) In this context,
the $U(1)$-bundle is understood as a magnetic monopole-bundle over
$G/H$ whose holomorphic sections give wavefunctions on the lowest
Landau level in $G/H$. Note that the Landau problem was originally
considered on ${\bf R}^2$ but it can naturally be extended to
higher dimensional curved (coset) spaces. When $H^{(i)}$ is a
non-abelian group, we have a non-abelian vector bundle over $G/H$.
Physically this corresponds to the presence of a non-abelian
background magnetic field.

There is a series of remarkable results in the study of the edge
excitations of quantum Hall droplets on the lowest Landau level in
$\cp^k$ \cite{KaraNair1,nair4,nair5}. Here we simply state these
results. In \cite{nair4} it is shown that an effective action for
the edge excitations in a $U(1)$ background magnetic field is
given by a chiral bosonic action in the limit of a large number of
edge states. The action can be interpreted as a generalization of
a chiral abelian Wess-Zumino-Witten (WZW) theory. With a
non-ablian $U(k)$ background magnetic field, the effective action
for the edge excitations leads to a chiral and gauged WZW theory
generalized to higher dimensions, also in the limit of a large
number of edge states \cite{nair5}. (For uses of fuzzy spaces in
the quantum Hall systems, see also \cite{HaseKim}.)

The gauge principle is probably the most important concept in
physics in a sense that it provides a unified view of all physical
interactions, including gravity. The gauge principle means the
invariance of physical quantities under local frame
transformations. As is well-known, fibre bundles are the
mathematical framework for the local or gauge symmetries. The
concept of fibre bundles is then useful in understanding the
geometrical and topological properties of gauge theories. Fibre
bundles also provide a natural setting for all physical fields.
Matter fields are sections of various vector bundles over
spacetime manifold, with the fibre being complex numbers or
spinors of the Lorentz group. Gauge fields are connections on
these vector bundles. Connections of tangent bundles over
spacetime give Christoffel symbols, which lead to the metric and
spin connections and, eventually, the theory of gravity.

As mentioned above, bundle structures naturally arise in fuzzy
spaces. Like in ordinary commutative spaces, gauge fields on fuzzy
spaces are defined by `fuzzy' covariant derivatives. In a fuzzy
$G/H$-space, derivative and coordinate operators obey the same
algebra $\underline{G}$, so they are identical. This is related to
the fact that co-adjoint and adjoint orbits are equivalent for a
compact semi-simple group $G$. One can then regard the covariant
derivatives as `covariant' coordinates on fuzzy spaces. In this
sense, the gauge fields are considered as fluctuations from fuzzy
spaces. Gauge theories on noncommutative spaces in general have
been received a lot of interest \cite{BKVrev,NC,Harvey,Hama}. This
is partly motivated by the discovery that noncommutative spaces
can arise as solutions in string and M-theories. The solutions are
known as D-branes or simply branes, corresponding to
non-perturbative objects in string theories \cite{Polch}. Later we
shall consider such objects in relation to fuzzy spaces.
Application of noncommutative geometry to gauge theories was in
fact initiated by Connes and others \cite{Connes1,Madore1,Landi1}.
Part of their motivation is to understand the standard model of
particle physics (and the involving Higgs mechanism) in a more
mathematical framework, namely, in terms of the spectral triple
$(\A , \H , \D)$ \cite{ConLot,Kastl1}. There is also a series of
developments in construction of gravitational theories in terms of
the spectral triple
\cite{Cham1,Dopli,KalWal,Kastl2,ChamCon,Landi3}.

Gauge fields which describe gravitational degrees of freedom ({\it
i.e.}, frame fields and spin connections) on fuzzy spaces are
particularly interesting, since they would offer a regularized
gravity theory as a novel alternative to the Regge calculus or
triangulation of spaces, which is essentially the only finite-mode
truncation of gravity, preserving the notion of diffeomorphism.
Gravitational fields on a fuzzy $G/H$-space are given by hermitian
($N\times N$) matrices. The matrices have an invariance under
$U(N)$ transformations which is usually imposed in any hermitian
matrix models. The matrix elements of functions on a fuzzy space
correspond to the coefficients in a harmonic expansion of
truncated functions on the corresponding commutative space. This
so-called matrix-function correspondence implies the $U(N)$
invariance as a fuzzy analog of the coordinate invariance or the
diffeomorphism.

The gauge group of the gravitational fields on commutative
Euclidean spacetime is given by a combination of translational and
rotational space-time symmetries on the tangent frame. In ordinary
flat space, this group is the Poincar\'{e} group. However, in the
$G/H$-space, the Poincar\'{e} group is replaced by the compact
semi-simple group $G$. The stabilizer $H$, which we consider as a
subgroup of $G$ in what follows, corresponds to the Lorentz group
so that the translations on $G/H$ are represented by
$\underline{G} - \underline{H}$. Theories of gravity on such an
even-dimensional (coset) space have been studied in connection
with topological gauge theories. For example, an action for
two-dimensional gravity is given by the Jackiw-Teitelboim action
\cite{jackiw}. One can also construct a physically more
interesting case, {\it i.e.}, an action for gravity on
four-dimensional spacetime, following Chang, MacDowell and
Mansouri \cite{CMM}. As mentioned earlier in the context of the
construction of fuzzy spaces, the group $G$ is seen as a compact
group embedded in $U(n)$. So one can consider the existence of
$U(k)$ ($k \le n$) such that $G \subseteq U(k) \subset U(n)$. In
noncommutative spaces, gauge groups should contain a $U(1)$
element, otherwise one cannot properly define a noncommutative
version of curvature or field strength. A natural choice of the
gauge group on the fuzzy $G/H$-space is therefore the $U(k)$
group. It is based on these arguments that a
Chang-MacDowell-Mansouri (CMM) type action for gravity on
even-dimensional noncommutative spaces has been proposed by Nair
in \cite{NairNCg}. (For some of the other approaches to
noncommutative gravity, one may refer to \cite{Cham2}-\cite{Wali};
for a matrix model of gravity on fuzzy $S^2$ in particular, see
\cite{AbeNair1,Valt2}.) In \cite{AbeNair1} the CMM type action is
applied to fuzzy $S^2$ as well as fuzzy $\cp^2$ and actions for
gravity in terms of ($N\times N$) matrices are obtained. The
action on fuzzy $S^2$ reduces to the Jackiw-Teitelboim action on
$S^2$ in the large $N$ limit. We shall present these results in
chapter 4.

Fuzzy spaces are in principle constructed for any even-dimensional
symplectic manifolds. Restriction to the number of dimensions
should come from physical reasonings. One convincing reason is the
matrix model of M-theory or the M(atrix) theory proposed by Banks,
Fischler, Shenker and Susskind \cite{BFSS}. For a review of
M(atrix) theory, one may refer to \cite{Review1}. In M(atrix)
theory, nine dimensions out of eleven are described by $(N\times
N)$ matrices, being referred to the transverse directions. Brane
solutions are then described by fuzzy spaces as far as the
transverse directions are concerned. Solutions with matrix
configurations of $S^2$, $S^4$ and $\cp^2$ geometries are
known to exist \cite{KT1,Rey,CLT,Nair1}; they are respectively
called spherical membranes, spherical longitudinal five-branes and
longitudinal five branes of $\cp^2 \times S^1$ geometry. Note that
when the solutions involve the longitudinal directions, as opposed
to the transverse ones, they are called longitudinal branes. It is
known that there exist longitudinal five-brane solutions in
M(atrix) theory \cite{BD,GRT,BSS}. But brane solutions of
dimension higher than five are excluded due to energy
consideration \cite{BSS}. Details of these points are discussed in
chapter 5, where we also consider the emergence of longitudinal
seven-branes of $\cp^3 \times S^1$ geometry, introducing extra
potentials to the M(atrix) theory Lagrangian. For related analyses
on fuzzy spaces as brane solutions, one may refer to
\cite{TriVai,Bal3}.

There is another version of matrix model corresponding to type IIB
string theory proposed by Ishibashi, Kawai, Kitazawa and Tsuchiya
\cite{IKKT}. For a reveiw of this model, see \cite{Review2}. This
IIB matrix model also has solutions described by fuzzy spaces
\cite{Iso3,Kita1,Imai,Azuma1,Kita2}. Fuzzy spaces, or finite
dimensional matrix realization of spacetime, are suitable for
numerical simulations. There is a series of numerical studies on
certain fuzzy spaces appearing in a generalized IIB matrix model
\cite{ABNN}. For a different type of simulation, see also
\cite{Xavi,Panero}.

In terms of M(atrix) theory, the number of dimensions for fuzzy
spaces arising as transverse branes is restricted to 2, 4, 6 and
8. (We omit odd dimensions here because they do not lead to a
symplectic structure, but they may be possible as shown in
\cite{nair1}.) When the dimension is higher than four, we are
faced with higher dimensional brane solutions. These can be
interpreted either as extended physical objects along the lines of
a brane-world scenario \cite{RanSun}, or as bundles over
four-dimensional spacetime. In the former case, extra dimensions
are somehow allowed to exist and one can use Kaluza-Klein type
compactification to discuss their effects on spacetime.  In the
latter case, the extra dimensions are relevant to internal
symmetries or a fibre. A typical example is Penrose's twsitor
space $\cp^3$ which is an $S^2$-bundle over (compact) spacetime
$S^4$ \cite{penrose}. In this context, fuzzy $\cp^3$ is quite
interesting in application to physics. (It has also been useful in
construction of fuzzy $S^4$ \cite{Abe1}.) For a recent development
in connection with this idea, one may refer to \cite{LechSae}.

\vspace{0.3cm}

The rest of this dissertation is organized as follows. In chapter
2, we briefly review construction of fuzzy $\cp^k$ ($k=1,2,\cdots
$), following \cite{NairRev1,NairRev2}. We rephrase known results
such that relation to geometric quantization is transparent. In
this chapter, we also present algebraic construction of fuzzy
$\cp^k$ by use of creation and annihilation operators on a Hilbert
space \cite{BKVrev}. We follow the presentation given in an
appendix of \cite{Abe1}. In chapter 3, we review construction of
fuzzy $S^4$, following also \cite{Abe1}. Chapter 4 is devoted to
application of fuzzy spaces to theories of gravity. We shall
obtain Chang-MacDowell-Mansouri type matrix models for gravity,
following the work of \cite{NairNCg,AbeNair1}. Chapter 5 deals
with application of fuzzy spaces to M(atrix) theory based on a
recent work \cite{Abe2}. Finally, in chapter 6 we present brief
conclusions.

\chapter{Construction of fuzzy $\cp^k$}

\section{Hilbert space}

A finite dimensional Hilbert space $\H_N$ for fuzzy $\cp^k =
SU(k+1)/U(k)$ ($k=1,2,\cdots$) is given by holomorphic sections of
a complex line bundle over $\cp^k$. As discussed in section 1.2,
the holomorphic sections of the complex line bundle should
correspond to a unitary irreducible representation (UIR) of
$G=SU(k+1)$. Representation of $SU(k+1)$ $(k \ge 2)$ is given by a
general form $(p,q)$ ($p,q = 0,1,2,\cdots $) if we use a standard
tensor method. Notion of holomorphicity in the representation of
$G$ can be realized by totally symmetric part of the
representation, {\it i.e.}, $(n,0)$, where $n$ is the rank of the
representation ($n=1,2,\cdots $). The other totally symmetric
representation $(0,n)$ corresponds to antiholomorphic part of the
$SU(k+1)$ representation and the $(p,p)$-representation gives real
representation. For $SU(2)$ (corresponding to $k=1$), the
representation is given by a single component, say $(p)$, so there
is no real representation. (Because of this, the $SU(2)$
representation is sometimes called pseudo-real.) The dimension of
$\H_N$ is then determined by that of the $(n,0)$-representation
for $SU(k+1)$; \beq N^{(k)} \equiv \dim (n,0) =
\frac{(n+k)!}{k!~n!}  . \label{2-1a}\eeq Consequently, matrix
algebra of fuzzy $\cp^k$ is realized by $N^{(k)} \times
N^{(k)}$-matrices. Operators or matrix functions on fuzzy $\cp^k$
are expressed by linear combinations of $N^{(k)} \times
N^{(k)}$-matrix representations of the algebra of $SU(k+1)$ in the
$(n,0)$-representation. Let $L_A$, with $A=1,2,\cdots, k^2+ 2k =
\dim SU(k+1)$, denote such matrix representations. We need to
impose extra constraints on them otherwise the Hilbert space is
defined simply on $R^{k^2 +2k}$ without any information of
$\cp^k$. As we shall discuss later, such extra constraints can be
imposed at an algebraic level in terms of $L_A$ but, in order to
construct $\H_N$ along a program of geometric quantization, we
would rather consider a holomorphic line bundle on $\cp^k$ first
and implement the relevant extra constraints in it.

To begin with, we write down a holomorphic $U(1)$ bundle
$\Psi^{(n)}_{m}$ as \beqar \Psi^{(n)}_{m} (g) &=& \sqrt{N^{(k)}}
\D^{(n,0)}_{m N^{(k)}} (g) \, , \label{2-1a1}\\
 \D^{(n,0)}_{m N^{(k)}} (g) &=&
\< (n,0), m | \hat{g} | (n,0), N^{(k)} \> \label{2-1a2} \eeqar
where $|(n,0), m \>$ ($m=1,2,\cdots, N^{(k)}$) denote the states
on the Hilbert space $\H_N$, $|(n,0), N^{(k)} \>$ is the highest
or lowest weight state, $g$ is an element of $G=SU(k+1)$ and
$\hat{g}$ is a corresponding operator acting on these states.
$\D^{(n,0)}_{m N^{(k)}} (g)$ is known as Wigner $\D$-functions for
$SU(k+1)$ in the $(n,0)$-representation. The lower indices label
the states of this representation, allowing us to interpret the
$\D$-functions as matrix elements. As mentioned in chapter 1, $G$
acts on the Hilbert space as right translation. Let $R_A$ denote
the right-translation operator on $g$; \beq
 R_A ~g~ =~ g~ t_A \label{2-1b1}
\eeq where $t_A$ are the generator of $G$ in the fundamental
representation $(1,0)$. The element $g$ is given by $g = \exp (i
t_A \th^A) $ with continuous parameters $\th^A$. We now consider
the splitting of $t_A$'s to those of $U(k)=SU(k) \times U(1)$
subalgebra and the rest of them, {\it i.e.}, those relevant to
$\cp^k$. Let $t_{j}$ ($j=1,2,\cdots, k^2$) and $t_{k^2 +2k}$
denote the generators of $U(k) \subset SU(k+1)$, $t_{k^2 +2k}$
being a $U(1)$ element of the $U(k)$, and let $t_{\pm i}$
$(i=1,2,\cdots, k)$ denote the rest of $t_A$'s. One can consider
$t_{\pm i}$ as a combination of rasing-type ($t_{+ i}$) and
lowering-type ($t_{-i}$) operators acting on the states of $\H_N$.
Choosing $|(n,0), N^{(k)} \>$ to be the lowest weight state, we
then find \beqar R_{j} \D^{(n,0)}_{m N^{(k)}} (g) &=& 0
~~ (j=1,2,\cdots , k^2) , \label{2-1b2} \\
R_{k^2 + 2k} \D^{(n,0)}_{m N^{(k)}} (g) &=& -
\frac{nk}{\sqrt{2k(k+1)}} \D^{(n,0)}_{m N^{(k)}} (g) \, ,
\label{2-1b3} \\
R_{-i} \D^{(n,0)}_{m N^{(k)}} (g) &=& 0 \, . \label{2-1b4} \eeqar
Equations (\ref{2-1b2}) and (\ref{2-1b3}) indicate that
$\Psi^{(n)}_{m}(g) \sim \D^{(n,0)}_{m N^{(k)}} (g)$ is a $U(1)$
bundle over $\cp^k$. One can also check that under the $U(1)$
transformations, $g \rightarrow gh$ with $h= \exp (i t_{k^2+2k}
\th)$, $\th \equiv \th^{k^2 +2k}$, $\Psi^{(n)}_{m}(g)$ transforms
as \beq \Psi^{(n)}_{I}(g) \rightarrow \Psi^{(n)}_{m}(gh) =
\Psi^{(n)}_{m}(g) \exp \left( -i\frac{nk}{\sqrt{2k(k+1)}} \th
\right) \, .\label{2-1b5} \eeq  Note that we use the fact that the
states in the $(n,0)$-representation is constructed by products of
the states in the $(1,0)$-representation. We also use a
conventional choice of $t_{k^2+2k}$ as \beq t_{k^2 + 2k} =
\frac{1}{\sqrt{2k(k+1)}} \diag (1,1,\cdots, 1, -k) \,
.\label{2-1b6} \eeq

In terms of geometric quantization, equation (\ref{2-1b4})
corresponds to the polarization condition on a prequantum $U(1)$
bundle. The Hilbert space is therefore constructed as sections of
the holomorphic $U(1)$ bundle $\Psi^{(n)}_{m}$. The
square-integrability of $\H_N$ is guaranteed by the orthogonality
condition of the Wigner $\D$ function; \beq
 \int d \mu (g) \D^{* (R)}_{m,k} (g) \D^{(R')}_{m',k'} (g) =
 \del^{RR'} \frac{\del_{mm'}\del_{kk'}}{\dim R} \label{2-1b7}
\eeq where $\D^{*(R)}_{m,k} (g)= \D^{(R)}_{k,m} (g^{-1})$, $R$
denotes the representation of $G=SU(k+1)$, and $d \mu(g)$ is the
Haar measure of $G=SU(k+1)$ normalized to unity; $\int d \mu(g) =
1$. The orthogonality condition of our interest is given by \beq
 \int d \mu (g) \D^{* (n,0)}_{m,N^{(k)}} (g) \D^{(n,0)}_{m',N^{(k)}} (g) =
 \frac{\del_{mm'}}{N^{(k)}} \, . \label{2-1b8}
\eeq The normalization factor $\sqrt{N^{(k)}}$ in (\ref{2-1a1}) is
determined by this relation, which also provides a natural
definition of the inner product of $\Psi^{(n)}_{m}$. Note that the
integrand is invariant under $U(k)$, so we may use the Haar
measure of $SU(k+1)$ for the integration over $\cp^k$.

The K\"{a}hler two-form (or, equivalently, the symplectic
structure) of $\cp^k$ in terms of $g$ is obtained as follows. As
in (\ref{2-1b1}), $g \in G=SU(k+1)$ is considered as a $(k+1)
\times (k+1)$ matrix. In order to obtain coordinates on
$\cp^k=SU(k+1)/U(k)$ out of $g$, we need to impose the
identification $g \sim gh $ where $h \in H=U(k)$. Such subgroup
elements $h$ can be represented by $t_{k^2 +2k}$ in (\ref{2-1b6})
and
 \beq h_{SU(k)} = \left(%
\begin{array}{cc}
  h_k & 0 \\
  0 & 1 \\
\end{array}%
\right) \label{2-1c1} \eeq where $h_k$ is a $(k\times k)$-matrix.
The coordinates on $\cp^k$ are then defined by matrix elements
$g_{\al, k+1}$ ($\al = 1,2,\cdots , k+1$). Since $g^\dag g =1$, we
have $g^{*}_{k+1, \al} g_{\al , k+1} =1$. We now introduce the
notation $u_{\al}\equiv g_{\al, k+1}$, $\bar{u} \cdot u =1$. In
terms of $u_\al$'s, the Wigner $\D$-functions (\ref{2-1a2}) are
written in a form of $\D^{(n)} \sim u_{\al_1} u_{\al_2} \cdots
u_{\al_n}$. Homogeneous complex coordinates of $\cp^k$ are defined
by $Z =(z_1, z_2, \cdots , z_k)^{T}$ with $Z \sim \la Z$, where
$\la$ is a nonzero complex number and $T$ denotes transposition of
the vector or $(1 \times k)$-matrix. $u_{\al}$'s are related to
$Z$ by
 \beq
u_{\al} = \frac{1}{\sqrt{1+ \bz \cdot z}} (1 , z_1, z_2, \cdots,
z_k)^{T}.
 \label{2-1c2} \eeq
Using $u_{\al}$, one can construct a one form
 \beqar A &=& - i ~ u^{*}_{\al} d u_{\al}
\nonumber \\
 &=& - \frac{i}{2} \left(
 \frac{\bu \cdot du - d \bu \cdot u}{\bu \cdot u} \right)
\label{2-1c3} \eeqar where $\bu \cdot du = \bu^\al du_\al$, etc.
The K\"{a}hler two-form can be identified with $d A \sim
du^{*}_{\al} du_{\al}$, since it is closed but it is not exact.
Explicitly, the K\"{a}hler form $\Om$ is written as \beqar \Om &=&
 - i \left( \frac{d \bar{u} \cdot du}{\bar{u} \cdot u} -
\frac{d \bar{u} \cdot u  \bar{u} \cdot du}{(\bar{u} \cdot u)^2}
\right) \nonumber\\
&=& - i \left( \frac{d \bz_i dz_i}{1 + \bz \cdot z} - \frac{d \bz
\cdot z \bz \cdot dz_i}{(1 + \bz \cdot z)^2} \right).
\label{2-1c4}\eeqar

The one-form (\ref{2-1c3}) is also expressed as \beq A ~ =~ i~
\sqrt{\frac{2k}{k+1}} ~ \tr (t_{k^2 +2k} g^{-1} d g).
\label{2-1d1} \eeq This form suggests a general way to obtain a
symplectic structure for a co-adjoint orbit defined by $\{g t
g^{-1}:  g \in G \}$ where $t \in \underline{G}$, with
$\underline{G}$ denoting the algebra of a compact and semi-simple
group $G$. Namely, we start from a one-form, $A \sim \tr (t g^{-1}
d g)$, and then the symplectic two-form $\Om_s$ is given by $\Om_s
= d A$. When $t$ has degeneracy as in (\ref{2-1b6}), the
co-adjoint orbit is called degenerate. In our case, the stabilizer
$H_t$, defined by $[H_t , t] = 0$, $H_t \subset G$ as in chapter
1, becomes the $U(k)$ subgroup of $G=SU(k+1)$. While $t$ does not
have degeneracy, the stabilizer becomes the maximal torus of $G$.
In this case, the co-adjoint orbit also has K\"{a}hler structure
and $dA$ gives its K\"{a}hler form.

\section{Symbols and star products}

We define the symbol of a matrix operator $A_{ms}$ ($m,s =
1,2,\cdots, N^{(k)}$) on the Hilbert space of fuzzy $\cp^k$ by
 \beqar \< \hat{A} \>
&\equiv & \sum_{ms} ~ \D_{m,N^{(k)}}^{(n,0)} (g) ~ A_{ms} ~
\D^{*(n,0)}_{s,N^{(k)}}(g) \nonumber \\
&=& \< (n,0), N^{(k)}| \hat{g}^{T} \hat{A} \hat{g}^{*}
|(n,0),N^{(k)} \>
 \label{2-2a1} \eeqar
The star product of fuzzy $\cp^k$ is defined by $\< \hat{A}
\hat{B} \> \equiv \< \hat{A} \> * \< \hat{B} \>$. From
(\ref{2-2a1}), $\< \hat{A} \hat{B} \>$ can be written as \beqar \<
\hat{A}\hat{B} \> \! &=& \!\! \sum_{msr} A_{mr} B_{rs} ~
\D^{(n,0)}_{m,N^{(k)}} (g)
\D^{*(n,0)}_{s,N^{(k)}}(g) \nonumber \\
\! &=& \! \! \! \sum_{msrr'p} \! \D_{m,N^{(k)}}^{(n,0)}(g) A_{mr}
\D_{r,p}^{*(n,0)}(g) ~ \D_{r',p}^{(n,0)}(g)  B_{r's}
\D_{s,N^{(k)}}^{*(n,0)} (g) \label{2-2a2} \eeqar where we use the
relation
\beq
	\sum_{p} \D_{r,p}^{*(n,0)}(g) \D_{r',p}^{(n,0)}(g)=
	\del_{rr'} \, .
	\label{2-addition01}
\eeq
In the sum over $p = 1,2, \cdots, N^{(k)}$ on the
right hand side of (\ref{2-2a2}), the term corresponding to
$p=N^{(k)}$ gives the product $\< \hat{A} \>  \< \hat{B} \>$. The
terms corresponding to $p < N^{(k)}$ may be expressed in terms of
the raising operators $R_{+i}$ ($i=1,2,\cdots, k$) as
 \beq \D_{r',p}^{(n,0)}(g) = \sqrt{ \frac{(n-s)!}{n! ~
i_1 ! i_2 ! \cdots i_k !} } R_{+1}^{i_1}R_{+2}^{i_2} \cdots
R_{+k}^{i_k} \D_{r',N^{(k)}}^{(n,0)}(g) \label{2-2a3} \eeq where
$s = i_1 + i_2 + \cdots +i_k$ and the state $|(n,0),p \>$ is
specified by \beq R_{k^2 +2k} \D_{r',p}^{(n,0)}(g) ~=~ \frac{-nk
+sk + s}{\sqrt{2k(k+1)}} ~ \D_{r',p}^{(n,0)}(g) \, . \label{2-2a4}
\eeq Since $R_{+i} \D_{s,-n}^{*(n)} = 0$, we can also write \beqar
\sum_{r's} \left[ R_{+i} \D_{r',N^{(k)}}^{(n,0)}(g) \right] B_{r'
s}  \D_{s,N^{(k)}}^{*(n,0)}(g)
 &=& \sum_{r's}
  \left[ R_{+i}  \D_{r',N^{(k)}}^{(n,0)}(g)
 B_{r' s}  \D_{s,N^{(k)}}^{*(n,0)}(g) \right] \nonumber \\
 &=&  R_{+i} \< \hat{B} \> \, .
 \label{2-2a5} \eeqar The conjugate of (\ref{2-2a3}) can be written in
terms of $R_{-i}$ by use of the relation $R^{*}_{+i} = - R_{-i}$.
Combining (\ref{2-2a3})-(\ref{2-2a5}), we can express
(\ref{2-2a2}) as \beqar \< \hat{A} \hat{B} \> &=& \sum_{s=0}^{n}
(-1)^s \frac{(n-s)!}{n! s!} \sum^{n}_{i_1
 + i_2 + \cdots + i_k =s} \frac{s!}{_1 ! i_2 !
\cdots i_k !} \nonumber\\
&& ~~ \times  R_{-1}^{i_1}R_{-2}^{i_2} \cdots R_{-k}^{i_k}~ \<
\hat{A} \> ~ R_{+1}^{i_1}R_{+2}^{i_2}
\cdots R_{+k}^{i_k}~ \< \hat{B} \> \nonumber \\
&\equiv & \< \hat{A} \> *  \< \hat{B} \> . \label{2-2a6} \eeqar
This is a general expression for the star product of matrix
functions on fuzzy $\cp^k$. The term corresponding to $s=0$ gives
the ordinary product $ \< \hat{A} \>   \< \hat{B} \> $ and the
successive terms are suppressed by powers of $n$ as $n \rightarrow
\infty$.

This form of star product, first obtained by Karabali and Nair in
\cite{nair4}, is suitable for the discussion of large $n$ (or $N =
N^{(k)}$) limit. For example, the symbol of the commutator of
matrix functions is given by \beqar \< \, [\hat{A},\hat{B}] \, \>
&=& -\frac{1}{n} \sum_{i=1}^{k} \left( R_{-i} \< \hat{A}\> R_{+i}
\<\hat{B} \> - R_{-i} \<\hat{B} \> R_{+i} \<\hat{A} \> \right)
 ~ +~ \O (1/n^2) \nonumber\\
&=& \frac{i}{n} \{ \< \hat{A} \> , \< \hat{B} \> \} ~ + ~ \O
(1/n^2) \label{2-2b1} \eeqar where the term involving the actions
of $R_{\pm i}$'s on the symbols can be proven to be the Poisson
bracket on $\cp^k$. For detailed description, see
\cite{nair4,NairRev1}. The relation (\ref{2-2b1}) shows an
explicit correspondence between the algebra of Poisson brackets
for functions on $\cp^k$ and the algebra of commutation relations
for functions on fuzzy $\cp^k$ in the large $n$ limit, indicating
that the construction of fuzzy spaces is essentially the same as
the quantization of symplectic manifolds.

From (\ref{2-1b8}), the trace of a matrix operator $A$ can be
expressed as \beqar \Tr A &=& \sum_{m} A_{mm} ~= ~ N^{(k)} \int d
\mu (g) \D_{m,
N^{(k)}}^{(n,0)} A_{mm'} \D_{m',N^{(k)}}^{*(n,0)} \nonumber\\
&=&  N^{(k)} \int d \mu(g) \< \hat{A} \> \, . \label{2-2b2} \eeqar
The trace of the product of two matrices $A$, $B$, is also given
by \beq \Tr AB ~=~ N^{(k)} \int d \mu(g) \< \hat{A} \> *\< \hat{B}
\>  \, . \label{2-2b3} \eeq

\section{Large $N$ limit}

In this section, following \cite{NairRev1,nair5}, we briefly
review the large $n$ limit of the symbol for an arbitrary matrix
function $f(L_A)$, where $L_A$ ($A =1,2, \cdots, k^2+2k$) are, as
before, the $N^{(k)} \times N^{(k)}$-matrix representations of the
algebra of $SU(k+1)$ in the $(n,0)$-representation. From
(\ref{2-2a1}), the symbol of $L_B A$, $A$ being an arbitrary
$N^{(k)} \times N^{(k)}$-matrix, is given by $\< \hat{L}_B \hat{A}
\>  = \<N | \hat{g}^{T} \hat{L}_B A \hat{g}^{*} | N \>$, where
$|N\> \equiv |(n,0), N^{(k)} \>$. We now express the factor
$\hat{g}^{T} \hat{L}_B \hat{g}^{*}$ as \beqar \hat{g}^{T}
\hat{L}_B \hat{g}^{*} \! &=&\!\! S_{BC}(g) \hat{L}_C
 \nonumber\\
\! &=& \!\!\! \hf (S_{B+i} \hat{L}_{-i} + S_{B-i} \hat{L}_{+i}) +
S_{Bj}
\hat{L}_{j} + S_{Bk^2 +2k} \hat{L}_{k^2 +2k} \, , \label{2-3a1}\\
S_{BC}(g) \! & \equiv & \! 2 \, \tr(g^{T} t_B g^* t_C ).
\label{2-3a2} \eeqar Note that, in terms of $\hat{L}_C$ acting on
$\<N|$ from the right, the relations (\ref{2-1b2})-(\ref{2-1b4})
can be expressed as \beq \< N| \hat{L}_{j}~ = ~ \< N| \hat{L}_{+i}
~=~ 0 ~ , ~~~ \< N| \hat{L}_{k^2 +2k} = -\frac{nk}{\sqrt{2k(k+1)}}
\< N| ~. \label{2-3a3} \eeq The symbol $\< \hat{L}_B \hat{A} \>$
is then written as \beqar \< \hat{L}_B \hat{A} \> &=& S_{Bk^2 +2k}
\< N | \hat{L}_{k^2 +2k} \hat{g}^{T} \hat{A} \hat{g}^{*} |N \> +
\hf S_{B +i} \<N| \hat{L}_{-i} \hat{g}^{T} \hat{A} \hat{g}^{*} | N
\>
\nonumber\\
&=& \L_B \< \hat{A} \> \, , \label{2-3a4}\\
\L_B & \equiv &
 - \frac{nk}{\sqrt{2k(k+1)}} S_{B k^2 +2k} + \hf S_{B
+i}  \tilde{R}_{-i} \label{2-3a5} \eeqar where $\tilde{R}_{-i}$ is
defined by $\tilde{R}_{-i} g^{T} = \hat{L}_{-i} g^{T}$.

Assuming $\hat{A}$ as the $N^{(k)}$-dimensional identity matrix
${\bf 1}$, we find that the symbol $\< L_B \>$ is dominated by the
quantity $S_{B k^2 + 2k}(g)$ in the large $n$ limit. One can in
fact check that $-S_{B k^2 +2k}$ satisfy the algebraic constraints
for the coordinates of $\cp^k$ which are, as we shall see later,
given in (\ref{2-13})-(\ref{2-15}).

By taking $\hat{A}$ itself as a product of $\hat{L}_A$'s, we can
by iteration express symbols for any products of $L_A$'s as \beq
\< \hat{L}_{A_1} \hat{L}_{A_2} \cdots \hat{L}_{A_s} \> ~ =~ \<
\L_{A_1} \L_{A_2} \cdots \L_{A_s} \cdot {\bf 1} \> \label{2-3a7}
\eeq where $s=1,2, \cdots$. Thus symbols of any matrix functions
$f(\hat{L}_A)$ become the corresponding functions of $S_{A k^2
+2k}$, $\< f(\hat{L}_A)\> \approx f(S_{A k^2 +2k})$, in the large
$n$ limit.

\section{Algebraic construction}

In this section, we present construction of fuzzy $\cp^k$
($k=1,2,\cdots$) in the framework of the creation-annihilation
operators \cite{gro2,bal1}. The coordinates $Q_A$ of fuzzy ${\bf
CP}^k$ can be defined in terms of $L_A$ as \beq Q_A =
\frac{L_A}{\sqrt{C^{(k)}_2}} ~,\label{2-1}\eeq satisfying the
following two constraints \beqar
Q_A ~Q_A &=& {\bf 1} \, , \label{2-2} \\
d_{ABC}~Q_{A}~ Q_{B} & = & c_{k,n}~ Q_C \label{2-3} \eeqar where
$d_{ABC}$ is the totally symmetric symbol of $SU(k+1)$,
$C^{(k)}_2$ is the quadratic Casimir of $SU(k+1)$ in the
$(n,0)$-representation \beq C^{(k)}_2 = \frac{n~ k~ (n+k+1)}{2~
(k+1)} \label{2-4} \eeq and $N^{(k)}$ is the dimension of
$SU(k+1)$ in the $(n,0)$-representation given in (\ref{2-1a}).

In order to determine the coefficient $c_{k,n}$ in (\ref{2-3}), we
now notice that the $SU(k+1)$ generators in the
$(n,0)$-representation can be written by \beq \La_A = a_i^\dag ~
(t_A)_{ij}~ a_j \label{2-6} \eeq where $t_A$
($A=1,2,\cdots,k^2+2k$) are the $SU(k+1)$ generators in the
fundamental representation with normalization $\tr(t_{A}t_{B})=\hf
\del_{AB}$ and $a_i^\dag$, $a_i$ ($i=1,\cdots,k+1$) are the
creation and annihilation operators acting on the states of $\H_N$
which are spanned by \beq |~n_1, n_2, \cdots , n_{k+1}~ \ket =
(a_1^\dag)^{n_1}(a_2^\dag)^{n_2}\cdots
(a_{k+1}^\dag)^{n_{k+1}}~|~0~\ket \label{2-7} \eeq with the
following relations \beqar a_i^\dag a_i ~|~n_1, n_2, \cdots ,
n_{k+1} ~ \> &=&
(n_1+n_2+\cdots+n_{k+1})~|~n_1,n_2, \cdots , n_{k+1} ~\> \nonumber \\
&=& n~|~n_1, n_2, \cdots , n_{k+1}~ \> \, , \label{2-8}\\ a_i ~|~
0~\ket &=& 0 \, . \label{2-9} \eeqar Notice that the condition
(\ref{2-9}) corresponds to the polarization condition in the
context of geometric quantization.

Using the completeness relation for $t_A$'s \beq
(t_A)_{ij}~(t_A)_{kl}= \hf \left(~\del_{il}~\del_{jk}
~-~\frac{1}{k+1}~ \del_{ij}~\del_{kl}~\right) \label{2-10} \eeq
and the commutation relation $[a_i,a_j^\dag]=\del_{ij}$, we can
check $\La_A \La_A = C^{(k)}_2$, where the creation and
annihilation operators act on the states of the form (\ref{2-7})
from the left. We also find \beqar d_{ABC}~\La_B ~ \La_C &=&
(k-1)\left(\frac{n}{k+1} + \hf \right) ~a^\dag_i~(t_A)_{ij}~a_j
\nonumber \\ &=& (k-1)\left(\frac{n}{k+1} + \hf \right) ~\La_A \,
. \label{2-11}\eeqar  Representing $\La_A$ by $L_A$, we can
determine the coefficient $c_{k,n}$ in (\ref{2-3}) by \beq c_{k,n}
~=~ \frac{(k-1)}{\sqrt{C_2^{(k)}}} \left( \frac{n}{k+1} + \hf
\right). \label{2-12} \eeq For $k \ll n$, we have \beq
c_{k,n}~\longrightarrow ~~ c_k~=~\sqrt{\frac{2}{k(k+1)}}~ (k-1)
\label{2-13}\eeq and this leads to the constraints for the
coordinates $q_A$ of $\cp^k$ \beqar
q_A ~q_A &=& 1 \, , \label{2-14} \\
d_{ABC}~q_{A}~ q_{B} & = & c_{k}~ q_C \, . \label{2-15} \eeqar

The second constraint (\ref{2-15}) restricts the number of
coordinates to be $2k$ out of $k^2+2k$. For example, in the case
of $\cp^2=SU(3)/U(2)$ this constraint around the pole of $A=8$
becomes $d_{8BC}q_8 q_B = \frac{1}{\sqrt{3}} q_C$. Normalizing the
8-coordinate to be $q_8 = -2$, we find the indices of the
coordinates are restricted to 4, 5, 6, and 7 with the conventional
choice of the generators of $SU(3)$ as well as with the definition
$d_{ABC}=2 \tr(t_A t_B t_C + t_A t_C t_B)$.

\subsection{Matrix-Function Correspondence}

The matrix-function correspondence for fuzzy $\cp^k$ can be
expressed by \beq N^{(k)}\times N^{(k)}= \sum^{n}_{l=0} \dim (l,l)
\label{2-ten1} \eeq where $\dim (l,l)$ is the dimension of
$SU(k+1)$ in the $(l,l)$-representation. This expression indicates
that the number of matrix elements coincides with the number of
coefficients in an expansion series of truncated functions on
$\cp^k = SU(k+1)/U(k)$. We need the real $(l,l)$-representation in
order to have an expansion of scalar functions on $\cp^k$.
Symbolically the correspondence is written as \beq (n,0)~
\bigotimes ~(0,n) = \bigoplus_{l=0}^{n} ~ (l,l) \label{2-ten2}
\eeq in terms of the dimensionality of $SU(k+1)$. The
left-hand-side of (\ref{2-ten2}) can be interpreted from the fact
that $\La_A = a_{i}^\dag (t_A)_{ij} a_j \sim a_{i}^\dag a_j$
transforms like $(n,0) \otimes (0,n)$. The right-hand-side of
(\ref{2-ten2}), on the other hand, can be interpreted by a usual
tensor analysis, {\it i.e.}, $\dim (l,l)$ is the number of ways to
construct tensors of the form
$T^{i_1,i_2,\cdots,i_l}_{j_1,j_2,\cdots,j_l}$ such that the tensor
is traceless and totally symmetric in terms of $i,j = 1,2, \cdots,
k+1$.

\chapter{Construction of fuzzy $S^4$}

\section{Introduction to fuzzy $S^4$}

As we have witnessed for more than a decade, the idea of fuzzy
$S^2$ \cite{mad1} has been one of the guiding forces for us to
investigate fuzzy spaces. For example, as discussed in the
previous chapter, fuzzy complex projective spaces $\cp^k$
($k=1,2,\cdots$) are successfully constructed in the same spirit
as the fuzzy $S^2$. From physicists' point of view, it is of great
interest to obtain a four-dimensional fuzzy space. The
well-defined fuzzy $\cp^2$ is not suitable for this purpose, since
$\cp^2$ does not have a spin structure \cite{bal1}. Construction
of fuzzy $S^4$ is then physically well motivated. (Notice that
fuzzy spaces are generally obtained for compact spaces and that
$S^4$ is the simplest four-dimensional compact space that allows a
spin structure.) Since $S^4$ naturally leads to ${\bf R}^4$ at a
certain limit, the construction of fuzzy $S^4$ would also shed
light on the studies of noncommutative Euclidean field theory.

There have been several attempts to construct fuzzy $S^4$ from a
field theoretic point of view \cite{gro1,oco1,dol1} as well as
from a rather mathematical interest \cite{ram1,ram2,naka},
however, it would be fair to say that the construction of fuzzy
$S^4$ has not yet been satisfactory. In \cite{ram1,ram2}, the
construction is carried out with a projection from some matrix
algebra (which in fact coincides with the algebra of fuzzy
$\cp^3$) and, owing to this forcible projection, it is advocated
that fuzzy $S^4$ obeys a non-associative algebra. Associativity is
recovered in the commutative limit, however, non-associativity
limits the use of fuzzy $S^4$ for physical models.
(Non-associativity is not compatible with unitarity of the algebra
for symmetry operations in any physical models.) Further,
non-associativity is not compatible with the definition of fuzzy
spaces (\ref{1-1}) in which the algebra of fuzzy spaces is given
by the algebra of finite dimensional matrices. In
\cite{oco1,dol1}, fuzzy $S^4$ is alternatively considered in a way
of constructing a scalar field theory on it, based on the fact
that $\cp^3$ is a $\cp^1$ (or $S^2$) bundle over $S^4$. While the
resulting action leads to a correct commutative limit, it is, as a
matter of fact, made of a scalar field on fuzzy $\cp^3$. Its
non-$S^4$ contributions are suppressed by an additional term.
(Such a term can be obtained group theoretically.) The action is
interesting but the algebra of fuzzy $S^4$ is still unclear. In
this sense, the approach in \cite{oco1,dol1} is related to that in
\cite{ram1,ram2}. Either approach uses a sort of brute-force
method which eliminates unwanted degrees of freedom from fuzzy
$\cp^3$. Such a method gives a correct counting for the degrees of
freedom of fuzzy $S^4$, but it does not clarify the construction
of fuzzy $S^4$ {\it per se}, as a matrix approximation to $S^4$.
This is precisely what we attempt to do in this chapter. Note that
the term ``\,fuzzy $S^4$\,'' is also used, mainly in the context
of M(atrix) theory, {\it e.g.}, in \cite{holi,kim1}, for the space
developed in \cite{CLT}. This space actually obeys the constraints
for fuzzy $\cp^3$. We shall discuss this point later in section
5.5.

In \cite{naka}, the construction of fuzzy $S^4$ is considered
through fuzzy $S^2 \times S^2$. This allows one to describe fuzzy
$S^4$ with some concrete matrix configurations. However, the
algebra is still non-associative and one has to deal with
non-polynomial functions on fuzzy $S^4$. Since those functions do
not naturally become polynomials on $S^4$ in the commutative
limits, there is no proper matrix-function correspondence. The
matrix-function correspondence is a correspondence between
functions on fuzzy spaces and truncated functions on the
corresponding commutative spaces. In the case of fuzzy $\cp^k$,
the fuzzy functions are represented by full ($N \times
N$)-matrices, so the product of them is given by matrix
multiplication which leads to associativity of the algebra for
fuzzy $\cp^k$. As we have seen in (\ref{2-2a6}), the star products
of fuzzy $\cp^k$ reduce to ordinary commutative products of
functions (or symbols) on $\cp^k$ in the large $N$ limit. In this
case, one may check the matrix-function correspondence by the
matching between the number of matrix elements and that of
truncated functions. This matching, however, is not enough to
warrant the matrix-function correspondence of fuzzy $S^4$; further
we need to confirm the correspondence between the product of fuzzy
functions and that of truncated functions. In order to do so, it
is important to construct fuzzy $S^4$ with a clear matrix
configuration (which should be different from the proposal in
\cite{naka}).

The plan of this chapter is as follows. In section 3.2, following
Medina and O'Connor in \cite{oco1}, we propose construction of
fuzzy $S^4$ by use of the fact that $\cp^3$ is an $S^2$ bundle
over $S^4$. We shall obtain fuzzy $S^4$, imposing a further
constraint on fuzzy $\cp^3$. The extra constraint is expressed as
an algebraic constraint such that it enables us to describe the
algebra of fuzzy $S^4$ in terms of the algebra of $SU(4)$ in the
$(n,0)$-representation. The emerging algebra is not a subalgebra
of fuzzy $\cp^3$ since the algebra of fuzzy $\cp^3$ is defined
globally by $\underline{SU(4)}$ with the algebraic constraints
given in (\ref{2-2}) and (\ref{2-3}) for $k=3$. The algebra of
fuzzy $S^4$ is obtained from $\underline{SU(4)}$ as well with the
extra constraint on top of these fuzzy $\cp^3$ constraints. As
mentioned in chapter 1, the algebra of fuzzy $S^4$ is consequently
given by a subset of $\underline{SU(4)}$, preserving closure and
associativity of the algebra. The structure of algebra becomes
clearer in the commutative limit which will be considered in terms
of homogeneous coordinates of $\cp^3$. With these coordinates we
shall explicitly show that the extra constraint for fuzzy $S^4$
has a correct commutative limit. The idea of constructing fuzzy
spaces from another by means of an additional constraint was in
fact first proposed by Nair and Randjbar-Daemi in obtaining fuzzy
$S^3 / {\bf Z}_2$ from fuzzy $S^2 \times S^2$ \cite{nair1}. Our
construction of fuzzy $S^4$ provides another example of such
construction.

In section 3.3, we show the matrix-function correspondence of
fuzzy $S^4$. After a brief review of the case of fuzzy $S^2$, we
shall present different calculations of the number of truncated
functions on $S^4$. We then show that this number agrees with the
number of degrees of freedom for fuzzy $S^4$. This number turns
out to be {\it a sum of absolute squares}, and hence we can choose
a block-diagonal matrix configuration for functions on fuzzy
$S^4$. This block-diagonal form is also induced from the structure
of fuzzy functions. The star products are determined by matrix
products of such functions and naturally reduce to commutative
products, similarly to what happens in fuzzy $\cp^3$. This leads
to the precise matrix-function correspondence of fuzzy $S^4$. Of
course, such a matrix realization of fuzzy $S^4$ is not the only
one that leads to the correspondence; there are a number of ways
related to the ways of allocating the absolute squares to form any
block-diagonal matrices. Our construction is, however, useful in
comparison with the fuzzy $\cp^3$.

The fact that ${\bf CP}^3$ is an $S^2$ bundle over $S^4$ can be
seen by a Hopf map, $S^7 \rightarrow S^4$ with the fiber being
$S^3$. One can derive the map, noticing that the $S^4$ is the
quaternion projective space. In the same reasoning, octonions
define a Hopf map, $S^{15} \rightarrow S^8$ with its fiber being
$S^7$, giving us another fact that ${\bf CP}^7$ is a ${\bf CP}^3$
bundle over $S^8$. Following these mathematical facts, in section
3.4, we apply our construction to fuzzy $S^8$ and outline its
construction.


\section{Construction of fuzzy $S^4$}

We begin with construction of fuzzy $\cp^3$. The algebraic
construction of fuzzy ${\bf CP}^k$ ($k=1,2,\cdots$) is generically
given in section 2.4; here we briefly rephrase it in the case of
$k=3$. The coordinates $Q_A$ of fuzzy ${\bf CP}^3$ can be defined
by \beq Q_A = \frac{L_A}{\sqrt{C^{(3)}_2}} \label{3-01}\eeq where
$L_A$ are $N^{(3)}\times N^{(3)}$-matrix representations of
$SU(4)$ generators in the $(n,0)$-representation. The coordinates
satisfy the following constraints:
\beqar Q_A ~Q_A &=& {\bf 1} \, , \label{3-02} \\
d_{ABC}~Q_{A}~ Q_{B} & = & c_{3,n}~ Q_C \, . \label{3-03} \eeqar
As discussed before, in the large $n$ limit these constraints
become constraints for the coordinates on $\cp^3$ as embedded in
${\bf R}^{15}$. In (\ref{3-01})-(\ref{3-03}), $C^{(3)}_2$, ${\bf
1}$, $d_{ABC}$ and $c_{3,n}$ are all defined in chapter 2,
including the relation \beq N^{(3)}=\frac{1}{6}(n+1)(n+2)(n+3).
\label{3-04}\eeq

We now consider the decomposition, $SU(4) \rightarrow SU(2) \times
SU(2) \times U(1)$, where the two $SU(2)$'s and one $U(1)$ are
defined
by \beq\left(%
\begin{array}{cc}
\underline{SU(2)} & 0 \\
0 & 0 \\
\end{array}%
\right)  , \qquad \left(%
\begin{array}{cc}
0 & 0 \\
0 & \underline{SU(2)} \\
\end{array}%
\right)  , \qquad \left(%
\begin{array}{cc}
1 & 0 \\
0 & -1 \\
\end{array}%
\right) \label{3-m1} \eeq in terms of the ($4 \times 4$)-matrix
generators of $SU(4)$ in the fundamental representation. (Each
$\underline{SU(2)}$ denotes the algebra of $SU(2)$ group in the
($2\times 2$)-matrix representation.) As we shall see in
subsections 3.2.1 and 3.3.1, functions on $S^4$ are functions on
$\cp^3=SU(4)/U(3)$ which are invariant under transformations of $H
\equiv SU(2) \times U(1)$, $H$ being relevant to the above
decomposition of $SU(4)$. In order to obtain functions on fuzzy
$S^4$, we thus need to require
\begin{equation}\label{3-m2} [ \F, L_\al ] = 0
\end{equation}
where $\F$ denote matrix-functions of $Q_A$'s and $L_\al$ are
generators of $H$ represented by $N^{(3)} \times
N^{(3)}$-matrices. Construction of fuzzy $S^4$ can be carried out
by imposing the additional constraint (\ref{3-m2}) onto the
functions on fuzzy $\cp^3$. What we claim is that the further
condition (\ref{3-m2}) makes the functions $\F(Q_A)$ become
functions on fuzzy $S^4$. This does not mean that fuzzy $S^4$ is a
subset of fuzzy $\cp^3$. Notice that $Q_A$'s are defined in ${\bf
R}^{15}$ ($A=1,\cdots, 15$) with the algebraic constraints
(\ref{3-02}) and (\ref{3-03}). While locally, say around the pole
of $A=15$ in (\ref{3-03}), one can specify the six coordinates of
fuzzy $\cp^3$, globally they are embedded in ${\bf R}^{15}$.
Equation (\ref{3-m2}) is a global constraint in this sense. So the
algebra of fuzzy $S^4$ is given by a subset of
$\underline{SU(4)}$. The emerging algebraic structure of fuzzy
$S^4$ will be clearer when we consider the commutative limit of
our construction.

\subsection{Commutative limit}

As shown in section 2.3, in the large $n$ limit we can approximate
$Q_A$ to the commutative coordinates on $\cp^3$; \beq Q_A ~
\approx ~ \p_A = - 2 ~ \tr (g^{\dag}t_{A} g t_{15}) \label{3-07}
\eeq which indeed obey the following constraints for $\cp^3$ \beq
\p_A~ \p_A = 1 \, , \qquad d_{ABC}~\p_A~ \p _B =
\sqrt{\frac{2}{3}}~ \p_C \, . \label{3-08}\eeq Algebraic
constraints for $\cp^k$ are in general given in
(\ref{2-13})-(\ref{2-15}). In (\ref{3-07}), $t_A$'s are the
generators of $SU(4)$ in the fundamental representation and $g$ is
a group element of $SU(4)$ given as a ($4 \times 4$)-matrix.
Truncated functions on $\cp^3$ are then written as \beq
f_{\cp^3}(u, {\bar u})\sim f^{i_1 i_2 \cdots i_l}_{j_1 j_2 \cdots
j_l} {\bar u}_{i_1} {\bar u}_{i_2} \cdots {\bar u}_{i_l}
u_{j_1}u_{j_2}\cdots u_{j_l} \label{3-09} \eeq where
$l=0,1,2,\cdots ,n$, $u_{j}=g_{j 4}$, ${\bar u}_{i}=(g^\dag)_{4
i}$ and ${\bar u}_{i} u_{i} = 1$ ($i, j= 1,2,3,4$). One can
describe $\cp^3$ in terms of four complex coordinates $Z_i$ with
the identification $Z_i \sim \la Z_i$ where $\la$ is a nonzero
complex number ($\la \in {\bf C} - \{0\}$). Following Penrose and
others \cite{penrose}, we now write $Z_i$ in terms of two spinors
$\om$, $\pi$ as \beq Z_i = ( \om_a , \pi_{\adot})=
(x_{a\adot}\pi_{\adot}, \pi_\adot) \label{3-010} \eeq where
$a=1,2$, $\adot = 1,2$ and $x_{a\adot}$ can be defined with the
coordinates $x_\mu$ on $S^4$ via $x_{a \adot} = ({\bf 1}x_4 - i
\vec{\si} \cdot \vec{x})$, $\vec{\si}$ being $(2\times 2)$ Pauli
matrices. The scale invariance $Z_i \sim \la Z_i$ leads to the
invariance $\pi_\adot \sim \la \pi_\adot $. The $\pi_\adot$'s then
describe a $\cp^1 = S^2$. This shows the fact that $\cp^3$ is an
$S^2$ bundle over $S^4$, or Penrose's projective twistor space.
Note that, as in (\ref{2-1c2}), we can parametrize $u_i$ of
(\ref{3-09}) by the homogeneous coordinates $Z_i$, {\it i.e.},
$u_i = \frac{Z_i}{\sqrt{Z \cdot \bZ}}$.

Functions on $S^4$ can be considered as functions on $\cp^3$ which
satisfy \beq \frac{\d}{\d \pi_{\adot}} f_{\cp^3}(Z,\bZ) =
\frac{\d}{\d \bar{\pi}_{\adot}} f_{\cp^3}(Z,\bZ) = 0 \, .
\label{3-011} \eeq This implies that $f_{\cp^3}$ are further
invariant under transformations of $\pi_\adot$, ${\bar
\pi}_\adot$. In terms of the four-spinor $Z$, such transformations
are expressed by \beq Z \rightarrow e^{i t_\al \th_\al} Z
\label{3-012} \eeq where $t_\al$ represent the algebra of
$H=SU(2)\times U(1)$ defined previously in regard to the
decomposition of $SU(4)$ in (\ref{3-m1}). The coordinates $\p_A$
in (\ref{3-07}) can be written by $\p_A(Z, \bZ)\sim \bZ_i
(t_A)_{ij} Z_j$. Under an infinitesimal ($\th_\al \ll 1$)
transformation of (\ref{3-012}), the coordinates $\p_A(Z, \bZ)$
transform as \beq \p_A \rightarrow \p_A + \th_\al \, f_{\al
AB}\,\p_B \label{3-013} \eeq where $f_{ABC}$ is the structure
constant of $SU(4)$. The constraint (\ref{3-011}) is then
rewritten as \beq f_{\al AB} ~ \p_B ~\frac{\d}{\d \p_A} ~
f_{\cp^3} = 0 \label{3-014} \eeq where $f_{\cp^3}$ are seen as
functions of $\p_A$'s rather than that of $(Z, \bZ)$. Note that
$\p_A$'s in (\ref{3-014}) are defined by (\ref{3-07}), {\it i.e.},
they are globally defined on ${\bf R}^{15}$.

From the relation $\p_A \sim \bZ_i (t_A)_{ij} Z_j$, we find
$f_{\al AB}\,\p_B \sim \bZ_i([t_A ,t_\al])_{ij} Z_j$ where $t_\al$
are the generators of $H=SU(2)\times U(1)\subset SU(4)$ as before.
The constraint (\ref{3-011}) or (\ref{3-014}) is then realized by
$[t_A , t_\al]=0$, which can be considered as a commutative
implementation of the fuzzy constraint (\ref{3-m2}). Specifically,
we may choose $t_\al = \left\{ t_1, t_2, t_3,
\sqrt{\frac{2}{3}}t_8 + \sqrt{\frac{1}{3}}t_{15} \right\}$ in the
conventional choices of the generators of $SU(4)$ in the
fundamental representation. The constraint $[t_A , t_\al]=0$ then
restricts $A$ to be $A=8,13,14,$ and $15$. This is, of course, a
local analysis. The constraint $[t_A , t_\al]=0$ does globally
define $S^4$ as embedded in ${\bf R}^{15}$ similarly to how we
have defined $\cp^3$. The number of $\cp^3$ coordinates $\p_A$ is
locally restricted to be six because of the algebraic constraints
in (\ref{3-08}). On top of these, the constraint $[t_A , t_\al]=0$
further restricts the number of coordinates to be four, which is
correct for the coordinates on $S^4$.

Functions on $S^4$ are polynomials of $\p_A=- 2 \tr (g^{\dag}t_{A}
g t_{15})$ which obey $[t_A, t_\al]=0$. Products of functions are
determined by the products of such $t_A$'s. Extension to the fuzzy
case is essentially done by replacing $t_A$ with $L_A$, where
$L_A$ is the matrix representation of the algebra of $SU(4)$ in
the totally symmetric $(n,0)$-representation. The algebra of fuzzy
$S^4$ naturally becomes associative in the commutative limit,
while associativity of fuzzy $S^4$, itself, will be discussed in
the next section, where we shall present a concrete matrix
configuration of fuzzy $S^4$ so that the associativity is
obviously seen. Even without any such matrix realizations, we can
extract another property of the algebra from the condition
(\ref{3-m2}), that is, closure of the algebra; since functions on
fuzzy $S^4$ are represented by matrices, it is easily seen that
products of such functions also obey the condition (\ref{3-m2}).
In what follows, we shall clarify these points in some detail.

\section{Matrix-function correspondence}

In this section, we examine our construction of fuzzy $S^4$ by
confirming its matrix-function correspondence. To show a
one-to-one correspondence, one needs to say two things: (a) a
matching between the number of matrix elements for fuzzy $S^4$ and
the number of truncated functions on $S^4$; and (b) a
correspondence between the product of functions on fuzzy $S^4$ and
that on $S^4$. It is now suggestive to take a moment to review how
(a) and (b) are fulfilled in the case of fuzzy $S^{2}=SU(2)/U(1)$.
Let $\D_{mn}^{(j)}(g)$ be Wigner $\D$-functions for $SU(2)$. As we
have discussed in section 2.1, these are the spin-$j$ matrix
representations of an $SU(2)$ group element $g$;
$\D_{mn}^{(j)}(g)=\< jm| \hat{g} |jn \>$ ($m,n=-j,\cdots,j$).
Functions on $S^2$ can be expanded in terms of particular Wigner
$\D$-functions, $\D^{(j)}_{m0}(g)$, which are invariant under a
$U(1)$ right-translation operator acting on $g$. For definition of
such an operator, see (\ref{2-1b1}). Since the state $|j0\ket$ has
no $U(1)$ charge, right action of the $U(1)$ operator, $R_3$, on
$g$ makes $\D^{(j)}_{m0}(g)$ vanish, $R_{3}\D^{(j)}_{m0}(g)=0$; in
fact one can choose any fixed value ($m=-j,\cdots,j$) for this
$U(1)$ charge. The $\D$-functions are essentially the spherical
harmonics,
$\D^{(l)}_{m0}=\sqrt{\frac{4\pi}{2l+1}}(-1)^{m}Y^{l}_{-m}$, and so
a truncated expansion can be written as $f_{S^2} = \sum_{l=0}^{n}
\sum_{m=-l}^{l} f^{l}_{m} \D^{(l)}_{ml}$. The number of
coefficients $f^{l}_{m}$ are counted by $\sum_{l=0}^{n}(2l+1) =
(n+1)^{2}$. This relation implements the condition (a) by defining
functions on fuzzy $S^2$ as $(n+1)\times (n+1)$ matrices. The
product of truncated functions at the same level of $n$ is also
expressed by the same number of coefficients. Therefore, the
product corresponds to $(n+1)\times (n+1)$ matrix multiplication.
This implies the condition (b). One can show an exact
correspondence of products, following the general lines in section
2.2. Let $f_{mn}$ ($m,n=1,\cdots,n+1$) be an element of matrix
function-operator $\hat{f}$ on fuzzy $S^2$. As in (\ref{2-2a1}),
we define the symbol of the function as \beq \bra \hat{f} \ket =
\sum_{m,n} f_{mn} \D^{*(j)}_{mj}(g) \D^{(j)}_{nj}(g)
\label{3-s2-1}\eeq where $ \D^{*(j)}_{mj}(g)=
\D^{(j)}_{jm}(g^{-1})$. We here consider $|jj \>$ as the highest
weight state. The star product of fuzzy $S^2$ is defined by
$\bra\hat{f}\hat{g}\ket = \bra \hat{f}\ket
* \bra \hat{g} \ket$. From (\ref{3-s2-1}), we can write \beqar \bra
\hat{f}\hat{g} \ket &=&
\sum_{mnl}f_{mn}g_{nl} \D^{*(j)}_{mj}(g) \D^{(j)}_{lj}(g) \nonumber\\
&=& \sum_{mnkrl}f_{mn}g_{kl} \D^{*(j)}_{mj}(g) \D^{(j)}_{nr}(g)
\D^{*(j)}_{kr}(g)\D^{(j)}_{lj}(g)\label{3-s2-2}\eeqar where we use
the orthogonality of $\D$-functions $\sum_{r} \D^{(j)}_{nr}(g)
\D^{*(j)}_{kr}(g) = \delta_{nk}$. Let $R_-$ be the lowering
operator in right action, we then find
\beq
	R_-  \D^{(j)}_{mn}(g) =
	\sqrt{(j+n)(j-n+1)} \D^{(j)}_{m n-1}(g) \, .
	\label{3-additional01}
\eeq
By iteration,
(\ref{3-s2-2}) may be rewritten as \beq \bra \hat{f}\hat{g} \ket =
\sum_{s=0}^{2j}(-1)^s \frac{(2j-s)!}{s!(2j)!} ~ R^{s}_{-} \bra
\hat{f}\ket ~ R^{s}_{+} \bra\hat{g} \ket ~ \equiv \bra \hat{f}\ket
*\bra\hat{g} \ket \label{3-s2-3}\eeq where we use the relation
$R_{-}^{*}= - R_+$. In the large $j$ limit, the term with $s=0$ in
(\ref{3-s2-3}) dominates and this leads to an ordinary commutative
product of $\bra\hat{f}\ket$ and $\bra\hat{g}\ket$. By
construction, the symbols of functions on fuzzy $S^2$ can be
regarded as commutative functions on $S^2$. The expression
(\ref{3-s2-3}) therefore shows a one-to-one correspondence between
the product of fuzzy functions and the product of truncated
functions on $S^2$.

From (\ref{3-s2-2}) and (\ref{3-s2-3}), it is easily seen that the
square-matrix configuration, in addition to the orthogonality of
the $\D$-functions or of the states $|jm \>$, is the key
ingredient for the condition (b) in the case of fuzzy $S^2$.
Associativity of the star product is direct consequence of this
matrix configuration. Suppose the number of truncated functions on
some compact space is given by an absolute square. Then, following
the above procedure, one may establish the matrix-function
correspondence. As shown in (\ref{2-ten1}), this is true for fuzzy
$\cp^k$ in general. In the case of fuzzy $\cp^3$, the absolute
square appears from \beqar
N^{(3)}\times N^{(3)}&=& \sum^{n}_{l=0} \dim (l,l), \label{3-c1} \\
\dim (l,l)&=& \frac{1}{12}(2l+3)(l+1)^2(l+2)^2\label{3-c2} \eeqar
where $\dim (l,l)$ is the dimension of $SU(4)$ in the real
$(l,l)$-representation. This arises from the fact that functions
on $\cp^3=SU(4)/U(3)$ can be expanded by $\D^{(l,l)}_{M {\bf
0}}(g)$, Wigner $\D$-functions of $SU(4)$ in the
$(l,l)$-representation ($l=0,1,2,\cdots$). Here, $g$ is an element
of $SU(4)$. The lower index $M$ $(M=1,\cdots , \dim(l,l))$ labels
the state in the $(l,l)$-representation, while the index ${\bf 0}$
represents any suitably fixed state in this representation. Like
in (\ref{3-s2-1}), the symbol of fuzzy $\cp^3$ is defined in terms
of $\D^{(n,0)}_{I N^{(3)}}(g)$ and its complex conjugate, where
$\D^{(n,0)}_{I N^{(3)}}(g)=\< (n,0),I | g |(n,0), N^{(3)} \>$ are
the $\D$-functions belonging to the symmetric
$(n,0)$-representation. The states of fuzzy $\cp^3$ are expressed
by $|(n,0),I \>$. The index $I$ ($I = 1,2,\cdots ,
\dim(n,0)=N^{(3)}$) labels these states and the index $N^{(3)}$
indicates the highest or lowest weight state.  Notice that one can
alternatively express the states by $\p_{i_1 i_2 \cdots i_n}$
where the sequence of $i_{m}=\{1,2,3,4\}$ $(m=1,\cdots,n)$ is in a
totally symmetric order.

We now return to the conditions (a) and (b) of fuzzy $S^4$. In the
following subsections, we present (i) different ways of counting
the number of truncated functions on $S^4$, (ii) a one-to-one
matrix-function correspondence for fuzzy $S^4$, and (iii) a
concrete matrix configuration for functions on fuzzy $S^4$. In
(ii), the condition (a) is shown; we find the number of matrix
elements for fuzzy $S^4$ agrees with the number calculated in (i).
The condition (b) is also shown in (ii) by considering the symbols
and star products on fuzzy $S^4$ in the commutative limit. In
(iii), we confirm the one-to-one correspondence by proposing a
block-diagonal matrix realization of fuzzy $S^4$. Along these
arguments, it will become clear that the algebra of fuzzy $S^4$ is
closed and associative.

\subsection{Ways of Counting}

A direct counting of the number of truncated functions on $S^4$
can be made in terms of the spherical harmonics $Y_{l_1 l_2 l_3
m}$ on $S^4$ with a truncation at $l_1=n$ \cite{naka}; \beq
N^{S^4}(n)=
\sum^{n}_{l_1=0}\sum^{l_1}_{l_2=0}\sum^{l_2}_{l_3=0}(2l_3 +1) =
\frac{1}{12}(n+1)(n+2)^2(n+3) \, . \label{3-r5} \eeq

Alternatively, one can count $N^{S^4}(n)$ by use of a tensor
analysis. The number of truncated functions on $\cp^3$ is given by
the totally symmetric and traceless tensors $f^{i_1 \cdots
i_l}_{j_1 \cdots j_l}$ ($i,j=1,\cdots,4$) in (\ref{3-09}). Now we
split the indices into $i=a,\adot$ ($a=1,2$, $\adot =3,4$), and
similarly for $j=b, \bdot$. The additional constraint
(\ref{3-011}) for the extraction of $S^4$ from $\cp^3$ means that
the tensors are independent of any combinations of $\adot \,$'s in
the sequence of $i\,$'s. In other words, in terms of the
transformation (\ref{3-012}), $Z \rightarrow e^{it_\al \th_\al}
Z$, functions on $S^4$ are invariant under the transformations
involving $(t_\al)_{\adot_1 \adot_2}$ where $t_\al$ are the
($4\times 4$) matrix representations of the algebra of $H=SU(2)
\times U(1)$. There are $N^{(2)}(l)=\hf(l+1)(l+2)$ ways of having
a symmetric order $i_1,i_2,\cdots,i_l$ for $i= \{1,2,\adot \}$
($\adot =3,4$). This can be regarded as an $N^{(2)}(l)$-degeneracy
due to an $S^2$ internal symmetry for the extraction of $S^4$ out
of $\cp^3 \sim S^4 \times S^2$. This internal symmetry is relevant
to the above $(t_\al)_{\adot_1 \adot_2}$-transformations. Since
the number of truncated functions on $\cp^3$ is given by
(\ref{3-c2}), the number of those on $S^4$ may be calculated by
\beq N^{S^4}(n)=\sum^{n}_{l=0}\frac{\dim(l,l)}{N^{(2)}(l)}=
\sum^{n}_{l=0}\frac{1}{6}(l+1)(l+2)(2l+3) =
\frac{1}{12}(n+1)(n+2)^2(n+3) \label{3-c9} \eeq which reproduces
(\ref{3-r5}). This is also in accordance with a corresponding
calculation in the context of $S^4=SO(5)/SO(4)$ \cite{oco1,dol1}.

\subsection{One-to-one matrix-function correspondence}

As mentioned earlier in this section, the states of fuzzy $\cp^3$
can be denoted by $\p_{i_1 i_2 \cdots i_n}$ where the sequence of
$i_{m}=\{1,2,3,4\}$ $(m=1,\cdots,n)$ is in a totally symmetric
order. Let $(\hat{F})_{IJ}$ ($I,J=1,2,\cdots, N^{(3)}$) denote a
matrix-function on fuzzy $\cp^3$. Matrix elements of the function
$\hat{F}$ on fuzzy $\cp^3$ can be defined by $\bra I | \hat{F} | J
\ket$, where we denote $\p_{i_1 \cdots i_n}=|i_{1} \cdots i_{n}
\ket \equiv |I \ket$. We need to find an analogous matrix
expression $(\hat{F}^{S^4})_{IJ}$ for a function on fuzzy $S^4$.
We now consider the states on fuzzy $S^4$ in terms of $\p_{i_1 i_2
\cdots i_n}$. Splitting each $i$ into $a$ and $\adot$, we may
express $\p_{i_1 i_2 \cdots i_n}$ as \beq \p_{i_1 i_2 \cdots
i_n}=\{  \p_{\adot_1 \adot_2 \cdots \adot_n} ~,~ \p_{a_1 \adot_1
\cdots \adot_{n-1}} ~,~ \cdots ~,~ \p_{a_1 \cdots a_{n-1}
\adot_1}~,~ \p_{a_1 a_2 \cdots a_n}  \} \, . \label{3-c10} \eeq
From the analysis in the previous section, one can obtain the
states corresponding to fuzzy $S^4$ by imposing an additional
condition on (\ref{3-c10}), {\it i.e.}, the invariance under the
transformations involving any $\adot_m$ $(m=1,\cdots,n)$.
Transformations of the states on fuzzy $S^4$, under this
particular condition, can be considered as follows. On the set of
states $\p_{\adot_1 \adot_2 \cdots \adot_n}$, which are $(n+1)$ in
number, the transformations must be diagonal because of
(\ref{3-011}), but we can have an independent transformation for
each state. (The number of the states is $(n+1)$, since the
sequence of $\adot_m = \{3,4\}$ is in a totally symmetric order.)
Thus we get $(n+1)$ different functions proportional to identity.
On the set of states $\p_{a_1 \adot_1 \cdots \adot_{n-1}}$, we can
transform the $a_1$ index (to $b_1 = \{1,2\}$ for instance),
corresponding to a matrix function $f_{a_1, b_1}$ which have $2^2$
independent components. But we can also choose the matrix $f_{a_1,
b_1}$ to be different for each choice of $(\adot_1 \cdots
\adot_{n-1})$ giving $2^2 \times n$ functions in all, at this
level. We can represent these as $f_{a_1, b_1}^{(\adot_1 \cdots
\adot_{n-1})}$, the extra composite index $(\adot_1 \cdots
\adot_{n-1})$ counting the multiplicity. Continuing in this way,
we find that the set of all functions on fuzzy $S^4$ is given by
\beqar (\hat{F}^{S^4})_{IJ} &=& \{ f^{(\adot_1 \cdots \adot_n)}
~\hat{\del}_{\adot_1 \cdots \adot_n , \bdot_1 \cdots \bdot_n} ,~
f_{a_1 , b_1}^{(\adot_1 \cdots \adot_{n-1})} ~\hat{\del}_{\adot_1
\cdots \adot_{n-1} , \bdot_1 \cdots \bdot_{n-1}} , \nonumber\\&&
~~~~~~~ f_{a_1 a_2 , b_1 b_2}^{(\adot_1 \cdots \adot_{n-2})}
~\hat{\del}_{\adot_1 \cdots \adot_{n-2} , \bdot_1 \cdots
\bdot_{n-2}} ,~ \cdots \cdots ,~ f_{a_1 \cdots a_n , b_1 \cdots
b_n}  \} \nonumber\\\label{3-c11}\eeqar where we split $i_m$ into
$a_m$, $\adot_m$ and $j_m$ into $b_m$, $\bdot_m$. Each of the
operators $\hat{\del}_{\adot_1 \cdots \adot_m , \bdot_1 \cdots
\bdot_m}$ indicates an identity operator such that the
corresponding matrix is invariant under transformations from
$\{\adot_1 \cdots \adot_m \}$ to $\{\bdot_1 \cdots \bdot_m \}$.
The structure in (\ref{3-c11}) shows that $\hat{F}^{S^4}$ is
composed of $(l+1) \times (l+1)$-matrices ($l=0,1,\cdots,n$), with
the number of these matrices for fixed $l$ being $(n+1-l)$. Thus
the number of matrix elements for fuzzy $S^4$ is counted by \beq
N^{S^4}(n)=\sum^{n}_{l=0} (l+1)^2 (n+1-l) =
\frac{1}{12}(n+1)(n+2)^2(n+3) \, . \label{3-c12} \eeq

This relation satisfies the condition (a). In order to show the
precise matrix-function correspondence, we further need to show
the condition (b), the correspondence of products. We carry out
this part in analogy with the case of fuzzy $S^2$ in
(\ref{3-s2-1})-(\ref{3-s2-3}). The symbol of the function
$\hat{F}$ on fuzzy $\cp^3$ can be defined as \beq \< \hat{F} \>
~=~ \sum_{I,J} \< N | g | I \> ~(\hat{F})_{IJ}~ \< J | g | N \>
\label{3-sym1} \eeq where $|N \> \equiv |(n,0), N^{(3)}\>$ is the
highest or lowest weight state of fuzzy $\cp^3$ and $\< J | g | N
\>$ denotes the previous $\D$-function, $\D^{(n,0)}_{J
N^{(3)}}(g)$. The symbol of a function on fuzzy $S^4$ is defined
in the same way except that $(\hat{F})_{IJ}$ is replaced with
$(\hat{F}^{S^4})_{IJ}$ in (\ref{3-sym1}). We now consider the
product of two functions on fuzzy $S^4$. As we discussed above, a
function on fuzzy $S^4$ can be described by $(l+1)\times
(l+1)$-matrices. From the structure of $\hat{F}^{S^4}$ in
(\ref{3-c11}), we are allowed to treat these matrices
independently. The product is then considered as a set of matrix
multiplications. This leads to a natural definition of the product
preserving closure, since the product of functions also becomes a
function, retaining the same structure as in (\ref{3-c11}). The
star product of fuzzy $S^4$ is written as \beq \<
\hat{F}^{S^4}\hat{G}^{S^4} \> = \sum_{IJK}
(\hat{F}^{S^4})_{IJ}(\hat{G}^{S^4})_{JK} \< N |g|I \> \< K |g| N
\> \equiv \< \hat{F}^{S^4} \> * \< \hat{G}^{S^4} \>
\label{3-sym2}\eeq where the product
$(\hat{F}^{S^4})_{IJ}(\hat{G}^{S^4})_{JK}$ is given by the set of
matrix multiplications. This fact, along with the orthogonality of
the $\D$-functions, leads to associativity of the star products.

The symbols and star products of fuzzy $S^4$ can be obtained from
those of fuzzy $\cp^3$ by simply replacing the function operator
$\hat{F}$ with $\hat{F}^{S^4}$. So the correspondence between
fuzzy and commutative products on $S^4$ can be shown in the large
$n$ limit as we have seen in section 2.2. We can in fact directly
check this correspondence even at the level of finite $n$ from the
following discussion.

Let us consider functions on $S^4$ in terms of the homogeneous
coordinates on $\cp^3$, $Z_i = (\om_a, \pi_\adot) =
(x_{a\adot}\pi_{\adot}, \pi_\adot)$, as in (\ref{3-010}).
Functions on $S^4$ can be constructed from $x_{a\adot}$ under the
constraint (\ref{3-011}), which implies that the functions are
independent of $\pi_\adot$ and $\bpi_\adot$. Expanding in powers
of $x_{a\adot}$, we may express the functions by the following set
of terms; $\{1, x_{a\adot}, x_{a_1 \adot_1}x_{a_2 \adot_2},
  x_{a_1 \adot_1}x_{a_2 \adot_2}x_{a_3 \adot_3} , \cdots \}$, where the
indices $a$'s and $\adot$'s are symmetric in their order. Owing to
the extra constraint (\ref{3-011}), one can consider that all the
factors involving $\pi_\adot$ and $\bpi_\adot$ can be absorbed
into the coefficients of these terms. By iterative use of the
relations, $x_{a\adot} \pi_\adot = \om_a$ and its complex
conjugation, the above set of terms can be expressed in terms of
$\om$'s and $\bom$'s as \beq 1 ~,~ \underbrace{\left(
\begin{array}{c}
\bom_{a_1} \\
\om_{b_1} \\
\end{array}
\right)}_{2\times 2}~,~\underbrace{\left(%
\begin{array}{c}
\bom_{a_1}\bom_{a_2} \\
\bom_{a_1} \om_{b_1} \\
\om_{b_1} \om_{b_2} \\
\end{array}%
\right)}_{3\times 3}~,~ \underbrace{\left(%
\begin{array}{c}
\bom_{a_1}\bom_{a_2}\bom_{a_3} \\
\bom_{a_1}\bom_{a_2}\om_{b_1} \\
\bom_{a_1}\om_{b_1} \om_{b_2} \\
\om_{b_1} \om_{b_2}\om_{b_3} \\
\end{array}%
\right)}_{4\times 4} ~,~ \cdots \cdots \label{3-c4} \eeq where the
indices $a$ and $b$ are used to distinguish $\bom$ and $\om$.
Because the indices are symmetric, the number of independent terms
in each column should be counted as indicated in (\ref{3-c4}).

Notice that even though functions on $S^4$ can be parametrized by
$\om$'s and $\bom$'s, the overall variables of the functions
should be given by the coordinates on $S^4$, $x_\mu$, instead of
$\om_a=\pi_\adot x_{a\adot}$. The coefficients of the terms in
(\ref{3-c4}) need to be chosen accordingly. For instance, the term
$\om_a$ with a coefficient $c_a$ will be expressed as $c_a \om_a =
c_a \pi_\adot x_{a\adot} \equiv h_{a\adot} x_{a\adot}$, where
$h_{a\adot}$ is considered as some arbitrary set of constants. We
now define truncated functions on $S^4$ in the present context.
Functions on $S^4$ are generically expanded in powers of $\bom_a$
and $\om_b$ ($a=1,2$ and $b = 1,2$) \beq f_{S^4}(\om, \bom) \sim
f^{a_1 a_2\cdots a_{\al}}_{b_1 b_2 \cdots b_{\bt}}
\bom_{a_1}\bom_{a_2} \cdots \bom_{a_{\al}}
\om_{b_1}\om_{b_2}\cdots \om_{b_{\bt}} \label{3-c3} \eeq where
$\al, \bt = 0,1,2,3,\cdots$ and the coefficients $f^{a_1 a_2\cdots
a_{\al}}_{b_1 b_2 \cdots b_{\bt}}$ should be understood as
generalizations of the above-mentioned $c_a$. The truncated
functions on $S^4$ may be obtained by putting an upper bound for
the value $(\al + \bt)$. We choose this by setting $\al +\bt \le
n$. In (\ref{3-c4}), this choice corresponds to a truncation at
the column which is to be labelled by $(n+1)\times (n+1)$. In
order to count the number of truncated functions in (\ref{3-c3}),
we have to notice the following relation between $\om_a$ and
$\bom_a$ \beq \bom_{a}\om_{a} \sim x_\mu x_\mu = x^2 .
\label{3-c5} \eeq Using this relation, we can contract $\bom_a$'s
in (\ref{3-c4}). For example, we begin with the contractions
involving $\bom_{a_1}$ with all terms in (\ref{3-c4}), which yield
the following new set of terms \beq 1 ~,~ \left(
\begin{array}{c}
\bom_{a_2} \\
\om_{b_1} \\
\end{array}
\right)~,~\left(%
\begin{array}{c}
\bom_{a_2}\bom_{a_3} \\
\bom_{a_2} \om_{b_1} \\
\om_{b_1} \om_{b_2} \\
\end{array}%
\right)~,~ \cdots \cdots \label{3-c6} \eeq The coefficients for
the terms in (\ref{3-c6}) are independent of those for
(\ref{3-c4}), due to the scale invariance $\bpi_\adot \pi_\adot
\sim |\la |^2 $ ($\la \in {\bf C} - \{0\}$) in the contracting
relation (\ref{3-c5}). Consecutively, we can make similar
contractions at most $n$-times. The total number of truncated
functions on $S^4$ is then counted by \beq N^{S^4}(n) \equiv
\sum_{l=0}^{n} \left[ 1^2 + 2^2 + \cdots + (l+1)^2 \right]=
\frac{1}{12}(n+1)(n+2)^2(n+3) \label{3-c7} \eeq which indeed
equals to the previously found results in (\ref{3-r5}) and
(\ref{3-c9}).

From (\ref{3-c4})-(\ref{3-c7}), we find that all the coefficients
in $f_{S^4}(\om, \bom)$ correspond to the number of the matrix
elements for $\hat{F}^{S^4}$ given in (\ref{3-c12}). Further,
since any products of fuzzy functions do not alter their structure
in (\ref{3-c11}), such products correspond to commutative products
of $f_{S^4}(\om, \bom)$'s. This leads to the precise
correspondence between the functions on fuzzy $S^4$ and the
truncated functions on $S^4$ at any level of truncation.

\subsection{Block-diagonal matrix realization of fuzzy $S^4$}

We have analyzed the structure of functions on fuzzy $S^4$ and
their products in some detail, however, we have not presented an
explicit matrix configuration for those fuzzy functions. But, by
now, it is obvious that we can use a block-diagonal matrix to
represent them, which naturally leads to associativity of the
algebra of fuzzy $S^4$. Let us write down the equation
(\ref{3-c12}) in the following form: \beqar N^{S^4}(n)&=& ~~ 1
\nonumber\\&& + 1+2^2 \nonumber\\&& + 1+2^2 +3^2 \nonumber\\&& +
1+2^2 +3^2 + 4^2 \nonumber\\&& + ~~~~\cdots \cdots \cdots
\nonumber\\&& + 1+2^2 +3^2 + 4^2 + \cdots +(n+1)^2 .
\label{3-r6}\eeqar If we locate all the squared elements
block-diagonally, then the dimension of an embedding matrix is
given by \beq \sum_{l=0}^{n} \left[ 1 + 2 + \cdots + (l+1) \right]
=\frac{1}{6}(n+1)(n+2)(n+3) = N^{(3)} . \label{3-r8} \eeq
Coordinates of fuzzy $S^4$ are then represented by these
$N^{(3)}\times N^{(3)}$ block-diagonal matrices, $X_A$, which
satisfy \beq X_A X_A \sim {\bf 1} \label{3-r9} \eeq where ${\bf
1}$ is the $N^{(3)}\times N^{(3)}$ identity matrix and $A=1,2,3,4$
and $5$, four of which are relevant to the coordinates of fuzzy
$S^4$. The fact that $N^{S^4}$ is a sum of absolute squares does
not necessarily warrant associativity of the algebra. (Every
integer is a sum of squares, $1+1+\cdots +1$, but this does not
mean any linear space of any dimension is an algebra.) It is the
structure of $\hat{F}^{S^4}$ as well as the matching between
(\ref{3-c7}) and (\ref{3-c12}) that lead to these block-diagonal
matrices $X_A$.

Of course, $X_A$ are not the only matrices that describe fuzzy
$S^4$. Instead of diagonally locating every block one by one, we
can also put the same-size blocks into a single block, using
matrix multiplication or matrix addition. Then, the final form has
a dimension of $\sum_{l=0}^{n}(l+1)=\hf (n+1)(n+2)= N^{(2)}$. This
implies an alternative description of fuzzy $S^4$ in terms of
$N^{(2)}\times N^{(2)}$ block-diagonal matrices, $\overline{X}_A$,
which are embedded in $N^{(3)}$-dimensional square matrices and
satisfy $\overline{X}_A \overline{X}_A \sim \overline{{\bf 1}}$,
where $\overline{{\bf 1}}=diag(1,1,\cdots,1,0,0,\cdots,0)$ is an
$N^{(3)}\times N^{(3)}$ diagonal matrix, with the number of $1$'s
being $N^{(2)}$. Our choice of $X_A$ is, however, convenient in
comparison with fuzzy $\cp^3$. The number of $1$'s in $X_A$ is
$(n+1)$. This corresponds to the dimension of an $SU(2)$
subalgebra of $SU(4)$ in the $N^{(3)}(n)$-dimensional matrix
representation. (Notice that fuzzy $S^2 =SU(2)/U(1)$ is
conventionally described by $(n+1)\times (n+1)$ matrices in this
context.) Using the coordinates $X_A$, we can then confirm the
constraint in (\ref{3-m2}), {\it i.e.}, \beq [\F(X), L_\al ] = 0
\label{3-r10}\eeq where $\F(X)$ are matrix-functions of $X_A$'s
and $L_\al$ are the generators of $H=SU(2)\times U(1) \subset
SU(4)$, represented by $N^{(3)}\times N^{(3)}$ matrices. If both
$\F(X)$ and $\G(X)$ commute with $L_\al$, so does $\F(X)\G(X)$.
Thus, there is closure of such `functions' under multiplication.
This indicates that fuzzy $S^4$ follows a closed and associative
algebra.

\section{Construction of fuzzy $S^8$}

We outline construction of fuzzy $S^8$ in a way of reviewing our
construction of fuzzy $S^4$. As mentioned in section 3.1, $\cp^7$
is a $\cp^3$ bundle over $S^8$. We expect that we can similarly
construct fuzzy $S^8$ by factoring out fuzzy $\cp^3$ from fuzzy
$\cp^7$.

The structure of fuzzy $S^4$ as a block-diagonal matrix has been
derived, based on the following two equations \beqar N^{S^4}(n)
&=& \sum_{l=0}^{n} \left( N^{(1)}(l)\right)^2 ~
 N^{(1)}(n-l) \, ,\label{3-e1} \\
N^{(3)}(n) &=& \sum_{l=0}^{n} N^{(1)}(l) ~ N^{(1)}(n-l)
\label{3-e2} \eeqar where $N^{(k)}(l) = \frac{(l+k)!}{k! ~ l!}$ as
in (\ref{2-1a}). Fuzzy $S^8$ analogs of these equations are \beqar
N^{S^8}(n) &=&
\sum_{l=0}^{n} N^{S^4}(l) ~ N^{(3)}(n-l) \, , \label{3-e3} \\
N^{(7)}(n) &=& \sum_{l=0}^{n} N^{(3)}(l) ~ N^{(3)}(n-l)
\label{3-e4} \eeqar where $N^{S^8}(n)$ is the number of truncated
functions on $S^8$, which can be calculated in terms of the
spherical harmonics as in the case of $S^4$ in (\ref{3-r5});
\beqar N^{S^8}(n) &=&
\sum^{n}_{a=0}\sum^{a}_{b=0}\sum^{b}_{c=0}\sum^{c}_{d=0}
\sum^{d}_{e=0}\sum^{e}_{f=0}\sum^{f}_{g=0}(2g +1) \nonumber \\ &=&
\frac{1}{4\cdot 7!}(n+1)(n+2)(n+3)(n+4)^2(n+5)(n+6)(n+7) \, .
\nonumber \\
\label{3-55} \eeqar This number is also calculated by a tensor
analysis as in (\ref{3-c9}); \beqar N^{S^8}(n) &=&
\sum^{n}_{l=0}\frac{\dim(l,l)}{N^{(6)}(l)} \nonumber \\
&=&\sum^{n}_{l=0}\frac{1}{7!}(2l+7)(l+1)(l+2)(l+3)(l+4)(l+5)(l+6)
\nonumber \\ &=& \frac{n+4}{4}\,\frac{(n+7)!}{7!~n!}
\label{3-56}\eeqar where $\dim(l,l)$ is the dimension of $SU(8)$
in the $(l,l)$-representation, {\it i.e.},
$\dim(l,l)=\frac{1}{7!~6!}(2l+7)\left( (l+1)(l+2)\cdots
(l+6)\right)^2$. Calculations from (\ref{3-e1}) to (\ref{3-56})
are carried out by use of {\it Mathematica}.

Equations (\ref{3-e3}) and (\ref{3-e4}) indicate that fuzzy $S^8$
is composed of $N^{(3)}(l)$ dimensional block-diagonal matrices of
fuzzy $S^4$ ($l=0,1,\cdots,n$), with the number of these matrices
for fixed $l$ being $N^{(3)}(n-l)$. Thus fuzzy $S^8$ is also
described by a block-diagonal matrix whose embedding square matrix
has a dimension $N^{(7)}(n)$. Notice that we have a nice
matryoshka-like structure for fuzzy $S^8$, namely, a fuzzy-$S^8$
box is composed of a number of fuzzy-$S^4$ blocks and each of
those blocks is further composed of a number of fuzzy-$S^2$
blocks. Fuzzy $S^8$ is then represented by $N^{(7)}\times N^{(7)}$
block-diagonal matrices $X_A$ which satisfy $X_A X_A \sim {\bf 1}$
($A=1,2,\cdots,9$), where ${\bf 1}$ is the $N^{(7)}\times N^{(7)}$
identity matrix. Similarly to the case of fuzzy $S^4$, fuzzy $S^8$
should also obey a closed and associative algebra.

Let us now consider the decomposition \beq SU(8) \rightarrow SU(4)
\times \underbrace{SU(4) \times U(1)}_{= H^{(4)}} \label{3-e5}
\eeq where the two $SU(4)$'s and one $U(1)$ are defined similarly
to (\ref{3-m1}) in terms of the generators of $SU(8)$ in the
fundamental representation. Noticing the fact that the number of
$1$-dimensional blocks in the coordinate $X_A$ of fuzzy $S^8$ is
$N^{(3)}(n)$, we find $[X_A \, , \, L_\al]=0$ where $L_\al$ are
now the generators of $H^{(4)}$ represented by $N^{(7)} \times
N^{(7)}$ matrices. This is in accordance with the statement that
functions on $S^8$ are functions on $\cp^7 = SU(8)/U(7)$ which are
invariant under transformations of $H^{(4)}=SU(4)\times U(1)$.
Coming back to the original idea, we can then construct fuzzy
$S^8$ out of fuzzy $\cp^7$ by imposing the particular constraint
$[\F, L_\al] = 0$, where $\F$ are matrix-functions of coordinates
$Q_A$ on fuzzy $\cp^7$, $Q_A$ being defined as in (\ref{2-1}) for
$k=7$. This constraint is imposed on the function $\F(Q_A)$, on
top of the fuzzy $\cp^7$ constraints for $Q_A$, so that it becomes
a function on fuzzy $S^8$, that is, a polynomial of $X_A$'s.

Following the same method, we may construct higher dimensional
fuzzy spheres \cite{ram2,kim2,dop}. But we are incapable of doing
so as far as we utilize bundle structures analogous to $\cp^3$ or
$\cp^7$. This is because, as far as complex number coefficients
are used, there are no division algebra allowed beyond octonions.
The fact that $\cp^7$ is a $\cp^3$ bundle over $S^8$ is based on
the fact that octonions provide the Hopf map, $S^{15} \rightarrow
S^8$ with its fiber being $S^7$. Since this map is the final Hopf
map, there are no more bundle structures available to construct
fuzzy spheres in a direct analogy with the constructions of fuzzy
$S^8$, $S^4$ and $S^2$.

\chapter{Matrix models for gravity}

From this chapter on, applications of fuzzy spaces to physical
models will be discussed.

\section{Introduction to noncommutative gravity}

As mentioned in section 1.3, noncommutative spaces can arise
as solutions in string and M-theories. Fluctuations of brane
solutions are described by gauge theories on such spaces and, with
this motivation, there has recently been a large number of papers
dealing with gauge theories, and more generally field theories, on
noncommutative spaces (see, {\it e.g.},
\cite{BKVrev,NC,Harvey,Hama}). There is also an earlier line of
development in close connection with Connes' original idea, using
the spectral triple and the so-called `spectral actions'
\cite{ConLot}-\cite{Landi3}.

Even apart from their string and M-theory connections,
noncommutative spaces are interesting for other reasons. Many of
the noncommutative spaces recently discussed have an underlying
Heisenberg algebra for different coordinates. Lie algebra
structures are more natural from a matrix model point of view;
these typically lead to noncommutative analogues of compact spaces
and, in particular, fuzzy spaces. Because these spaces are
described by finite dimensional matrices, the number of possible
modes for fields on such spaces is limited and so one has a
natural ultraviolet cutoff. We may think of such field theories as
a finite mode approximation to commutative continuum field
theories, providing, in some sense, an alternative to lattice
gauge theories. Indeed, this point of view has been pursued in
some recent work (see, {\it e.g.}, \cite{BGY},
\cite{Iso1}-\cite{Ydri2}). While lattice gauge theories may be
most simply described by standard hypercubic lattices, gravity is
one case where the noncommutative approach can be significantly
better. This can provide a regularized gravity theory preserving
the various desirable symmetries, which is hard to do with
standard lattice versions. It would be an interesting alternative
to the Regge calculus, which is essentially the only
finite-mode-truncation of gravity known with the concept of
coordinate invariance built in. A finite-mode-truncation is not
quantum gravity, but it can give a formulation of standard gravity
where questions can be posed and answered in a well defined way.

Partly with this motivation, a version of gravity on
noncommutative spaces has been suggested by Nair in
\cite{NairNCg}. This led to an action for even dimensional, in
particular four-dimensional, noncommutative spaces generalizing
the Chang-MacDowell-Mansouri approach used for commutative
four-dimensional gravity \cite{CMM}. In this chapter, we shall
consider the case of fuzzy $S^2$ in some detail, setting up the
required structures, eventually obtaining an action for gravity in
terms of ($N\times N$)-matrices. The large $N$ limit of the action
will give the usual action for gravitational fields on $S^2$. We
also construct a finite-dimensional matrix model action for
gravity on fuzzy $\cp^2$ and indicate how this may be generalized
to fuzzy $\cp^k$ ($k=1,2,\cdots $).

\section{Derivatives, vectors, etc.}

We shall primarily be concerned with fuzzy versions of coset
spaces of the form $G/H$ for some compact Lie group $G$, $H$ being
a subgroup of $G$. Most of our discussion will be based on $S^2 =
SU(2)/U(1)$. Functions on fuzzy $S^2$ are given by $(N\times
N)$-matrices with elements $f_{mn}$. As given in (\ref{3-s2-1}),
the symbol of these fuzzy functions are expressed as $\< \hat{f}
\> = \sum_{m,n} f_{mn} {\cal D}^{*(j)}_{mj} (g) {\cal
D}^{(j)}_{nj} (g)$, where $\D^{(j)}_{mk} (g)$ are Wigner
$\D$-functions for $SU(2)$ belonging to the spin-$j$
representation. The matrix dimension $N$ is given by $N=2j+1$. In
this way of representing functions, derivatives may be realized as
the right translation operators $R_a$ on $g$, \beq R_a \cdot
\D^{(j)}_{mk}(g) = \left[ \D^{(j)}\left(g~{t_a}\right)
\right]_{mk} \label{4-2} \eeq where $t_a =\sigma_a/2$, with
$\sigma_a$ being the Pauli matrices. In order to realize various
quantities, particulary an action, purely in terms of matrices, we
need to introduce a different but related way of defining
derivatives, vectors, tensors, etc., on a fuzzy coset space.

Let $g$ denote an element of the group $G$ and define
 \beq S_{Aa}~= ~ 2 ~\tr (g^{-1} t_A g t_a ) \label{4-3} \eeq
where $t_a$ and $t_A$ are hermitian matrices forming a basis of
the Lie algebra of $G$ in the fundamental representation. We
normalize these by $\tr (t_a t_b)= \hf \del_{ab}$, $\tr (t_A t_B
)= \hf \del_{AB}$. The distinction between upper and lower case
indices is only for clarity in what follows. For $SU(2)$, $a, A
=1,2,3$ and $S_{Aa}$ obey the relations \beqar
 S_{Aa} ~S_{Ab} &=& \delta_{ab}~, \nonumber\\
 S_{Aa} ~ S_{Ba} &=& \delta_{AB}~, \nonumber\\
 \ep_{ABC}~ S_{Aa}~ S_{Bb} &=& \ep_{abc} S_{Cc}~,\nonumber \\
 \ep_{abc}~ S_{Aa} ~S_{Bb} &=& \ep_{ABC} S_{Cc}~.
 \label{4-4} \eeqar

Let $L_A$ be the $(N\times N)$-matrix representation of the
$SU(2)$ generators, obeying the commutation rules $[L_A, L_B] = i
\ep_{ABC} L_C$. We then define the operators
 \beq {\cal K}_a ~= ~S_{Aa} L_A - \hf R_a \label{4-5} \eeq where
$R_a$ are the right translation operators, $R_a g = g t_a$. One
can think of them as differential operators \beq R_a ~=~ i
(E^{-1})^{i}_{a} \frac{\d}{\d \vf^i} \label{4-5a} \eeq in terms of
the group parameters $\vf^i$ and the frame field $E_{i}^{a}$,
satisfying
 \beq g^{-1} d g ~=~ (-it_a)~ E^{a}_{i} d\vf^i . \label{4-5b}\eeq
$R_a$ obey the commutation rules $[R_a ,R_b]= i \ep_{abc} R_c$. We
then find \beq \left[ {\cal K}_a , {\cal K}_b \right]~ =~ {i\over
4} \epsilon_{abc} ~ R_c \, . \label{4-6}
 \eeq Identifying the $U(1)$ subgroup generated by $t_3$ as the
$H$-subgroup, we define derivatives on fuzzy $S^2$ as ${\cal
K}_\pm = {\cal K}_1 \pm i {\cal K}_2$. Notice that this is a
hybrid object, being partially a matrix commutator and partially
something that depends on the continuous variable $g$. This is
very convenient for our purpose and in the end $g$ will be
integrated over anyway.

We now define a matrix-function $f$ on fuzzy $S^2$ with no
$g$-dependence. The derivative of $f$ is then defined as \beq
{\cal K}_\mu \cdot f ~\equiv ~\left[ {\cal K}_\mu , f \right]~ =~
S_{a\mu} [L_A ,f]
 \label{4-7} \eeq where $\mu = \pm$. Since $\left[ {\cal
K}_+ , {\cal K}_- \right] = \hf R_3$ from (\ref{4-6}), we find
$\left[ {\cal K}_+ , {\cal K}_- \right] \cdot f = 0$, consistent
with the expectation that derivatives commute when acting on a
function. Equation (\ref{4-7}) also shows that it is natural to
define a vector on fuzzy $S^2$ as \beq V_\mu ~=~ S_{A\mu}~ V_A
\label{4-8} \eeq where $V_A$ are three $(N\times N)$-matrices. On
a two-sphere, a vector should only have two independent
components, so this is one too many and $V_A$ must obey a
constraint. Notice that the quantity $[L_A, f]$ obeys the
condition $L_A [L_A,f] + [L_A,f]L_A =0$, since $L_A L_A$ is
proportional to the identity matrix. This suggests that the
correct constraint for a general vector is $L_A V_A + V_A L_A =0$.
In the large $N$ limit, $L_A$ become proportional to $x_A$, the
commutative coordinates of the two-sphere as embedded in ${\bf
R}^3$ (with $x_A x_A = 1$). So the condition $x\cdot V=0$ is
exactly what we need to restrict the vectors to directions
tangential to the sphere. We may thus regard $L_A V_A + V_A L_A
=0$ as the appropriate fuzzy version. As we shall see below this
constraint will also emerge naturally when we define integrals on
fuzzy $S^2$. Using \beq [R_a , S_{Ab} ] ~=~ i \epsilon_{abc}
S_{Ac} \, , \label{4-9} \eeq we find \beq \left[ {\cal K}_+ ,
{\cal K}_- \right] \cdot V_\pm ~=~ \pm\hf ~ V_\pm\label{4-10} \eeq
which is consistent with the Riemann curvature of $S^2$; $R^+_{+-
\, +} = - R^-_{+- \, -}= \hf$. Higher rank tensors may also be
defined in an analogous way with several $S_{Aa}$'s, {\it i.e.},
$T_{{\mu_1}{\mu_2}\cdots{\mu_r}} \equiv S_{A_1 \mu_1}S_{A_2
\mu_2}\cdots S_{A_r \mu_r}~ T_{A_1 A_2\cdots A_r}$.

We now turn to a definition of `integration' on fuzzy $S^2$. We
will only need, and will only define, integration of the fuzzy
analogue of an antisymmetric rank-2 tensor or a two-form. Such a
quantity has components of the form $W_{+-}= (S_{A+} S_{B-}-
S_{A-} S_{B+} ) W_{AB}$. From the properties of $S_{Aa}$, we find
$S_{A+} S_{B-}- S_{A-} S_{B+}= -2i\epsilon_{ABC} S_{C3}$.
Integration of $W_{+-}$ over $g$ (with the trace of the matrix
$W_{AB}$) will give zero. To get a nonzero integral we must
introduce a density factor $\rho$. Such a factor must commute with
$R_3$ to be properly defined on $SU(2)/U(1)$ and must give nonzero
upon $g$-integration with $S_{C3}$. The only choice is $\rho =
{1\over 3} S_{K3} L_K$, up to normalization, which can be
determined by $\int_g S_{K3} S_{C3} = 3 \del_{KC}$ where factor 3
corresponds to $\dim SU(2)$. The appearance of such a density
factor is actually very natural. If we consider a commutative
$S^2$ embedded in ${\bf R}^3$ with coordinates $x_A$, then $x_A =
S_{A3}$ in a suitable parametrization. The usual volume element is
oriented along $x_A = S_{A3}$ and so we can expect a factor $\rho
= {1\over 3} S_{K3}L_K$ in the fuzzy case. With the introduction
of the factor $\rho$, we can consider an `integral' of the form
$\int_g \Tr (\rho W)$. However, if we consider $\int_g \Tr (\rho W
f)$ where $f$ is a function, we do not have the expected cyclicity
property since $[\rho ,f]\neq 0$ in general. Cyclicity property
can be obtained if we symmetrize the factors inside the trace
except the density factor $\rho$. Gathering these points, we now
define an `integral' over fuzzy $S^2$, denoted by $\slashint$, as
follows:
 \beq \slashint A_1 A_2 \cdots A_l ~=~ \int_g ~ \Tr ~ \left[ \rho
 ~{1\over l}\sum_{cycl.} (A_1 A_2 \cdots A_l ) \right]
 \label{4-11}
\eeq where $A_1, A_2, \cdots ,A_l$ are functions, vectors,
tensors, etc., such that the product is an antisymmetric rank-2
tensor (of the form $W_{+-}$), {\it i.e.}, a fuzzy analogue of a
two-form on $S^2$. The summation in (\ref{4-11}) is taken over
cyclic permutations of the arguments. Note that we can express
such a two-form as $A_1 A_2 \cdots A_l = (-2i) \ep_{ABC} S_{C3}
(A_1 A_2 \cdots A_l)_{AB}$. So the integral is further written as
\beqar \slashint A_1 A_2 \cdots A_l &=& (-2i) \ep_{ABC}~ \Tr~
\left[ L_C
\frac{1}{l} \sum_{cycl.} (A_1 A_2 \cdots A_l)_{AB} \right] \nonumber \\
&=& (-2i) \ep_{ABC}~ \STr~ \left[ L_C  A_1 A_2 \cdots A_l
\right]_{AB} \label{4-11a} \eeqar where $\STr$ is the symmetrized
trace over the $(N \times N)$ matrices inside the bracket, the
lower indices $A$, $B$ being assigned to some of the matrices in
$A_1$, $A_2$, etc.

In a similar fashion, we now consider a fuzzy analogue of an
exterior derivative, in particular, the analogue of a two-form
corresponding to the curl of a vector $V_\mu = S_{A\mu}V_A$, $\mu
=\pm$. Since we have defined ${\cal K}_{\pm}$ as derivatives on
fuzzy $S^2$, a fuzzy analogue of such a term can be given by
\beqar dV
&\equiv& [{\cal K}_+ ,V_-] - [{\cal K}_- ,V_+] \nonumber\\
 &=& \left( S_{A+}S_{B-} - S_{A-}S_{B+} \right) [L_A,V_B]
   ~-2S_{C3} V_C \nonumber\\ &=& (-2i) ~S_{C3}
   \left( \ep_{ABC} [L_A,V_B] -i V_C \right) \, .\label{4-12} \eeqar
If $h$ is a function on fuzzy $S^2$, we also have
 \beqar V ~dh &\equiv& V_+ [{\cal K}_- ,h] - V_-
 [{\cal K}_+ ,h] \nonumber\\
 &=& (-2i) ~\ep_{ABC} ~S_{C3} V_A [L_B ,h] \, . \label{4-13} \eeqar
Using the definition of the integral (\ref{4-11}) we find \beqar
\slashint dV~h &=& (-2i) {1\over 2}~ \Tr ~\biggl[~ L_K \left\{
\ep_{ABC}
 [L_A,V_B] -i V_C\right\}~h  \nonumber \\
 && ~~~~~~ +\left\{ \ep_{ABC} [L_A,V_B] -i
 V_C\right\} L_K
 ~h~ \biggr] \int_g {1\over 3} S_{K3}S_{C3}
 \nonumber\\
 &=& (-2i) {1\over 2}~ \Tr~ \biggl[~ L_C \left\{ \ep_{ABC}
 [L_A,V_B] -i
 V_C\right\}~h  \nonumber \\
 && ~~~~~~~~~~~~~~~~
 + \left\{ \ep_{ABC} [L_A,V_B] -i V_C\right\} L_C ~h ~\biggr]
 \label{4-14} \eeqar
where we used $\int_g S_{K3} S_{C3} = 3\delta_{KC}$. Similarly we
have \beq \slashint V~dh ~=~ (-2i)
 {1\over 2}~\Tr~ \biggl[ \ep_{ABC} (L_C V_A +V_A L_C)
 [L_B,h]\biggr] \, .
 \label{4-15} \eeq
By using cyclicity of the trace for the finite dimensional
matrices $L_A, V_B, h$, etc., we find that the desired partial
integration property \beq \slashint ~dV~h ~ =~ \slashint ~ V~dh
\label{4-16} \eeq holds if $V_A$ obey the constraint \beq L_A V_A
+ V_A L_A ~=~ 0 \, . \label{4-17} \eeq This relation has been
introduced earlier based on geometric properties of $S^2$. We have
now justified this relation as a correct constraint for vectors on
fuzzy $S^2$, based on integration properties. When $V_A$ are gauge
fields, this constraint will have to be slightly modified for
reasons of gauge invariance. The relevant constraint is shown in
(\ref{4-20a}).

\section{Action for gravity on fuzzy $S^2$}

We are now in a position to discuss actions for gravity on fuzzy
$S^2$. As mentioned in chapter 1, we follow the proposal of
\cite{NairNCg} for the action of gravity on noncommutative $G/H$
space, where the gravitational fields ({\it i.e.}, frame fields
$e_\mu$ and spin connections $\Om_\mu$) are described by $U(k)$
gauge fields, with $U(k)$ being specified by $G \subseteq U(k)$.
In our case, the gauge group is then chosen as $U(2)$ and the
gauge fields are written as \beqar
 {\cal A}_\mu &=& {\cal A}_\mu^a I^a \nonumber\\
&=& e^+_\mu I^+ + e^-_\mu I^- +\Omega^3_\mu I^3 +\Omega^0_\mu I^0
. \label{4-18} \eeqar The components $(\Omega^0_\mu ,
~\Omega^3_\mu,~e^{\pm}_{\mu} )$ are vectors on fuzzy $S^2$ as
defined in the previous section. The upper indices of these
vectors correspond to components for the Lie algebra of $U(2)$,
$(I^0, I^3, I^\pm)$, form the $(2\times 2)$-representation of
$U(2)$. Specifically, in terms of the Pauli matrices $\sigma_i$,
$I^0 =\hf {\bf 1}$, $I^3 =\hf \sigma_3$, $I^\pm = \hf ( \sigma_1
\pm i \sigma_2)$. ${\cal A}_\mu$ is thus a vector on fuzzy $S^2$
which also takes values in the Lie algebra of $U(2)$. This $U(2)$
is the group acting on the upper indices of ${\cal A}_\mu$ or the
tangent frame indices. Notice that, with $L_A$, $R_a$ and the
$I$'s, we have three different actions for $SU(2)$. In terms of
${\cal A}_\mu$ we now define a field strength $F_{\mu \nu}$ as
\beq [{\cal K}_\mu + {\cal A}_\mu
 , {\cal K}_\nu + {\cal A}_\nu ] ~=~ {i\over 4}
 \ep_{\mu\nu\alpha} R_\alpha ~+~ F_{\mu\nu} \, . \label{4-19} \eeq

In our description, gravity is parametrized in terms of deviations
from $S^2$. The vectors $e^{\pm}_{\mu}$ are the frame fields for
this and $\Om_\mu^{\al}$ ($\al=0,3$) are the spin connections. As
opposed to the commutative case, there can in general be a
connection for the $I^0$ component, since we need the full $U(2)$
to form noncommutative gauge fields. One can expand $F_{\mu\nu}$
as \beq
 F_{\mu\nu}= F_{\mu\nu}^0 ~I^0 ~+~ {\cal R}_{\mu\nu}^3 ~I^3
 ~+~ {\cal T}_{\mu\nu}^a ~I^a  \label{4-20} \eeq
where ${\cal T}_{\mu\nu}^a$ is the torsion tensor and ${\cal
R}_{\mu\nu}^3$ is of the form $R_{\mu\nu}(\Omega ) + 2 (e^+_\mu
e^-_\nu - e^-_\mu e^+_\nu )$ where $R_{\mu\nu}(\Omega )$ is the
Riemann tensor on commutative $S^2$. The expression for ${\cal
R}_{\mu\nu}^3$ is thus a little more involved for fuzzy $S^2$.

In defining an action, we shall use our prescription for the
integral. The gauging of ${\cal K}_\mu$ is equivalent to the
gauging $L_A \rightarrow L_A + {\cal A}_A$. Thus we must also
change our definition of $\rho$ to $\rho ={1\over 3} S_{K3} (L_K
+{\cal A}_K)$. The constraint (\ref{4-17}) is now replaced by \beq
(L_A +{\cal A}_A) (L_A +{\cal A}_A) = L_A L_A \, . \label{4-20a}
\eeq Note that ${\cal A}_A$ is expanded in terms of the $I^a$ as
in (\ref{4-18}). This constraint was first proposed in \cite{KNP}
as the correct condition to be used for gauge fields on fuzzy
$S^2$.

The data for gravity is presented in the form of the gauge field
${\cal A}_A$. Following the action suggested in \cite{NairNCg}, as
a generalization McDowell-Mansouri approach for commutative
gravity, we can express an action for gravity on fuzzy $S^2$ as
 \beq {\cal S}~ = ~ \alpha ~ \slashint ~\tr ~( Q F ) \label{4-21} \eeq
where $\tr$ denotes the trace over the $I$'s regarded as $(2\times
2)$-matrices. $F$ denotes a two-form on fuzzy $S^2$ corresponding
to the field strength; it is in general expressed by $F= F^a I^a$,
being in the algebra of $U(N) \otimes U(2)$. For higher
even-dimensional $G/H$-spaces, the actions are given in the
following form \cite{NairNCg}: \beq \S ~ \sim ~ \slashint ~ \tr ~
(QFF...F) \label{4-21a} \eeq where $Q$ is a combination of the
$I$'s which commutes with the $H$-subgroup of $G$. For the present
case, we can choose $Q =I^3$. However, unlike the case of four and
higher dimensions, the term involving $F^0$ in $\slashint \tr (I^3
F)$ vanishes, which would be the fuzzy analogue of the statement
that the two-dimensional Einstein-Hilbert action $\int R \sqrt{g}$
is a topological invariant. As in the commutative context, we may
need to use a Lagrange multiplier scalar field $\eta$ to obtain
nontrivial actions. In the present case, the analogous action is
given by \beq \S ~=~ \al ~ \slashint ~\tr ~ (I^3 \eta F )
\label{4-22} \eeq where $\eta = \eta^0 I^0 + \eta^3 I^3 + \eta^+
I^+ + \eta^- I^-$, $(\eta^0 ,\eta^3, \eta^\pm)$ being scalar
functions on fuzzy $S^2$. Using the decomposition (\ref{4-20}) for
the field strength, we can simplify this expression as \beq
 \S ~=~ -i~{\alpha \over 2}~\Tr ~ \biggl[ I^3 \eta \left[~(L_C+{\cal A}_C)
F_C + F_C (L_C+{\cal A}_C)~ \right]\biggr] \label{4-22a} \eeq
where $F_C$ is defined as follows:
\beqar
	&& \!\!\!\!\!\!\!\!  F_C  \equiv  F_C^0 I^0
	+ F_C^3 I^3 +
	F_C^+ I^+ + F_C^- I^-  \, ,
	\label{4-22b} \\
	&& \!\!\!\!\!\!\!\!  F^0_C = \hf
	\left\{[L_A,\Om^0_B] + \hf (\Om^0_A \Om^0_B + \Om^3_A
	\Om^3_B)
	+(e^{+}_A e^{-}_B + e^{-}_A e^{+}_B)\right\}\epsilon_{ABC}
	-{i\over 2} \Omega^0_C  \, ,
	\label{4-23} \\
	&& \!\!\!\!\!\!\!\!  F^3_C = \hf \left\{ [L_A,\Om^3_B] + \hf (\Om^0_A \Om^3_B
	+\Om^3_A \Om^0_B) +(e^{+}_A e^{-}_B - e^{-}_A
	e^{+}_B)\right\}
	\epsilon_{ABC}
	-{i\over 2} \Omega^3_C  \, ,
	\label{4-24} \\
	&& \!\!\!\!\!\!\!\!  F^-_C = \hf \left\{[L_A,e^-_B] + \hf e^-_A (\Om^0_B +
	\Om^3_B )
	-\hf (\Om^0_B-\Om^3_B) e^-_A\right\} \epsilon_{ABC}
	-{i\over 2}e^-_C \, ,
	\label{4-25} \\
	&& \!\!\!\!\!\!\!\!  F^+_C = \hf \left\{[L_A,e^+_B] + \hf e^+_A (\Om^0_B -
	\Om^3_B )
	-\hf (\Om^0_B +\Om^3_B) e^+_A\right\}\epsilon_{ABC}
	-{i\over 2} e^+_C \, .
	\label{4-26}
\eeqar

Equation ({\ref{4-22a}), with (\ref{4-22b})-(\ref{4-26}), is the
action for gravity on fuzzy $S^2$. They are expressed entirely in
terms of finite dimensional $(N\times N)$-matrices, $L_A,
e^{\pm}_A$, and $\Om_A^{\al}$ ($\al =0,3$). As mentioned earlier,
this action provides a new regularization scheme for gravity, and,
in principle, we can calculate many interesting physical
quantities, correlation functions in particular, from
(\ref{4-22a})-(\ref{4-26}) by analyzing it as a matrix model.

In what follows, we shall analyze the action (\ref{4-22a}) a bit
further and discuss its commutative limit. Variations of the
action with respect to $\et$'s provide four equations of motion,
{\it i.e.}, \beq {\cal F}^a ~ \equiv ~ \left[(L_C+{\cal A}_C) F_C
+ F_C (L_C+{\cal A}_C)\right]^a =0\label{4-27} \eeq for $a = 0,~3,
\pm$. The components $a =\pm$ correspond to the vanishing of
torsion. ${\cal F}^3$ is not quite the Riemann tensor associated
with $\Om^3$, due to the $e^+e^-$-term. The vanishing of ${\cal
F}^3$ shows that the Riemann tensor is proportional to the
$e^+e^-$-term.

There are also equations of motion associated with the variation
of the $e^\pm ,~\Om^3 ,~\Om^0$, which are equations coupled to
$\eta$'s. We do not write them out here, they can be easily worked
out from the expressions (\ref{4-23})-(\ref{4-26}) for the
$F_C$'s. Notice however that one solution of such equations of
motion is easy to find. The variation of the action with respect
to the $e^\pm ,~\Om^3 ,~\Om^0$ is of the form \beq
 \del \S ~=~
 -i{\alpha \over 2} ~\Tr ~ \biggl[ ~I^3 ~ \eta ~
\delta ~ \left[(L_C+{\cal A}_C) F_C + F_C (L_C+{\cal A}_C)\right]
~ \biggr] \, .\label{4-28} \eeq This evidently shows that $\eta
=0$ is a solution.

The equations for the connections $e^\pm ,~\Om^3 ,~\Om^0$ in
(\ref{4-27}) are also solved by setting all $F_{\mu\nu}$ to zero.
This corresponds to the pure gauge solutions, {\it i.e.}, the
choice of ${\cal A}_\mu = S_{B\mu} {\cal A}_B$, ${\cal A}_B = i
U^{-1} [\hat{L}_B, U]$ where $U$ is a matrix which is an element
of $U(N)\otimes U(2)$, and $\hat{L}_B$ is viewed as $L_B \otimes
{\bf 1}$. In other words, it is an element of $U(2)$ with
parameters which are $(N\times N)$-matrices. This solution
corresponds to the fuzzy $S^2$ itself.

\section{Commutative limit}

We now consider the commutative limit of the action (\ref{4-22a})
by taking the large $N$ limit. The matrices $L_A$'s are matrix
representations of the generators of $SU(2)$ in the spin
$n/2$-representation. The matrix dimension $N$ is then given by
$N=n+1$. We introduce the states of fuzzy $S^2$, $|\al \>$
$(\al=0,1,\cdots,n)$, characterized by $\<z | \al \> = 1, z,
\cdots, z^n$ for each $\al$. The operators $\L_A$ acting on such
states can be expressed as \cite{nair2}:
 \beqar \L_+ &=& \frac{n+2}{2} \p_+ + z^2 \frac{\d}{\d z} ~,~~~ \p_+
~ =~ \frac{2z}{1+z\bz}~, \nonumber \\
\L_- &=& \frac{n+2}{2} \p_- - \frac{\d}{\d z} ~, ~~~ \p_- ~=~
 \frac{2\bz}{1+z\bz} ~,\nonumber \\
\L_3 &=& \frac{n+2}{2} \p_3 + z \frac{\d}{\d z} ~ , ~~~ \p_3 ~ =~
   \frac{1-z\bz}{1+z\bz}  \label{4-30} \eeqar
where $\p$'s are the coordinates on $S^2$, obtained by usual
stereographic projection on a complex plane. Note that $\L_A$'s
correspond to those obtained in (\ref{2-3a5}) for $k=1$. Using
such a Hilbert space, we can consdier the vectors ($\Om$'s and
$e$'s) as functions of $z, \bz$. Large $n$ limits of the matrix
operator $L_A$ and the commutator $[L_A, \Om_B]$ can then be given
by the following replacements: \beqar
 L_A & \longrightarrow & \frac{n+2}{2} \p_A \, , \label{4-31} \\
 \left[L_A, \Om_B\right] & \longrightarrow & \frac{1}{n+1}
 \left\{\frac{n+2}{2}\p_A , \Om_B \right\} \nonumber \\
 &=&  \frac{n+2}{2}\frac{1}{n+1}(1+z\bz)^2(\d \p_A \bd
 \Om_B -\bd \p_A \d \Om_B) \nonumber \\
 &=& \frac{n+2}{2} \frac{1}{n+1}~k_A ~\Om_B \label{4-31'}
 \eeqar
where $\d=\frac{\d}{\d z}$, $\bd=\frac{\d}{\d \bz}$ and the
operators $k_A$ are defined in terms of a Poisson bracket $k_A
\Om_B \equiv \left\{\p_A , \Om_B \right\} = (1+z\bz)^2(\d \p_A \bd
\Om_B -\bd \p_A \d \Om_B)$ with \beq k_+ ~=~ 2( z^2 \d + \bd) \, ,
~~ k_- ~=~ -2( \d + \bz^2 \bd) \, , ~~
 k_3 ~=~ 2( z \d - \bz \bd) \, . \label{4-31a} \eeq
Notice that $\hf k_A$ satisfy the $SU(2)$ algebra; \beq
\left[\frac{k_+}{2},\frac{k_-}{2}\right]=2\frac{k_3}{2} \, ,~~
\left[\frac{k_3}{2},\frac{k_+}{2}\right]=\frac{k_+}{2} \, ,~~
\left[\frac{k_3}{2},\frac{k_-}{2}\right]=-\frac{k_-}{2} \, .
\label{4-31b} \eeq Actions of $k_A$'s on $\p_B$'s can be
calculated as \beqar && k_+ \p_+=0 \, ,~~~ k_+\p_-=4\p_3
\, ,~~~ k_+\p_3=-2\p_+ \, , \nonumber \\
&& k_-\p_+ = -4\p_3 \, ,~~~ k_-\p_-=0
\, ,~~~ k_-\p_3=2\p_- \, , \label{4-31c} \\
&& k_3\p_+ = 2\p_+ \, ,~~~ k_3\p_-=-2\p_- \, ,~~~ k_3\p_3=0 \, .
\nonumber \eeqar The replacement of commutator with Poisson
bracket in (\ref{4-31'}) is analogous to the passage from the
quantum theory to the classical theory, $1/(n+1)$ serving as the
analogue of Planck's constant. As we have discussed, this
correspondence can be best seen by geometric quantization of
$S^2$. Notice also that, because of (\ref{4-31}), the term $L_A$
dominates in the expression of $L_A +{\cal A}_A$ for large values
of $n$.

It is instructive to consider the large $n$ limit of one of the
terms in the action, say the term involving $\eta^0$, in some
detail. Denoting this term as $\S [\eta^0]$ and using
(\ref{4-31}), (\ref{4-31'}), we find
\beqar
	\S[\eta^0] \!\! &=& \!\!
	-i\frac{\al}{2} \left( \frac{n+2}{2}\right)
	~\ep_{ABC} \nonumber \\
	&&~~~ \Tr ~ \biggl[ \et^0 \left(
	 \frac{n+2}{2(n+1)}\p_C k_A \Om^3_B + \p_C (e^+_A e^-_B -
	e^-_A e^+_B) \right) + \O \left(\frac{1}{n}\right)
	 \biggr] \nonumber \\
	 &\approx & \!\!
	-i \frac{\bt}{n+1}~\ep_{ABC}
	\, \Tr ~ \biggl[\et^0  \p_C \left( \hf ~ k_A \Om^3_B + (e^+_A
	e^-_B - e^-_A e^+_B) \right) + \O \left(\frac{1}{n}\right) \biggr]
	 \label{4-32}
\eeqar
where $\bt = \frac{\al}{2} \left(
\frac{n+2}{2}\right) (n+1)$. This will be taken as an
$n$-independent constant. In carrying out these simplifications,
it is useful to keep in mind that the $\Om_A$ obey the constraint
\beq \p_A~\Om_A ~+~ \Om_A~\p_A \approx ~2~ \phi_A ~\Om_A ~\approx
~0 \label{4-38} \eeq which is a natural reduction of the
constraint for vectors on fuzzy $S^2$ as shown in (\ref{4-17}).
Since $\hf k_A$ can serve as derivative operators on $S^2$, we can
define $k_A \Om_B$ as \beq k_A \Om_B ~ = ~ 2 \frac{\d}{\d \p^A}
\Om_B ~\equiv ~2~ \d_A \Om_B \, . \label{4-39} \eeq As in the
general case given in (\ref{2-2b2}), the trace over ($N\times
N$)-matrices can be replaced by the integral over $z$ and $\bz$;
\beq \frac{1}{n+1}~\Tr ~\longrightarrow ~ \int \frac{dz d\bz} {\pi
(1+ z \bz)^2} ~ \equiv ~ \int_{z, \bz} \, . \label{4-40} \eeq We
can now rewrite (\ref{4-32}) as \beq \S[\et^0] ~ \approx ~ -i \bt
\ep_{ABC} \int_{z, \bz} \et^0  \p_C  \left[ \d_A \Om^3_B + (e^+_A
e^-_B - e^-_A e^+_B) \right] \, . \label{4-41} \eeq

Similar results can be obtained for the rest of $\et$'s. With a
simple arrangement of notation, (\ref{4-41}) and the analogous
formulae for the other $\eta$'s, we recover the commutative action
\beq \S ~\sim~
 \ep_{AB}~\int_{z, \bz} \et F_{AB} \label{4-42}
\eeq where $\et (z,\bz)$ is the Lagrange multiplier and $F_{AB}(z,
\bz)$ is the Riemann tensor on $S^2$. This action is known as the
two-dimensional Jackiw-Teitelboim action on $S^2$ \cite{jackiw}.
We have therefore checked that, in the large $N$ limit, the matrix
action (\ref{4-22a}) for gravity reduces to a corresponding
commutative action.

\section{Generalizations}

Even though we have derived the matrix action (\ref{4-22a}) via
our definitions of ${\cal K}_\mu$, the final result is simple and
can be interpreted more directly. The key quantity that enters in
the action is the combination $L_A + {\cal A}_A^a I^a$. We can
write this as \beqar
L_A + {\cal A}_A^a I^a &=& D_A^a I^a~\equiv~D_A ~ , \nonumber\\
D_A^0 &=&  L_A + {\cal A}_A^0 ~ , \nonumber\\
D_A^a &=& {\cal A}_A^a ~~~~~ (a \neq 0)  \label{4-43} \eeqar where
$a$ denotes the full $U(2)$ indices $(\pm, 0, 3)$. The key
ingredient is thus a set of $(N\times N)$ hermitian matrices
$D_A^a$. The definition of the curvatures is seen to be \beqar
\bigl[ D_A, D_B \bigr]&=& \bigl[ D_A^a
I^a , D_B ^b I^b \bigr]~= ~i \ep_{ABC} D_C^cI^c ~+~ F_{AB}^cI^c\nonumber\\
&=& i \epsilon_{ABC} D_C ~+~ F_{AB} . \label{4-44} \eeqar The
action (\ref{4-22a}) is then given by \beqar {\cal S} &=& -i
{\alpha \over 2}~ \Tr ~\biggl[ I^3\eta ~ \ep_{ABC}\left( D_C
F_{AB} +
F_{AB}D_C\right) \biggr]\nonumber\\
&=& -2i \al ~ \Tr ~ \biggl[ I^3\eta~ \left( \ep_{ABC} D_A D_B D_C
- i D^2\right) \biggr] . \label{4-45} \eeqar The constraint on the
the $D$'s is $D_A D_A = L_A L_A$. It is only in this constraint
that the restriction to the sphere arises. Notice that for this
particular case, we could absorb the factor of $I^3$ inside the
trace into the field $\eta$.

The general structure is as thus follows. We start with an
irreducible finite dimensional representation of the Lie algebra
of $SU(2)\times U(1)$ given by the $I^a$ with the commutation
relation $[I^a, I^b ]= i f^{abc} I^c$. Specifically, here we have
$f^{abc} =\epsilon^{abc}$ for $a,b,c = 3, \pm$ and zero otherwise.
We then construct the combinations $D_A = D_A^a I^a$ where the
$D_A^a$ are arbitrary hermitian matrices of some given dimension
$N=n+1$. Using the same $SU(2)$ structure constants we define the
curvatures by $F_{AB} = [D_A , D_B] - i f_{ABC} D_C$. This does
not make any reference to the sphere yet. We restrict to the
sphere by imposing the constraint $D_A D_A = L_A L_A$. The action
is then constructed in terms of $F_{AB}$ as in (\ref{4-45}).

We can use this structure to generalize to $SU(3)$, which will
apply to the case of gravity on fuzzy $\cp^2$. Let $I^a$, $a=1,
2,...,8$ be a set of $(3\times 3)$- matrices forming a basis of
the Lie algebra of $SU(3)$, with the commutation rules $[I^a ,I^b
]=i f^{abc} I^c$. We include $I^0 = {1\over \sqrt{6}} {\bf 1}$ to
make up the algebra of $U(3)$. Let $L_A$ denote an irreducible
representation of the $SU(3)$ algebra in terms of $(N \times
N)$-matrices, with $[L_A , L_B ]= if_{ABC}L_C$. Note that $N$ is
restricted by $N \equiv N^{(2)}=(n+1)(n+2)/2$ ($n=1,2,\cdots$) as
in (\ref{2-1a}). The dynamical variables are then given by $D_A^a$
which are a set of arbitrary $(N\times N)$-matrices. (There are 72
matrices since $A=1,2,...,8$ and $a= 0,1,...,8$.) The curvatures
are defined by $F_{AB} =[ D_A , D_B] -if_{ABC} D_C $, $D_A = D_A^a
I^a$. As the constraints to be obeyed by the $D$'s, we choose
\beqar
D_A D_A &=& L_A L_A \, , \label{4-46} \\
d_{ABC} D_B D_C &=& \left( \frac{n}{3} + \hf \right) D_A
\label{4-46'} \eeqar where the constant in (\ref{4-46'}) is given
by the relation (\ref{2-11}) for $k=2$. The continuum limit of
these conditions gives $\cp^2$ as an algebraic surface in ${\bf
R}^8$ and they have been used to construct noncommutative, and
particularly fuzzy, versions of $\cp^2$ \cite{Nair1,GroStr,bal2}.
Following the construction of the action given in \cite{NairNCg}
and our general discussion in section 4.3, we can write the action
for gravity on fuzzy $\cp^2$ as
\beqar
	\S &=& \alpha ~\Tr ~\biggl[
	I^8\left( D_A F_{KL} F_{MN}+
	F_{KL} F_{MN} D_A\right) \biggr] f_{KLB} f_{MNC} d_{ABC}\nonumber\\
	&=& \al ~\Tr ~\Biggl[ I^8 ~ \biggl( D_A \{[D_K, D_L]-if_{KLR}D_R\}
	\{[D_M, D_N]-if_{MNS}D_S\} \nonumber\\
	&&
	\hskip .5in
	+ \{[D_K, D_L]-if_{KLR}D_R\}\{[D_M,
	D_N]-if_{MNS}D_S\}
	D_A\biggr) \Biggr]
	\nonumber\\ && \hspace{3in}
	\times f_{KLB} f_{MNC} d_{ABC}\, .
	\label{4-47}
\eeqar
This action, along with the constraints (\ref{4-46}) and
(\ref{4-46'}), gives gravity on fuzzy $\cp^2$ as a matrix model.
One can also check directly that the large $N$ limit of this will
reduce to the MacDowell-Mansouri version of the action for gravity
on commutative $\cp^2$.

It is clear that similar actions can be constructed for all
$\cp^k$ ($k=1,2,\cdots$). Notice that the quantity $f_{KLB}
f_{MNC} d_{ABC}$ is the fifth rank invariant tensor of $SU(3)$.
For $\cp^k$ we can use $k$ factors of $F$'s and one factor of $D$
and then contract indices with $\omega^{A_1 ...A_{2k+1}}$, the
invariant tensor of $SU(k+1)$ with rank $(2k+1)$. For an explicit
form of such a tensor, see (\ref{e2}).
Actions for gravity on
fuzzy $\cp^k$ are then written in a generalized form as
\beq \S =
\alpha \Tr \Biggl[ I^{((k+1)^2- 1)}~\biggl( D_{A_1}F_{A_2 A_3}
F_{A_4 A_5}...+ F_{A_2 A_3} F_{A_4 A_5}...D_{A_1} \biggr)\Biggr]
\omega^{A_1 ...A_{2k+1}} \, . \label{4-48}
\eeq
In the large $N$ limit, such an action will contain the Einstein term (in the
MacDowell-Mansouri form), but will also have terms with higher
powers of the curvature. The action (\ref{4-48}) has to be
supplemented by suitable constraints on the $D$'s, which may also
be taken as the algebraic constraints for fuzzy $\cp^k$ shown in
(\ref{2-2}) and (\ref{2-3}).

\chapter{Fuzzy spaces as brane solutions to M(atrix) theory}

\section{Introduction to M(atrix) theory}

There has been extensive interest in the matrix model of M-theory
or the M(atrix) theory since its proposal by Banks, Fischler,
Shenker and Susskind (BFSS) \cite{BFSS}. As mentioned in section
1.3, in M(atrix) theory 9 dimensions out of 11 are
described by $(N\times N)$-matrices, while the other dimensions
correspond to light-front coordinates. This structure arises as a
natural extension of matrix regularization of bosonic membranes in
light-front gauge. The ordinary time component and the extra
spatial direction, the so-called longitudinal one, emerge from the
light-front coordinates in M(atrix) theory. The longitudinal
coordinate is considered to be toroidally compactified with a
radius $R$. In this way, the theory can be understood in 10
dimensions. This is in accordance with one of the features of
M-theory, {\it i.e.}, as a strongly coupled limit of type {\rm
II}A string theory, since the radius $R$ can be related to the
string coupling constant $g$ by $R=g l_s$ where $l_s$ is the
string length scale. From 11-dimensional points of view, one can
consider certain objects which contain a longitudinal momentum
$N/R$ as a Kaluza-Klein mode. Partly from these observations it
has been conjectured that the large $N$ limit of M(atrix) theory
should describe M-theory in the large longitudinal momentum limit
or in the so-called infinite momentum frame (IMF). This BFSS
conjecture has been confirmed in various calculations, especially
in regard to perturbative calculations of graviton interactions
(see, {\it e.g.}, \cite{KT2,OY}), capturing another feature of
M-theory, {\it i.e.}, emergence of 11-dimensional supergravity in
the low energy limit. There also exits a related matrix model by
Ishibashi, Kawai, Kitazawa and Tsuchiya (IKKT) \cite{IKKT} which
corresponds to type {\rm II}B string theory. This IKKT model has
been investigated with a lot of attention as well.
For a review of this model, one may refer to \cite{Review2}.

Besides gravitons, M(atrix) theory does contain extended and
charged objects, namely, memberanes and 5-branes. The membrane in
matrix context appeared originally in the quantization of the
supermembrane a number of years ago by de Wit, Hoppe and Nicolai
\cite{deWit}. Membranes of spherical symmetry in M(atrix) theory
have been obtained in \cite{KT1,Rey}. As regards 5-branes, they
were obtained as longitudinal 5-branes or L5-branes
\cite{BD,GRT,BSS}. The L5-branes are named after the property that
one of their five dimensions coincides with the longitudinal
direction in M(atrix) theory. One may think of the existence of
transverse 5-branes as opposed to L5-branes, but it turns out that
there is no classically conserved charges corresponding to the
transverse 5-branes. Thus it is generally believed that the
L5-branes are the only relevant 5-branes in M(atrix) theory at
least in the classical level. In a modified M(atrix) theory, {\it
i.e.}, the so-called plane wave matrix theory \cite{BMN}, the
existence of transverse 5-branes is discussed at a quantum level
\cite{Mald1}. L5-branes with spherical symmetry in the transverse
directions have also been proposed in \cite{CLT}. Although this
spherical L5-brane captures many properties of M-theory, it is as
yet unclear how to include matrix fluctuations contrary to the
case of spherical membranes. The only other L5-brane that is known
so far is an L5-brane with $\cp^2$ geometry in the transverse
directions \cite{Nair1}. Matrix configuration of this L5-brane is
relevant to that of the fuzzy $\cp^2$ \cite{bal1}.

Fuzzy spaces are one of the realizations of noncommutative
geometry in terms of $(N\times N)$-matrices, hence, those extended
objects in M(atrix) theory are possibly described by the fuzzy
spaces as far as the transverse directions are concerned.
Following this idea, in the present chapter we shall consider
fuzzy complex projective spaces $\cp^k$ ($k=1,2,\cdots$) as
ans\"{a}tze to the extended objects or the brane solutions in
M(atrix) theory. This approach towards a solution to M(atrix)
theory was originally pursued by Nair and Randjbar-Daemi in
\cite{Nair1} which, among the other known brane solutions,
revealed the existence of the L5-brane of $\cp^2 \times S^1$
geometry. At this stage, we are familiar to the fact that fuzzy
$\cp^k$ are constructed in terms of matrix representations of the
algebra of $SU(k+1)$ in the $(n,0)$-representation under a certain
set of algebraic constraints. This fact makes it relatively
straightforward to include transverse fluctuations of branes with
$\cp^2$ (or $\cp^k$) geometry in comparison with the case of the
spherical L5-brane. This point is one of the advantages to
consider fuzzy $\cp^k$ as ans\"{a}tze for the brane solutions.
Note that fluctuations of branes are described by gauge fields on
noncommutative geometry. This means that the dynamics of the
extended objects in M(atrix) theory can be governed by gauge
theories on fuzzy spaces.

From a perspective of type IIA string theory, the gravitons,
membranes and L5-branes of M-theory are respectively relevant to
D0, D2 and D4 brane solutions. Type IIA string theory also
contains a D6 brane. The D6 brane is known to be a Kaluza-Klein
magnetic monopole of 11-dimensional supergravity compactified on a
circle and is considered to be irrelevant as a brane solution in
M(atrix) theory. Naively, however, since D6 branes are Hodge dual
to D0 branes in the same sense that D2 and D4 branes are dual to
each other, we would expect the existence of L7-branes in M(atrix)
theory. It is important to remind that fuzzy spaces can be
constructed only for compact spaces. If we parametrize branes by
fuzzy spaces, the transverse directions are also all compactified
in the large $N$ limit. As far as the capture of a Kaluza-Klein
mode in the scale of $N/R$ is concerned, one cannot distinguish
the longitudinal direction from the transverse ones. The gravitons
or the corresponding D0 branes of M-theory would possibly live on
the transverse directions in this case. Thus we may expect the
existence of L7-branes as a Hodge dual description of such
gravitons in an M-theory perspective. Construction of L7-branes
(or transverse D6-branes) has been suggested in
\cite{BSS,Taylor1}, however, such extended objects have not been
obtained in the matrix model. Besides the fact that no L7-brane
charges appear in the supersymmetry algebra of M(atrix) theory,
there is a crucial obstruction to the construction of L7-brane,
that is, as shown by Banks, Seiberg and Shenker \cite{BSS}, the
L7-brane states have an infinite energy in the large $N$ limit,
where the energy of the state is interpreted as an energy density
in the transverse directions. Indeed, as we shall discuss in the
next section, an L7-brane of $\cp^3 \times S^1$ geometry leads to
an infinite energy in the large $N$ limit and, hence, one cannot
make sense of the theory with such an L7-brane.

In order to obtain an L7-brane as a solution to M(atrix) theory,
it would be necessary to introduce extra potentials or fluxes to
the M(atrix) theory Lagrangian such that the brane system has a
finite energy as $N \rightarrow \infty$. Since M(atrix) theory is
defined on a flat space background, such an additional term
suggests the description of the theory in a nontrivial background.
The most notable modification of the M(atrix) theory Lagrangian
would be the one given by Berenstein, Maldacena and Nastase (BMN)
to describe the theory in the maximally supersymmetric
parallel-plane (pp) wave background \cite{BMN}. There has been a
significant amount of papers on this BMN matrix model of M-theory.
(For some of the earlier papers, see
\cite{BMNwork1}-\cite{BMNwork6}.) Another important approach to
the modification of BFSS M(atrix) theory is to introduce a
Ramond-Ramond (RR) field strength as a background such that it
couples to brane solutions. Specifically, one may have a RR 4-form
as an extra potential from a IIA string theory viewpoint. As shown
by Myers \cite{Myers}, the matrix equation of motion with this RR
flux allows fuzzy $S^2$ as a static solution, meaning that the
corresponding IIA theory has a spherical D2-brane solution. The RR
field strength is associated with a charge of this D2 brane. The
modified equation of motion also allows a diagonal matrix
configuration as a solution which corresponds to $N$ D0-branes,
with $N$ being the dimension of matrices. One may interpret these
solutions as bound states of a spherical D2-brane and $N$
D0-branes. From a D0-brane perspective, the RR field strength is
also associated with a D0-brane charge. So the extra RR flux gives
rise to a D-brane analog of a dielectic effect, known as Myers
effect. A different type of flux, {\it i.e.}, a RR 5-form which
produces bound states of $N$ D1-branes and a D5-brane with $\cp^2$
geometry has been proposed by Alexanian, Balachandran and Silva
\cite{Bal3} to describe a generalized version of Myers effect from
a viewpoint of IIB string theory. From a M(atrix) theory
perspective, the D5 brane corresponds to the L5-brane of $\cp^2
\times S^1$ geometry. In this chapter, we consider further
generalization along this line of development, namely, we consider
a general form for all possible extra potentials that allows fuzzy
$\cp^k$ as brane solutions or solutions of modified matrix
equations of motion. We find several such potentials for $k \le
3$.

The extra potentials we shall introduce in the consideration of a
possible L7 brane solution to M(atrix) theory are relevant to
fluxes on a curved space of $(\cp^3 \times S^1) \times \M_4$ where
$\M_4$ is an arbitrary four-dimensional manifold. We shall show
that one of the potentials can be interpreted as a 7-form flux in
M(atrix) theory. According to Freund and Rubin \cite{FR},
existence of a 7-form in 11 dimensional (bosonic) theories implies
compactification of 7 or 4 space-like dimensions. The existence of
the 7-form in M(atrix) theory is interesting in a sense that it
would lead to a matrix version of Freund-Rubin type
compactification.
This means that the introduction of the 7-form can also lead to
a physically interesting matrix model in four dimensions.
In hope of such a possibility, we also consider
compactification of M(atrix) theory down to fuzzy $S^4$
which can be defined in terms of fuzzy $\cp^3$ \cite{Abe1}.

The plan of the rest of this chapter is as follows. In the next
section, following \cite{Nair1}, we show that fuzzy $\cp^k$ ($k\le
4$) provide solutions to bosonic matrix configurations in M(atrix)
theory. Along the way we briefly review definitions and properties
of fuzzy $\cp^k$. We further discuss the energy scales of the
solutions and see that the energy becomes finite in the large $N$
limit only in the cases of $k=1,2$, corresponding to the membrane
and the L5-brane solutions in M(atrix) theory. In section 5.3, we
examine supersymmetry of the brane solutions for $k \le 3$. We
make a group theoretic analysis to show that those brane solutions
break the supersymetries in M(atrix) theory. Our discussion is
closely related to the previous analysis \cite{Nair1} in the case
of $k=2$. In section 5.4, we introduce extra potentials to the
M(atrix) theory Lagrangian which are suitable for the fuzzy
$\cp^k$ brane solutions. We consider the effects of two particular
potentials to the theory. These effects can be considered as
generalized Myers effects. We find a suitable form of potentials
for the emergence of static L7-brane solutions, such that the
potentials lead to finite L7-brane energies in the large $N$
limit. Section 5.5 is devoted to the discussion on possible
compactification models in non-supersymmetric M(atrix) theory. We
show that one of the extra potentials introduced for the presence
of L7-branes can be interpreted as a matrix-valued or `fuzzy'
7-form in M(atrix) theory.
Using the idea of Freund-Rubin type compactification,
this suggests the compactification down to 7 or 4 dimensions.
The compactification model down to 4 dimension is physically the more interesting
and we consider, as a speculative model of it, a compactified
matrix model on fuzzy $S^4$.

\section{Fuzzy $\cp^k$ as brane solutions to M(atrix) theory}

The M(atrix) theory Lagrangian can be expressed as
 \beq \L = \Tr \left(
\frac{1}{2R} {\dot X}_{I}^{2} + \frac{R}{4} [X_I , X_J ]^2 +
\th^{T} {\dot \th} + i R \th^{T} \Ga_I [X_I , \th] \right)
 \label{5-c1} \eeq
where $X_I$ ($I=1,2,\cdots, 9$) are hermitian $N\times N$
matrices, $\th$ denotes a 16-component spinor of $SO(9)$
represented by $N\times N$ Grassmann-valued matrices, and $\Ga_I$
are the $SO(9)$ gamma matrices in the 16-dimensional
representation. The Hamiltonian of the theory is given by \beq \H
= \Tr \left( \frac{R}{2} P_I P_I - \frac{R}{4} [X_I, X_J]^2 -i R
\th^{T} \Ga_I [X_I, \th ] \right) \label{5-c2}\eeq where $P_I$ is
the canonical conjugate to $X_I$; $P_I = \frac{\d \L}{\d
\dot{X}_I}$. As discussed in the introduction, we will be only
interested in those energy states that have finite energy in the
limit of the large longitudinal momentum $N/R$. Since the
Hamiltonian (\ref{5-c2}) leads to an infinite energy state in the
limit of $R \rightarrow \infty$, we will consider the large $N$
limit with a large, but fixed value for $R$. With this limit
understood, the theory is defined by (\ref{5-c1}) or (\ref{5-c2})
with a subsidiary Gauss law constraint \beq [ X_I , \dot{X}_I ] -
[ \th , \th^{T} ] = 0 \, . \label{5-c3}\eeq In this section, we
shall consider the bosonic part of the theory, setting the $\th$'s
to be zero. The relevant equations of motion for $X_I$ are given
by
 \beq \frac{1}{R} {\ddot X}_I - R [ X_J , [X_I , X_J]] = 0
 \label{5-c4} \eeq
 with a subsidiary constraint
 \beq [X_I , \dot{X}_I] = 0 \, .
 \label{5-c5} \eeq

We shall look for solutions to these equations, taking the
following ans\"{a}tze
 \beq X_I  =  \left\{
    \begin{array}{ll}
      r(t) Q_i & \mbox{for $I=i=1,2,\cdots, 2k$}
\\
      0  & \mbox{for $I=2k+1, \cdots, 9$}
    \end{array} \right. \label{5-c6} \eeq
where $Q_i$ denote the local coordinates of fuzzy $\cp^k =
SU(k+1)/U(k)$ ($k=1,2,\cdots$). Since $X_I$ are defined for
$I=1,2,\cdots, 9$, the ans\"{a}tze are only valid for $k \le 4$.

\subsection{Local coordinates of fuzzy $\cp^k$}

As discussed in section 2.4, fuzzy $\cp^k$ can be constructed in
terms of certain matrix generators of $SU(k+1)$ as embedded in
${\bf R}^{k^2 + 2k}$ under a set of algebraic constraints. Here we
shall briefly review such a construction. Let $L_A$
($A=1,2,\cdots, k^2 +2k=\dim SU(k+1)$ be $N^{(k)} \times
N^{(k)}$-matrix representations of the generators of $SU(k+1)$ in
the $(n,0)$-representation. The coordinates of fuzzy $\cp^k$ as
embedded in ${\bf R}^{k^2 + 2k}$ are parametrized by $Q_A = L_A
/\sqrt{C^{(k)}_2}$ where $C^{(k)}_2$ is the quadratic Casimir of
$SU(k+1)$ in the $(n,0)$-representation
\beq C^{(k)}_2 = \frac{n k
(n+k+1)}{2 (k+1)} \, . \label{5-a4}
\eeq
The matrix dimension is given by
\beq N^{(k)} = \frac{(n+k)!}{k!~n!} \sim n^k .
\label{5-a5}
\eeq
The fuzzy $\cp^k$ coordinates, as embedded in
${\bf R}^{k^2 + 2k}$, are then defined by the following two
constraints on $Q_A$:
\beqar
Q_A ~Q_A &=& {\bf 1} \, , \label{5-a2} \\
d_{ABC}~Q_{A}~ Q_{B} & = & c_{k,n}~ Q_C \label{5-a3}
\eeqar
where ${\bf 1}$ is the $N^{(k)} \times N^{(k)}$ identity matrix,
$d_{ABC}$ is the totally symmetric symbol of $SU(k+1)$ and the
coefficient $c_{k,n}$ is given by \beq c_{k,n} ~=~
\frac{(k-1)}{\sqrt{C_2^{(k)}}} \left( \frac{n}{k+1} + \hf
\right)\, . \label{5-a6} \eeq The first constraint (\ref{5-a2}) is
trivial due to the definition of $Q_A$. The second constraint
(\ref{5-a3}) is what is essential for the global definition of
fuzzy $\cp^k$ as embedded in ${\bf R}^{k^2 +2k}$. For $k \ll n$,
the coefficient $c_{k,n}$ becomes $c_{k,n} \rightarrow c_k =
\sqrt{\frac{2}{k(k+1)}} (k-1)$ and this leads to the constraints
for the coordinates $q_A$ of commutative $\cp^k$, {\it i.e.}, $q_A
~q_A = 1$ and $d_{ABC}q_{A} q_{B}  =  c_{k} q_C$. As discussed
earlier in section 2.4, the latter constraint restricts the number
of coordinates to be $2k$ out of $k^2+2k$. Similarly, under the
constraint (\ref{5-a3}), the coordinates of fuzzy $\cp^k$ are
effectively expressed by the local coordinates $Q_i$
($i=1,2,\cdots, 2k$) rather than the global ones $Q_A$
($A=1,2,\cdots, k^2+2k$).

We now consider the commutation relations of $Q_i$'s. By
construction they are embedded in the $SU(k+1)$ algebra.
We first split the generators $L_A$ of $SU(k+1)$ into $L_i \in
\underline{SU(k+1)}-\underline{U(k)}$ and $L_\al \in
\underline{U(k)}$, where $\underline{G}$ denotes the Lie algebra
of group $G$.  The indices $i=1,2, \cdots, 2k$ are then relevant to the
$\cp^k$ of our interest, while the indices $\al = 1,2,\cdots, k^2$
correspond to the $U(k)$ subgroup of $SU(k+1)$.
The $SU(k+1)$ algebra,
$[ L_A, L_B]= if_{ABC}L_C$ with the structure constant $f_{ABC}$,
is then expressed by the following set of commutation relations
\beqar \left[Q_i, Q_j \right] &=& i ~
\frac{c_{ij\al}}{\sqrt{C^{(k)}_2}}~Q_\al \, , \label{5-c7}\\
\left[Q_{\al}, Q_{\bt} \right] &=& i~
\frac{f_{\al\bt\ga}}{\sqrt{C^{(k)}_2}}~Q_{\ga} \, , \label{5-c8}\\
\left[Q_{\al}, Q_i \right] &=& i ~ \frac{f_{\al i
j}}{\sqrt{C^{(k)}_2}}~Q_j \label{5-c9} \eeqar
where we use $Q_A=
L_A/{\sqrt{C^{(k)}_2}}$ and denote $f_{ij\al}$ by $c_{ij\al}$ to
indicate that it is relevant to the commutators of $Q_i$'s.
$f_{\al\bt\ga}$ is essentially the structure constant of $SU(k)$
since the $U(1)$ part of the $U(k)$ algebra can be chosen such
that it commutes with the rest of the algebra. We can calculate
$c_{\al ij} c_{\bt ij}$ as \beq c_{\al ij} \, c_{\bt ij} \, = \,
f_{\al AB} f_{\bt AB} -f_{\al \ga \del} f_{\bt \ga \del} \,=\,
\del_{\al\bt} \label{5-c10}\eeq by use of the relations $f_{\al
AB} f_{\bt AB}= (k+1)\del_{\al\bt}$ and $f_{\al \ga \del} f_{\bt
\ga \del} = k \del_{\al\bt}$. Notice that the result (\ref{5-c10})
restricts possible choices of the $\cp^k$ indices $(i,j)$. For
example, in the case of $k=2$ we have $(i,j) = (4,5), (6,7)$ with
the conventional choice of the structure constant $f_{ABC}$ of
$SU(3)$. Similarly, in the case of $k=3$ we have
$(i,j)=(9,10),(11,12),(13,14)$. Under such restrictions, we can
also calculate $c_{ij\al} f_{j\al k}$ as
\beq c_{ij\al} \, f_{j\al
k} \, = \, c_{ij\al} \, c_{kj\al} \, = \, \del_{ik} \, .
\label{5-c11}
\eeq
In what follows, we shall use the symbol $c_{ij\al}$
rather than $f_{ij \al}$ to indicate that we are interested in
this peculiar subset of the $SU(k+1)$ algebra.

We can also classify the totally symmetric symbol $d_{ABC}$ as follows.
\beq
    d_{ABC} = \left\{ \begin{array}{ll}
    d_{ij \al} & \\
    d_{\al \bt \ga} & \\
    0 & \mbox{otherwise}
    \end{array}\right.
    \label{c11-1}
\eeq
Notice that symbols such as $d_{\al \bt i}$ and $d_{ijk}$ do vanish.
In relation to the construction of $\cp^k$, it is useful to know
the fact that the symbol $d_{ii\al}$, a subset of $d_{ij \al}$, is expressed
as $d_{ii \al_{k^2 + 2k}}$ and is identical regardless the index $i$.
Here the index $i$ is relevant to a local coordinate of $\cp^k$
and the index $\al_{k^2 + 2k}$ is a hypercharge-like index in
a conventional choice of $SU(k+1)$ generators.

The normalization of $Q_A$'s is taken as (\ref{5-a2}). Thus
traces of matrix products are expressed as
\beqar
    \Tr(Q_{A} Q_{B}) &=& \frac{N^{(k)}}{k^2 +2k} \del_{AB} \, , \label{c12} \\
    \Tr(Q_{i} Q_{i}) &=& \frac{2k}{k^2 +2k} N^{(k)}  \, , \label{c13} \\
    \Tr(Q_{\al} Q_{\al}) &=& \frac{k^2}{k^2 +2k} N^{(k)} \, .  \label{c14}
\eeqar
These relations are also useful in later calculations.

\subsection{Fuzzy $\cp^k$ solutions to M(atrix) theory}

Using (\ref{5-c7})-(\ref{5-c11}), we can easily
find that $[Q_j , [Q_i , Q_j]]= - Q_i / C^{(k)}_2$. Thus, with the
ans\"{a}tze (\ref{5-c6}), we can express the equation of motion
(\ref{5-c4}) becomes
\beq
    \left( \frac{\ddot{r}}{R} + \frac{R}{C^{(k)}_2} r^3 \right) Q_{i} = 0 \, .
    \label{c15}
\eeq
This means that the equation of motion is reduced to
an ordinary differential equation of $r(t)$.
Notice that the subsidiary constraint (\ref{5-c5}) is also satisfied with
the ans\"{a}tze (\ref{5-c6}).
The equation of motion therefore reduces to
\beq
    \ddot{r} + \frac{R^2}{C^{(k)}_2} r^3 = 0 \, .
\label{5-c14}
\eeq
A general solution to this equation is written
as \beq r(t)~ = ~A~ {\rm cn} \left( \al(t-t_0);\ka^2 =\hf \right)
\label{5-c15} \eeq
where $\al=\sqrt{R^2/{C^{(k)}_2}}$ and ${\rm
cn}(u; \ka)={\rm cn}(u)$ is one of the Jacobi elliptic functions,
with $\ka$ ($0 \le \ka \le 1$) being the elliptic modulus. $A$ and
$t_0$ are the constants determined by the initial conditions.
Using the formula
\beqar \frac{d}{du}{\rm cn} (u; \ka)&=& - \,{\rm
sn}(u; \ka) \, {\rm dn}(u;\ka) \nonumber\\ &=& - u +
\frac{1+4\ka^2}{3!}u^3 - \cdots ~\, , \label{5-c16}
\eeqar
we can express $\dot{r}$ as
\beq \dot{r} ~=~ -A\al ~ {\rm sn}(\al
(t-t_0)) {\rm dn}(\al (t-t_0)) \, . \label{5-c161}
\eeq
In the limit of large $N$ (or $n$), $\dot{r}$ is suppressed by $\dot{r}
\sim 1/n^2$. Thus the solution (\ref{5-c15}) corresponds to a static
solution in the large $N$ limit.

Evaluated on the fuzzy $\cp^k$, the potential energy of M(atrix) theory is
calculated as
\beqar
    V (rQ) &=&  - \Tr \left( \frac{R}{4} [ r Q_{i} , r Q_{j} ]^2 \right)
    \nonumber \\
    &=& \frac{R r^4}{4 C_{2}^{(k)}} \Tr ( Q_{\al} Q_{\al} ) \nonumber \\
    &=& \frac{k^2}{k^2 + 2k }\frac{R r^4}{4 C_{2}^{(k)}} N^{(k)}
    \, \sim \, n^{k-2} R r^4 \, .
    \label{c20}
\eeqar
From this result we can easily tell that for
$k=1,2$ we have finite energy states in the large $N$ limit. These
states respectively correspond to the spherical membrane and the
L5-brane of $\cp^2$ geometry in M(atrix) theory. By contrast, for
$k=3,4$ we have infinite energy states. Thus, although these may
possibly correspond to L7 and L9 brane solutions, they are
ill-defined and we {\it usually} do not consider such solutions in
M(atrix) theory.
The main purpose of the present chapter is to show that we can have L7 brane
solutions by introducing extra potentials to the M(atrix) theory
Lagrangian (\ref{5-c1}).
Notice that in this chapter we shall not consider the
case of $k=4$ or a 9-brane solution to M(atrix) theory.
The 9-branes are supposed to correspond to ``ends of the world'' which
describe gauge dynamics of the 9-dimensional boundary of M-theory.
Thus these are in general considered irrelevant as brane solutions to the theory.

\section{Supersymmetry breaking}

In this section, we examine supersymmetry of the fuzzy $\cp^k$
brane solutions in M(atrix) theory for $k \le 3$.
As in the previous section, we make an analysis, following
the argument of Nair and Randjbar-Daemi in \cite{Nair1}.

We have set the fermionic matrix variables $\th$ to be zero. In
this section, we now consider the supersymmetry transformations of
the brane solutions in M(atrix) theory. The supersymmetric
variation of $\th$ is given by \beq \del \th_r ~=~ \hf \left(
\dot{X}_I (\Ga_{I})_{rs} + [X_I , X_J] (\Ga_{IJ})_{rs} \right)
\ep_s ~+~ \del_{rs}\xi_s \label{5-s1} \eeq where $\ep$ and $\xi$
are 16-component spinors of $SO(9)$ represented by $N \times N$
matrices ($r,s =1,2,\cdots,16$) and $\Ga_I$'s are the
corresponding gamma matrices as before. $\Ga_{IJ}$ are defined by
$\Ga_{IJ}= \hf [\Ga_I, \Ga_J]$. With our ans\"{a}tze, the equation
(\ref{5-s1}) reduces to
\beq \del \th_r ~=~ \hf \left( \dot{r} Q_i
(\ga_i)_{rs} +  r^2 \frac{i c_{ij\al}}{\sqrt{C^{(k)}_2}}\, Q_\al
(\ga_{ij})_{rs} \right) \ep_s ~+~ \del_{rs} \xi_s \label{5-s2}
\eeq
where $\ga_i$'s are the gamma matrices of $SO(2k)$ under the
decomposition of $SO(9) \rightarrow SO(2k) \times SO(9-2k)$.
Accordingly, we here set $i=1,2,\cdots, 2k$ and $r,s=1,2,\cdots,
2^k$. For the static solution we make $\dot{r} \sim n^{-2}$
vanish. Indeed, if $\del \th \sim n^{-2}$, we have $\Tr (\del
\th^{T} \del \dot{\th}) \sim N^{(k)}n^{-4} \sim n^{k-4}$ and, for
$k=1,2$ and $3$, this term vanishes in the large $N$ limit. The
other term $\Tr (i R \del \th^{T} \Ga_I [X_I , \del\th])$ in the
Lagrangian vanishes similarly. Thus, for static solutions, the
condition $\del \th =0$ is satisfied when $c_{ij\al} Q_\al
\ga_{ij}$ becomes a ${\bf c}$-number in the $SO(2k)$ subspace of
$SO(9)$ such that the $\ep$-term can be cancelled by $\xi$ in
(\ref{5-s2}). In what follows, we examine this
Bogomol'nyi-Prasad-Sommerfield (BPS)-like condition
for $k=1,2,3$.

It is known that the spherical membrane solution breaks all
supersymmetries. Let us rephrase this fact by examining the BPS
condition ($\del \th =0$) for $k=1$. The 2-dimensional gamma
matrices are given by $\ga_1 = \si_1$ and $\ga_2 = \si_2$, where
$\si_i$ is the ($2\times 2$)-Pauli matrices. The factor
$c_{ij\al}Q_\al \ga_{ij}$ becomes proportional to $Q_3 \si_3$
where $Q_3$ is an $N^{(1)} \times N^{(1)}$ matrix representing the
$U(1)$ part of the $SU(2)$ generators in the spin-$n/2$
representation. Now the factor $\si_3$ is not obviously
proportional to identity in the $SO(2)$ subspace of $SO(9)$. Thus we
can conclude that the BPS condition is broken.

For $k=2$, we can apply the same analysis to the factor of
$c_{ij\al}Q_\al \ga_{ij}$. We use the conventional choice for the
structure constant of $SU(3)$ where the group elements are defined
by $g= \exp (i\th^a \frac{\la^a}{2})$ with the Gell-Mann matrices
$\la^a$ ($a=1,2,\cdots, 8)$. As discussed earlier, with this
convention the set of $(i,j)$ is restricted to $(i,j)=(4,5)$ or
$(6,7)$. The relevant $c_{ij\al}$'s are given by $c_{453}=1/2$,
$c_{458}=\sqrt{3}/2$, $c_{673}=-1/2$ and $c_{678}=\sqrt{3}/2$.
Introducing the usual 4-dimensional gamma matrices $\ga_i$
($i=4,5,6,7$) \beqar && \ga_4 = \left(%
\begin{array}{cc}
  0 & 1 \\
  1 & 0 \\
\end{array}%
\right), ~~~~ \ga_5 =\left(%
\begin{array}{cc}
 0 & -i\si_1 \\
  i\si_1 & 0 \\
\end{array}%
\right), \nonumber\\
&&
  \ga_6=\left(%
\begin{array}{cc}
  0 & -i\si_2 \\
  i\si_2 & 0 \\
\end{array}%
\right), ~~~ \ga_7 =\left(%
\begin{array}{cc}
 0 & -i\si_3 \\
  i\si_3 & 0 \\
\end{array}%
\right), \label{5-s3} \eeqar
 we can calculate the factor of interest as
 \beqar c_{45\al}Q_\al \ga_{45} &\sim &
 \left( Q_3+ \sqrt{3} Q_8 \right) \left(%
\begin{array}{cc}
  i\si_1 & 0 \\
  0 & -i\si_1 \\
\end{array}%
\right) , \nonumber\\
c_{67\al}Q_\al \ga_{67} &\sim &
 \left( -Q_3+ \sqrt{3} Q_8 \right) \left(%
\begin{array}{cc}
  i\si_1 & 0 \\
  0 & i\si_1 \\
\end{array}%
\right) \label{5-s4}
\eeqar
where $\ga_{ij} = \hf [\ga_i , \ga_j]$,
and $Q_3$, $Q_8$ are $N^{(2)} \times N^{(2)}$ matrices
representing diagonal parts of $SU(3)$ algebra in the
totally symmetric representation $(n,0)$.
In either case, it is impossible to make
the factor $c_{ij\al}Q_\al \ga_{ij}$ be proportional to identity
or zero in terms of the $(4\times 4)$-matrix which corresponds to
$\ga_i$'s.
This indicates that the brane solution corresponding to
$k=2$ breaks the supersymmetries of M(atrix) theory as originally
analyzed in \cite{Nair1}.

The same analysis is applicable to the case of $k=3$ and we can
show that the brane solution corresponding to $k=3$ also breaks
the supersymmetries. For the completion of discussion, we present
the factors $c_{ij\al}Q_\al \ga_{ij}$ for
$(i,j)=(9,10),(11,12),(13,14)$ in suitable choices of $c_{ij\al}$
and 6-dimensional gamma matrices: \beqar c_{9\,10\al}~Q_\al ~
\ga_{9 \,10} &\sim & \left( \sqrt{3} Q_3 + Q_8 + 2\sqrt{2} Q_{15}
\right)
 \left(%
\begin{array}{cccc}
  \si_1 & 0 & 0 & 0 \\
  0 & -\si_1 & 0 & 0 \\
  0 & 0 & \si_1 & 0 \\
  0 & 0 & 0 & -\si_1 \\
\end{array}%
\right), \nonumber\\
c_{11\,12\al}~Q_\al ~\ga_{11 \,12} &\sim & \left( - \sqrt{3} Q_3 +
Q_8 + 2\sqrt{2} Q_{15} \right)\left(%
\begin{array}{cccc}
  \si_1 & 0 & 0 & 0 \\
  0 & \si_1 & 0 & 0 \\
  0 & 0 & \si_1 & 0 \\
  0 & 0 & 0 & \si_1 \\
\end{array}%
\right), \nonumber \\
c_{13\,14\al}~Q_\al ~\ga_{13 \,14} &\sim & \left(- 2 Q_8 +
2\sqrt{2} Q_{15} \right)\left(%
\begin{array}{cc}
  {\bf 1} & 0 \\
  0 & -{\bf 1} \\
\end{array}%
\right) \label{5-s5} \eeqar
where $Q_3$, $Q_8$ and $Q_{15}$ are
the $N^{(3)} \times N^{(3)}$ matrices representing diagonal parts
of $SU(4)$ algebra in the $(n,0)$-representation. In the last
line, ${\bf 1}$ denotes the $4\times 4$ identity matrix.

\section{L7-branes and extra potentials in M(atrix) theory}

As we have seen in (\ref{c20}), the potential energy of a prospective
L7-brane with $\cp^3 \times S^1$ geometry is proportional to $n$,
leading to infinite energy in the large $N$ limit. In this section,
we introduce extra potentials to the bosonic part of the M(atrix) theory
Lagrangian so that the total potential energy of the L7-brane becomes
finite in the large $N$ limit.
From (\ref{5-c161}) we have found $\dot{r} \sim n^{-2}$. Thus the
kinetic energy of brane states with
$\cp^k \times S^1$ geometry is proportional to $\frac{N^{(k)}}{R} n^{-4}$.
Since the kinetic energy is suppressed by $n^{k-4}$, we
can consider the brane solution for any of $k=1,2,3$ as a static solution.
Consideration of potential energies will suffice for the
stability analysis of brane solutions.
In what follows, we first present a general form of the extra potentials
which is appropriate for our fuzzy $\cp^k$ brane solutions.
We then consider a few cases in detail, eventually obtaining a suitable form
of the extra potential for the emergence of L7-branes.

\subsection{Extra potentials: a general form}

We consider the following form of potentials.
\beqar
    F_{2s+1} (X) &=& F_{[ij]^{s} \al }
    \Tr ( X_{i_1} X_{j_1}X_{i_2} X_{j_2}\cdots X_{i_r} X_{j_r} X_{\al} )
    \label{e1}\\
    F_{[ij]^{s} \al } &=&  \tr ( [ t_{i_1}, t_{j_1}] [ t_{i_2}, t_{j_2}]
    \cdots [ t_{i_s}, t_{j_s}] t_{\al} )
    \label{e2}
\eeqar
where $t_A$ ($A = i, \al$) are the generators of $SU(k+1)$ in the
fundamental representation with normalization $\tr ( t_A t_B ) = \hf \del_{AB}$.
As discussed earlier, $t_i$'s (including $t_j$'s) correspond to
the elements of $\underline{SU(k+1)} - \underline{U(k)}$ $(i = 1,2,\cdots , 2k)$
and $t_\al$ correspond to the elements of a $\underline{U(k)}$ subalgebra
$( \al = 1,2, \cdots , k^2$).
In the above expressions, $s$ takes the value of $s=1,2,\cdots, k$ and
$X_i$'s represent arbitrary matrix coordinates which
are, eventually, to be evaluated by the fuzzy $\cp^k$ coordinates $X_i = r(t) Q_i$.
Notice that the number of $X$'s is odd. This corresponds to the fact
that $F_{[ij]^{s} \al }$ are related to the rank-$(2s+1)$ invariant tensors of $SU(k+1)$.
We shall consider this point further in the next section.
In the following, we rather
show the correctness of the general form $F_{2s+1}$ in (\ref{e1}) for fuzzy $\cp^k$
brane solutions in M(atrix) theory.
The M(atrix) theory Lagrangian with the extra potential $F_{2s+1}$ is given
by
\beqar
    \L^{(2s+1)} &=& \L - \la_{2s+1}  F_{2s+1} (X) \label{e3}\\
    \L &=& \Tr \left( \frac{\dot{X_I}^2}{2R} + \frac{R}{4} [X_I , X_J]^2
    \right)
    \label{e4}
\eeqar
where $\L$ is the bosonic part of the original
M(atrix) theory Lagrangian (\ref{5-c1}) and $\la_{2s+1}$ is a coefficient of
the potential $F_{2s+1}$. The matrix equations of motion
are expressed as
\beq
    \frac{1}{R} {\ddot X}_I - R [ X_J , [X_I , X_J]] +
    \la_{2s+1} \frac{\del}{\del X_I} F_{2s+1} = 0 \, .
    \label{e5}
\eeq
Thus, in order to show the correctness of the general form in (\ref{e1}), it is
sufficient to see whether the term $\frac{\del}{\del X_I} F_{2s+1}$ is
proportional to the $Q_i$ when $X_I$ is evaluated by the ans\"{a}tze (\ref{5-c6}).

\subsection{Modification with $F_3$: Myers effect}

For $s=1$, we have
\beq
    F_{[ij]\al} = \tr ( [ t_i , t_j ] t_\al ) = \frac{i}{2} c_{ij \al}
    \label{e6}
\eeq
where we use the normalization $\tr ( t_\al t_\bt ) = \hf \del_{\al \bt}$.
The potential $F_3 (X)$ is then written as
\beq
    F_3 (X) = \frac{i}{2} c_{ij \al} \Tr (X_{i} X_{j} X_\al ) \, .
    \label{e7}
\eeq
Since $c_{ij \al} \sim \ep_{ij \al}$,
the addition of $F_3$ to the M(atrix) theory Lagrangian essentially
leads to the so-called Myers effect from a viewpoint of IIA string theory \cite{Myers}.
Now we can calculate
\beqar
    \left. \frac{\del}{\del X_i} F_3 (X) \right|_{X=rQ} &=&
    \frac{i}{2} r^2 c_{ij \al} Q_j Q_\al \nonumber \\
    &=& \left( \frac{i r}{2} \right)^2  Q_i
    \label{e8}
\eeqar
where we use the relation (\ref{5-c10}).
Thus we find that the fuzzy $S^2$ remains the solution of M(atrix) theory
modified with the extra potential $F_3$.
As we shall see in a moment, generalizations along these lines can be
made for the potentials with higher ranks.

\subsection{Modification with $F_5$}

For $s=2$, we have
\beqar
    F_{[ij]^{2} \al} &=& \tr ( [ t_{i_1} , t_{j_1} ][ t_{i_2} , t_{j_2} ] t_\al )
    \nonumber \\
    &=& i c_{i_{1}j_{1} \al_{1}} i c_{i_{2}j_{2} \al_{2}} \tr ( t_{\al_1}t_{\al_2}t_{\al} )
    \nonumber \\
    &=& - \frac{1}{4} c_{i_{1}j_{1} \al_{1}} c_{i_{2}j_{2} \al_{2}} d_{\al_{1} \al_{2} \al }
    \label{e9}
\eeqar
where we use the fact that $t_{\al_1}$ and $t_{\al_2}$ are commutative;
these generators correspond to `diagonal' elements of a $U(2)$ algebra
in terms of its matrix representation.
The symbol $d_{\al_{1}\al_{2}\al}$ is called the totally symmetric symbol of
$SU(k+1)$ and is defined by $d_{\al \bt \ga} = 2 \tr( \{ t_\al , t_\bt \} t_\ga )$.
The potential $F_5 (X)$ is then written as
\beq
    F_5 (X) =
    - \frac{1}{4} c_{i_{1}j_{1} \al_{1}} c_{i_{2}j_{2} \al_{2}} d_{\al_{1} \al_{2} \al }
    \Tr ( X_{i_1} X_{j_1} X_{i_2} X_{j_2} X_{\al} ) \, .
    \label{e10}
\eeq
This is a natural generalization of the Myers term (\ref{e7}) to a higher rank.
Notice that $F_{5}$ exists for any $SU(k+1)$ with $k \ge 2$.
The variation of $F_5$ with respect to $X_{i_1}$ is expressed as
\beqar
    \left. \frac{\del}{\del X_{i_1}} F_5 (X) \right|_{X=rQ} &=&
    - \frac{1}{4} r^4 c_{i_{1}j_{1} \al_{1}} c_{i_{2}j_{2} \al_{2}} d_{\al_{1} \al_{2} \al }
    Q_{j_1}
    \underbrace{Q_{i_2}Q_{j_2}}_{\frac{i}{2} \frac{c_{i_{2}j_{2}\bt_{2}}}{\sqrt{C_{2}^{(k)}}} Q_{\bt_2}}
    Q_\al \nonumber \\
    &=& \left( \frac{i}{2} \right)^3 \frac{r^4}{\sqrt{C_{2}^{(k)}}} c_{i_{1}j_{1} \al_{1}}
    \underbrace{d_{ \al_{1} \al_{2} \al } Q_{j_1} Q_{\al_2} Q_\al }_{ c_{k,n} Q_{j_1} Q_{\al_1}}
    \nonumber \\
    &=& \left( \frac{i r}{2} \right)^4 \frac{c_{k,n}}{C_{2}^{(k)}} Q_{i_1}
    \label{e11}
\eeqar
where we evaluate the variation with the fuzzy $\cp^k$ ans\"{a}tze (\ref{5-c6}),
using the relations (\ref{5-a3}), (\ref{5-c7}) and (\ref{5-c10}).
The result (\ref{e11}) shows that the fuzzy $\cp^k$ ($k=2,3$) remain the solutions
of M(atrix) theory even if it is modified with the extra potential $F_5 (X)$.

In this case, the matrix equations of motion (\ref{e5}) become
\beq
    \left[  \frac{\ddot{r}}{R} +
    \frac{R}{C^{(k)}_{2}} r^3 \left( 1 + \frac{\la_5 r}{16 R} c_{k,n} \right) \right] Q_{i}
    = 0 \, .
    \label{e12}
\eeq
This matrix equation is then reduced to an equation of $r(t)$ as in the case
case of the pure bosonic M(atrix) theory.
We can easily carry out the evaluation of $F_5$ on the fuzzy $\cp^k$ ans\"{a}tze as
\beqar
    F_5 (rQ) &=& - \frac{1}{4} r^5
    c_{i_{1}j_{1} \al_{1}} c_{i_{2}j_{2} \al_{2}} d_{\al_{1} \al_{2} \al }
    \Tr ( Q_{i_1} Q_{j_1} Q_{i_2} Q_{j_2}Q_{\al} ) \nonumber\\
    &=& \left( \frac{i}{2} \right)^{4} \frac{r^5}{C^{(k)}_{2}}
    \underbrace{
    d_{ \al_1 \al_2 \al} \Tr ( Q_{\al_1} Q_{\al_2} Q_{\al} )
    }_{ c_{k,n} \Tr (Q_{\al} Q_{\al} )} \nonumber \\
    &=&
    \frac{k^2}{k^2 + 2k} \frac{r^5 c_{k,n}}{16 C^{(k)}_{2}} N^{(k)} ~ \sim ~ n^{k-2}  r^5
    \label{e13}
\eeqar
where we use the relation (\ref{c14}).
Notice that the $n$ dependence of (\ref{e13}) is the same as
that of the M(atrix) theory potential in (\ref{c20}).

\subsection{Modification with $F_7$}

Since $s \le k$ and we are interested in $k=1,2,3$,
the case of $s=3$ is allowed only for $k=3$. In this case, we have
\beqar
    F_{[ij]^{3} \al}
    &=& \tr ( [ t_{i_1} , t_{j_1} ][ t_{i_2} , t_{j_2} ][ t_{i_3} , t_{j_3} ] t_\al )
    \nonumber \\
    &=& - i c_{i_{1}j_{1} \al_{1}}  c_{i_{2}j_{2} \al_{2}} c_{i_{3}j_{3} \al_{3}}
    \tr ( t_{\al_1} t_{\al_2} t_{\al_3} t_{\al} )
    \label{e14}
\eeqar
where, as in the case of $F_5$, $t_\al$'s
are corresponding to `diagonal' generators of $U(3)$.
Thus they are commutative to each other.
Anticommutation relations of these are given by
\beq
    \{t_{\al} , t_{\bt} \} = d_{\al \bt \ga} t_\ga
    \label{e15}
\eeq
where the symmetric symbol $d_{\al \bt \ga}$ is that of $SU(k+1)$
but its indices refer only to a $U(3)$ subgroup.
Notice that a $U(1)$ element is included in this subgroup;
for $SU(4)$ (corresponding to $k=3$) the $U(1)$ element is
conventionally chosen by $t_{15}$ and this choice would be
used for any $SU(k+1)$ ($k \ge 3$).
Using (\ref{e15}), we then find
\beqar
    F_{[ij]^{3} \al} &=& - \frac{i}{4}
    c_{i_{1}j_{1} \al_{1}}  c_{i_{2}j_{2} \al_{2}} c_{i_{3}j_{3} \al_{3}}
    d_{\al_1 \al_2 \bt} d_{\bt \al_3 \al } \, ,
	\label{e16}\\
    F_7 (X) &=& F_{[ij]^{3} \al}  \Tr ( X_{i_1} X_{j_1} X_{i_2} X_{j_2} X_{i_3} X_{j_3} X_{\al} ) \, .
    \label{e17}
\eeqar
The variation of $F_7$ with respect to $X_{i_1}$ is then expressed as
\beqar
    \left. \frac{\del}{\del X_{i_1}} F_7 (X) \right|_{X=rQ} &=&
    - \frac{i}{4} r^6 c_{i_{1}j_{1} \al_{1}}  c_{i_{2}j_{2} \al_{2}} c_{i_{3}j_{3} \al_{3}}
    d_{\al_1 \al_2 \bt} d_{\bt \al_3 \al }
    Q_{j_1}
    \underbrace{Q_{i_2}Q_{j_2}}_{\frac{i}{2} \frac{c_{i_{2}j_{2}\bt_{2}}}{\sqrt{C_{2}^{(k)}}} Q_{\bt_2}}
    \underbrace{Q_{i_3}Q_{j_3}}_{\frac{i}{2} \frac{c_{i_{3}j_{3}\bt_{3}}}{\sqrt{C_{2}^{(k)}}} Q_{\bt_3}}
    Q_\al \nonumber \\
    &=& i \left( \frac{i}{2} \right)^4 \frac{r^6}{C_{2}^{(k)}} c_{i_{1}j_{1} \al_{1}}
    d_{\al_1 \al_2 \bt} d_{\bt \al_3 \al } Q_{j_1} Q_{\al_2} Q_{\al_3} Q_\al
    \nonumber \\
    &=& i \left( \frac{i}{2} \right)^4 \frac{r^6}{C_{2}^{(k)}} c_{i_{1}j_{1} \al_{1}}
    c_{k,n}^{~2} Q_{j_1} Q_{\al_1}
    \nonumber \\
    &=&
    \left( \frac{i r}{2} \right)^6 \frac{2 c_{k,n}^{~2}}{\sqrt{C_{2}^{(k)}}^3 } Q_{i_1}
    \label{e18}
\eeqar
where we use the relation (\ref{5-a3}), {\it i.e.},
$d_{ \al \bt \ga } Q_{\al}Q_{\bt}  = c_{k,n} Q_{\ga}$, twice.
Notice that the symmetric symbol $d_{\al \bt i}$ vanishes as discussed in (\ref{c11-1}).
The result (\ref{e18}) shows that the fuzzy $\cp^k$ ($k=3$) remains as a solution
to M(atrix) theory even if it is modified with the extra potential $F_7 (X)$.
Lastly we can evaluate $F_7$ on the fuzzy $\cp^k$ ans\"{a}tze as
\beqar
    F_7 (rQ) &=&
    - \frac{i}{4} r^6 c_{i_{1}j_{1} \al_{1}}  c_{i_{2}j_{2} \al_{2}} c_{i_{3}j_{3} \al_{3}}
    d_{\al_1 \al_2 \bt} d_{\bt \al_3 \al }
    \Tr (Q_{i_1}Q_{j_1}Q_{i_2}Q_{j_2}Q_{i_3}Q_{j_3}Q_{\al} )
    \nonumber \\
    &=&
    i \left( \frac{i}{2} \right)^5 \frac{r^7}{\sqrt{C_{2}^{(k)}}^3 }
    \underbrace{d_{\al_1 \al_2 \bt} d_{\bt \al_3 \al }
    \Tr (Q_{\al_1}Q_{\al_2}Q_{\al_3}Q_{\al} )}_{ c_{k,n}^{~2} \Tr (Q_{\bt}Q_{\bt})}
    \nonumber \\
    &=& - \frac{k^2}{k^2 + 2k } \frac{r^7 c_{k,n}^{~2}}{32 \sqrt{C_{2}^{(k)}}^3 } N^{(k)}
    \, \sim \, n^{k-3} r^7 \, .
    \label{e19}
\eeqar

\subsection{Emergence of L7-branes}

To recapitulate, we are allowed to include the extra potentials of the
form $F_{2s+1} (X)$ ($s \le k$, $k=1,2,3$) in the M(atrix) theory
Lagrangian as far as the brane solutions of $\cp^k$ geometry in the
transverse directions are concerned. Evaluated on the fuzzy $\cp^k$ ans\"{a}tze,
these extra potentials are expressed as
\beqar
    F_3 (rQ) &=& -\frac{k^2}{k^2 + 2k } \frac{r^3}{4 \sqrt{C_{2}^{(k)}}} N^{(k)}
    \, \sim \, n^{k-1} r^3
    \label{l1} \\
    F_5 (rQ) &=& \frac{k^2}{k^2 + 2k } \frac{r^5 c_{k,n}}{16 C_{2}^{(k)}} N^{(k)}
    \, \sim \, n^{k-2} r^5
    \label{l2} \\
    F_7 (rQ) &=& - \frac{k^2}{k^2 + 2k } \frac{r^7 c_{k,n}^{~2}}{32 \sqrt{C_{2}^{(k)}}^3 } N^{(k)}
    \, \sim \, n^{k-3} r^7
    \label{l3} \\
    V (rQ) &=& \frac{k^2}{k^2 + 2k }\frac{R r^4}{4 C_{2}^{(k)}} N^{(k)}
    \, \sim \, n^{k-2} R r^4
    \label{l4}
\eeqar
where we include the M(atrix) theory potential in (\ref{c20}).
As mentioned earlier, we consider a static solution. Thus the effective
Lagrangian for the static solution is given by
\beq
    \L_{eff} = - V_{tot} (r) = - V (rQ) - \la_3 F_3 (rQ) -
    \la_5 F_5 (rQ) - \la_7 F_7 (rQ) \, .
    \label{l5}
\eeq
From (\ref{l1})-(\ref{l4}), we can express $V_{tot} (r)$ as
\beqar
    V_{tot}(r) &=& \frac{k^2}{k^2 + 2k }\frac{R}{C_{2}^{(k)}} N^{(k)} v(r)
    ~ \sim ~ n^{k-2} R
    \label{l6}\\
    v(r) &=& \frac{r^4}{4} - \mu_3 r^3 + \mu_5 r^5 + \mu_7 r^7
    \label{l7}
\eeqar
where
\beq
    \mu_3 = \frac{\la_3}{4 R} \sqrt{C_{2}^{(k)}} \, , ~~
    \mu_5 = \frac{\la_5}{16 R} c_{k,n} \, , ~~
    \mu_7 = - \frac{\la_7}{32 R} \frac{c_{k,n}^{~2}}{\sqrt{C_{2}^{(k)}}}  \, .
    \label{l8}
\eeq

In the case of $k=1$, only $F_3$ exists and the potential $v(r)$ becomes
$v_3 (r) \equiv \frac{r^4}{4} - \mu_3 r^3$.
This potential is relevant to the Myers effect. In Myers' analysis
\cite{Myers}, the coefficient $\la_3$ is determined such that it
satisfies the equations of motion $\frac{\d v_3}{\d r} = r^3 - 3 \mu_3
r^2 = 0$.
Thus we have $\mu_3 \sim r/3 \sim 1$ ($r > 0$), or $\la_3 \sim R/n$.
Analogously, we may require $\la_5 \sim R$, $\la_7 \sim n R$ such that
$v(r) \sim 1$.
Notice that we demand $\mu_{5}, \mu_{7} > 0$
so that the potential $v(r)$ is bounded below; otherwise
the solutions become unphysical in the limit of large $r$.
We also demand  $\mu_3 > 0$ such that $v(r)$ always has a minimum at $r>0$;
regarding the range of $r$, we require $r > 0$ because
it describes a size of each brain solution.

The total potential $V_{tot} (\sim n^{k-2} R)$ becomes finite for $k=1,2$
in the large $n$ limit.
In this limit, the brane solutions corresponding to $k=1,2$ therefore exist
regardless the value of $v(r)$.
For $k=3$, however, $V_{tot}(r)$ diverges in the
large $n$ limit {\it unless} $v(r) = 0$.

To further investigate the case of $k=3$, we now consider the following
potential $v_7 (r)\equiv \frac{r^4}{4} - \mu_3 r^3 + \mu_7 r^7 $, without
a $F_5$ term.
The equation of motion for $r$ is given by
\beq
    \frac{\d v_7}{\d r} = 7 \mu_7 r^2
    \left( r^4 + \frac{r}{7 \mu_7} - \frac{3 \mu_3}{7 \mu_7} \right) = 0 \, .
    \label{l9}
\eeq
Denoting the nonzero solution by $r_*$, we now plug
this back to $v_7 (r)$; $v_7 (r_*) = \frac{r_{*}^{3}}{7} (
\frac{3r_*}{4} - 4 \mu_3 )$. If we fix $\mu_3$ as $\mu_3 = \frac{3}{16}
r_*$, $v_7(r_*)$ vanishes. In this case, $V_{tot} (r_*)$ becomes finite
in the large $n$ limit and the corresponding L7-branes are allowed to
present as a stable solution at the minimum $r=r_*$. The L7-branes exist
for a particular value of $\mu_3$. In this sense, the strength of the
$F_3$ flux can be considered as a controlling parameter for the
emergence of L7-branes. The same analysis applies to a potential without
$F_7$; $v_5 (r) \equiv  \frac{r^4}{4} - \mu_3 r^3 + \mu_5 r^5$. If we
consider the full potential $v(r)$ with nonzero $\mu_{2s+1}$ ($s=1,2,3$),
the existence of L7-branes can be similarly shown at the minimum of $v(r)$,
with two of the three $\mu_{2s+1}$ serving as the controlling
parameters.

There are few remarks on the existence of the L7-brane solutions.
Firstly, if we introduce fluctuations from the minima, the potential $v(r)$
becomes nonzero and consequently the total potential $V_{tot}(r)$
diverges in the large $n$ limit. In other words, fluctuations from the
stabilized L7-branes are suppressed.
Secondly, the involving extra potentials are expressed as
$F_{2s+1} (rQ) \sim  \Tr {\bf 1}$
where ${\bf 1}$ is the $N^{(3)} \times N^{(3)}$ identity matrix.
These can be regarded as constant matrix-valued potentials.
This fact suggests that the analysis in the previous section also holds
with $F_{2s+1}(rQ)$, preserving the L7-brane solutions non-supersymmetric.
Lastly, in terms of M(atrix) theory as a 11-dimensional theory,
the emergence of L7-branes and the suppression of their
fluctuations suggest a compactification of the theory down to 7 dimensions.
We shall discuss this point further in the next section.

\section{Compactification scenarios in M(atrix) theory}

As mentioned in the introduction, the existence of a 7-form suggests a
compactification of the 11-dimensional theory down to 7 or 4 dimensions.
In this section, we first show that the extra potential $F_7 (X)$ in (\ref{e14})
can be considered as a 7-form in M(atrix) theory.
We then discuss that the effective
Lagrangian (\ref{l5}) with $k=3$ can be used
for a compactification model of M(atrix) theory down to 7 dimensions.
We also consider a compactification scenario of M(atrix)theory
down to 4 dimensions by use of fuzzy $S^4$ which can be defined in
terms of fuzzy $\cp^3$ \cite{Abe1}.

\subsection{$F_{2s+1}$ as matrix differential forms: a cohomology analysis}

The general expression of $F_{2s+1}(X)$ in (\ref{e1}) is closely related
to differential $(2s+1)$-forms of $SU(k+1)$ ($s=1,2,\cdots , k$).
Differential forms of $SU(k+1)$ are in general constructed by the Lie algebra
valued one-form
\beq
    g^{-1}d g = -i t_{A} E_{A}^{a} d \th^a =-i t_{A} E_A
    \label{d1}
\eeq
where $g=\exp(-i t^a \th^a)$ is an element of
$SU(k+1)$, $\th^a$'s are continuous group parameters, $t_A$'s are
generators of $SU(k+1)$ in the fundamental representation with normalization
$\tr(t_A t_B)=\hf \del_{AB}$, and $E_A = E_{A}^{a} (\th) d \th^a$ are
one-form frame fields on $SU(k+1)$ ($a, A=1,2,\cdots, k^2 + 2k$).
The differential $(2s+1)$-forms $\Om^{(2s+1)}$ of $SU(k+1)$ are then
defined as
\beqar
    \Om^{(2s+1)} &=& \tr (g^{-1} d g)^{2s+1} \nonumber\\
    &=& (-i)^{2s+1} \tr(t_{A_1}t_{A_2}\cdots t_{A_{2s+1}})E_{A_1}
    \wedge E_{A_2} \wedge \cdots \wedge E_{A_{2s+1}}  \nonumber\\
    &=& F_{A_1 A_2 \cdots A_{2s+1}} E_{A_1} \wedge E_{A_2} \wedge \cdots \wedge E_{A_{2s+1}} \, ,
    \label{d2} \\
    F_{A_1 A_2 \cdots A_{2s+1}} &=& \frac{(-i)^{2s+1}}{2^s}
    \tr ( [ t_{A_1} , t_{A_2} ] [ t_{A_3} , t_{A_4} ]  \cdots [ t_{A_{2s-1}} , t_{A_{2s}} ]
    t_{A_{2s+1}} ) \, .
    \label{d3}
\eeqar
Notice that the invariant tensor $F_{A_1 A_2 \cdots A_{2s+1}}$
is essentially the same as the tensor $F_{[ij]^{s} \al}$
defined in (\ref{e2}).
The only difference, apart from proportionality coefficients,
is the index assignments.
A peculiar form in $F_{[ij]^{s} \al}$
arises from the fact that we are interested in algebraic properties
of $\cp^{k} = SU(k+1)/U(k)$ rather than the full $SU(k+1)$.
In other words, $F_{[ij]^{s} \al}$ is a subset of
the invariant tensor $F_{A_1 A_2 \cdots A_{2s+1}}$.
The possible number of such tensors is $k$ $( \ge s)$;
these tensors are called the Casimir invariants for the Lie group $SU(k+1)$.

Mathematically, it is known that
the differential $(2s+1)$-forms $\Om^{(2s+1)}$ of $SU(k+1)$
are elements of $\H^{2s+1} (SU(k+1), {\bf R})$, {\it i.e.},
the $(2s+1)$-th cohomology group of $SU(k+1)$ ($s=1,2,\cdots, k$) over the real numbers.
The Casimir invariants $F_{A_1 A_2 \cdots A_{2s+1}}$
are in one-to-one correspondence with cohomology classes for the Lie
group $SU(k+1)$. This correspondence is related to the so-called Weil
homomorphism between Casimir invariants and Chern classes.
For descriptions of these mathematical aspects of $\Om^{(2s+1)}$, one may refer to \cite{NairBook}.

From the above argument, we can interpret the potentials $F_{2s+1}(X)$ in (\ref{e1})
as matrix-valued differential forms, or as fuzzification of the
differential forms $\Om^{(2s+1)}$ in (\ref{d2}); the fuzzification may be
carried out by replacing $E_A$ with arbitrary matrices $X_A$.
In the following, we justify this statement by showing cohomology properties of
$F_{2s+1}(X)$ evaluated on fuzzy $\cp^k$.
In other words, we shall see that
$F_{2s+1}(X)$, evaluated on fuzzy $\cp^k$, can be considered as
matrix-valued forms that are closed but not exact.

As we have shown in (\ref{e8}), (\ref{e11}) and (\ref{e18}), variations of
$F_{2s+1} (X)$ $(s=1,2,3)$ with respect to $X_i$ are linear in $Q_i$
when $X$'s are evaluated on on the fuzzy $\cp^k$ ans\"{a}tze  $X_i =r(t)Q_i$.
Since $Q_i$ are traceless matrices, this corresponds to the fact that
$F_{(2s+1)}(rQ)$ are matrix-valued {\it closed} differential forms.

On the other hand, as shown in (\ref{l1})-(\ref{l3}),
$F_{2s+1} (rQ)$ $(s=1,2,3)$ are nonzero constants.
This arises from the fact that
$F_{(2s+1)}(rQ)$ are matrix-valued {\it non-exact} differential forms.
Notice that the non-exactness of an ordinary differential form, say $\Om^{(3)}$, can
be shown by $\int_{S^3} \Om^{(3)} \ne 0$, where the integration is taken
over $SU(2) = S^3$.
(If $\Om^{(3)}$ is exact, {\it i.e.}, $\Om^{(3)} = d \al$,
Stokes' theorem says $ \int_{S^3} \Om^{(3)} = \int_{\d S^3} \al$ where
$\d S^3$ is the boundary of $S^3$. Since $S^3$ is a compact manifold, $
\int_{\d S^3} \al = 0$. Thus $\Om^{(3)}$ cannot be exact. One can
similarly show the non-exactness of $\Om^{(2s+1)}$ in general, using the
fact that the volume element of $SU(k+1)$ can be constructed in terms of
the wedge products of $\Om^{(2s+1)}$'s.)
$F_3 (Q)$ is a fuzzy analogue of $ \int_{S^3} \Om^{(3)}$.
Thus the value of $F_3 (Q)$ in (\ref{l1})
corresponds to the nonzero volume element of a fuzzy version of $S^3$.
Locally, we may parametrize $S^3$ as $S^3 \approx \cp^1 \times S^1$.
Thus $F_3 (rQ)$ can also be seen as the volume element of a fuzzy version of
$\cp^1 \times S^1$.
Analogously, we can make a local argument to show
that $F_{2k+1} (rQ)$ ($k=2,3$) correspond to the volume elements of fuzzy
versions of $S^{2k+1} \approx \cp^k \times S^1$.
(Note that since $\cp^k
= S^{2k+1}/ S^1$, we can locally express $S^{2k+1}$ as $\cp^k \times
S^1$ in general.)
We can therefore interpret $F_{2s+1} (rQ)$ as
matrix versions or fuzzifications of $(2s+1)$-forms $\Om^{(2s+1)}$,
given that the invariant tensors $F_{A_1 A_2 \cdots A_{2s+1}}$ in (\ref{d3})
are restricted to the form of $F_{[ij]^{s} \al}$ defined in (\ref{e2}).

\subsection{Freund-Rubin type compactification}

The fact that we can interpret $F_{(7)}(rQ)$ as a 7-form
in M(atrix) theory is interesting in search for a compactification model
of M(atrix) theory. As mentioned in the introduction, according to
Freund and Rubin \cite{FR}, existence of a differential $d^{\prime}$-form in
$d$-dimensional theories suggests compactification of $(d-d^{\prime})$ or $d^{\prime}$
space-like dimensions ($d^{\prime} < d$). Usually the Freund-Rubin type
compactification is considered in 11-dimensional supergravity which
contains a 4-form. Although this compactification has a problem in
regard to the existence of chiral fermions, the Freund-Rubin
compactification of M-theory has been shown to avoid such a problem and
presumably provides a realistic model of M-theory in lower dimensions
\cite{Acha}. The presence of the above-mentioned 7-form then supports a
possibility of the Freund-Rubin type compactification in M(atrix) theory.
It is not clear at this point how the effective Lagrangian
(\ref{l5}) relates to compactified 7-dimensional supergravity in the low energy limit.
However, as discussed before, the Lagrangian (\ref{l5})
with $k=3$ does capture a desirable physical property for the
compactification of M(atrix) theory down to 7 dimensions.

In terms of the 11-dimensional M-theory, the potential $F_7 (rQ)$
corresponds to a flux on a curved space of $( \cp^3 \times S^1 )\times
\M_4$ geometry where $\M_4$ is some four-dimensional manifold.
The Freund-Rubin type compactification requires that the
manifold $\M_4$ be a positively curved Einstein manifold.
This suggests that we in fact have to describe $\M_4$
by some fuzzy spaces, say, fuzzy $\cp^2$ or fuzzy $S^4$ in the context of M(atrix) theory.
So far we have neglected the contributions from $\M_4$ in the
fuzzy $\cp^k$ brane solutions (\ref{5-c6}) where we squash irrelevant directions.
We can however include $\M_4$ contributions to the M(atrix) theory potential (\ref{l6})
such that they do not affect the existence condition for the L7-branes, namely, the
finiteness of $V_{tot} (r)$ in the large $n$ limit.
Notice that there is freedom to add an $n$-independent constant
to $V_{tot}(r)$.
Such a case is possible, for example, if we identify $\M_4$
with a relatively small-size fuzzy $S^4$.

It is known that fuzzy $S^4$ can be represented by block-diagonal matrices, with
their full matrix dimensions given by $N^{(3)}$ \cite{Abe1}.
Thus it is natural to parametrize $\M_4$ by fuzzy $S^4$
for $n$-independent modifications of the Lagrangian (\ref{l5}) with $k=3$.
Notice that one of the four dimensions in $\M_4$ represents
the time component in M(atrix) theory.
Thus a naive application of fuzzy $S^4$ to the geometry of $\M_4$
is not suitable for the framework of M(atrix) theory.
However, as in the case of the IKKT model \cite{IKKT}, one can consider
the time component in terms of a matrix.
As far as a matrix model building of M-theory in the large
$N$ limit is concerned, we may then parametrize $\M_4$ in terms of fuzzy $S^4$.
Along the line of these considerations, we can therefore
interpret the Lagrangian (\ref{l5}) with $k=3$ as an effective
Lagrangian for a compactification model of M(atrix) theory down to 7 dimensions.

\subsection{Emergence of fuzzy $S^4$}

Compactification of M(atrix) theory down to 4 dimensions is also
possible for the Freund-Rubin compactification in the presence of
the 7-form. We shall discuss this possibility by use of fuzzy
$S^4$. As shown in (\ref{3-m2}), functions on fuzzy $S^4$ can be
constructed from functions on fuzzy $\cp^3$ by imposing the
following constraint:
\beq
    [ \F( Q_i ) , Q_{\tilde{\al}} ] = 0
    \label{d4}
\eeq
where $\F( Q_i )$ are arbitrary polynomial
functions of the fuzzy $\cp^3$ coordinates $Q_i$
($i=1,2,\cdots, 6$ or, in a conventional choice of $SU(4)$ generators,
$i=9,10,\cdots, 14$).
The indices $\tilde{\al}$
in $Q_{\tilde{\al}}$ corresponds to the algebra of $\tilde{H} =
SU(2)\times U(1)$ in terms of the decomposition, $SU(4)
\rightarrow SU(2)\times SU(2) \times U(1)$, as considered in (\ref{3-m1}).
With an imposition of (\ref{d4}), the functions on fuzzy $\cp^3$, $\F( Q_i )$,
are reduced to functions on fuzzy $S^4$.

As analyzed in section 3.3, upon the imposition of (\ref{d4})
the fuzzy $\cp^3$ coordinates $Q_A$ become fuzzy $S^4$ coordinates
, say, $Y_\mu$
($\mu=1,2,3,4$). These are no longer represented by full
$N^{(3)} \times N^{(3)}$ matrices but by
$N^{(3)} \times N^{(3)}$ {\it block-diagonal} matrices.
The block-diagonal matrix $Y_\mu$ is composed of $(n+2-m)$ blocks of dimension $m$
for $m=1,2,\cdots, n+1$ and can be expressed as
\beq
    Y_\mu = \mbox{block-diag} ( \underbrace{1,1,\cdots, 1}_{n+1},
    \underbrace{\Box_2, \Box_2,\cdots, \Box_2}_{n}, \cdots,
    \Box_{n},\Box_n , \Box_{n+1} )
    \label{d6}
\eeq
where $\Box_{m}$ denotes a full $(m \times m)$ block matrix.
Notice that the matrix dimension of $Y_\mu$ remains as
\beq
    \sum_{m=1}^{n+1} (n+2-m)m = \frac{1}{6}(n+1)(n+2)(n+3)=N^{(3)} \, ,
    \label{d6-1}
\eeq
while the number of nonzero matrix elements becomes
\beq
    \sum_{m=1}^{n+1} (n+2-m) m^2 = \frac{1}{12}(n+1)(n+2)^2 (n+3) \equiv N^{S^4} \, .
    \label{d6-2}
\eeq
We can in fact show that the number $N^{S^4}$ corresponds to
the number of coefficients in a mode expansion of
truncated functions on $S^4$.
(For details of the correspondence between
fuzzy $S^4$ and truncated functions on $S^4$, see \cite{Abe1}.)
From the expression (\ref{d6}), we can easily tell that
$Y_\mu$ commute with $N^{(1)} \times N^{(1)}$ block
matrices where $N^{(1)}=n+1$ is the number of 1's in (\ref{d6}).
Furthermore, $Q_{\tilde{\al}}$ is in an $N^{(1)} \times N^{(1)}$
matrix representation of $\underline{SU(2)}$ in terms of the
decomposition of $SU(4)$ discussed in (\ref{3-m1}).
Thus, from the expression (\ref{d6}), we can check that
$Y_\mu$ indeed satisfies the constraint (\ref{d4}).

The configuration (\ref{d6}) may be the most
natural one in comparison with fuzzy $\cp^3$ but it is not the only one
that describes fuzzy $S^4$.
For example, we can locate the same-size blocks in a single block,
following some operation, say, matrix multiplication or matrix addition,
instead of diagonally locating each block one by one.
The dimension of the alternative matrix configuration is then given by
\beq
    \sum_{m=1}^{n+1} m = \hf (n+1)(n+2) = N^{(2)} \, .
    \label{d6-3}
\eeq
This means that fuzzy $S^4$ can also be described by
$N^{(2)} \times N^{(2)}$ block-diagonal matrices, say, $\tilde{Y}_\mu$.

We now consider an imposition of the constraint (\ref{d4})
on the effective Lagrangian (\ref{l5}) with $k=3$.
Since the potentials $F_{2s+1}(rQ)$ are proportional to the identity matrix,
they are not affected by the constraint (\ref{d4}) and
the local coordinates of fuzzy $\cp^3$ $Q_i$ are simply replaced by the
fuzzy $S^4$ coordinates $Y_\mu$ after the imposition of (\ref{d4}).
Corresponding matrix equations of motion become linear in $Y_\mu$.
Thus, as in the case of the L7-brane solutions, we can similarly
consider emergence of L5-branes with fuzzy $S^4$ geometry
as brane solutions to modified M(atrix) theories.
As before, the emergence of such L5-branes can be argued by requiring
that the potential energy of the branes at minima of the
total potential energy becomes finite.

In terms of the local coordinates of fuzzy $\cp^3$ $Q_i$, the M(atrix)
theory potential is calculated as
$\Tr \frac{Rr^4}{4} [ Q_i , Q_j]^2 = -
\frac{N^{(3)}}{15} \frac{Rr^4}{ C_{2}^{(3)}}$.
The sum of the extra potentials for the emergence of L7-branes has been
given by $\frac{N^{(3)}}{15} \frac{R r_{*}^{4}}{ C_{2}^{(3)}}$ where
$r_*$ represent a minimum of $v(r)$ in (\ref{l7}).
In terms of the local coordinates of fuzzy $S^4$ $Y_\mu$,
a matrix Lagrangian for the emergence of the spherical L5-branes is then expressed as
\beq
    \L_{S^4 \times S^1} = \Tr \left( \frac{\dot{r}^{2} Y_{\mu}^{2}}{2R}
    + \frac{R r^4}{4} [Y_\mu , Y_\nu]^2 + \frac{R r_{*}^{4}} {15 C^{(3)}_{2}}
    {\bf 1}_{N^{(3)}} \right)
    \label{d7}
\eeq
where we include the kinetic term which is zero for static solutions.
The value of $r_*$ is determined by the controlling parameters for the emergence
of the spherical L5-branes. For example, consider the potential $v(r)$
of the form $v_5 (r) = \frac{r^4}{4} - \mu_3 r^3 + \mu_5 r^5$ where
$\mu_3$, $\mu_5$ are given by (\ref{l8}) with $k=3$. In this case, the
controlling parameter is given by $\mu_3$ as discussed before. From
$\left. \frac{\d v_5}{\d r} \right|_{r_*} =0$ and $v_5 (r_*) =0$, we can
easily find $r_* = 8 \mu_3$. Notice that $r_*$ is independent of $n$
since $\mu_3$ is an $n$-independent parameter.

In order to obtain compactification of M(atrix) theory down to 4
dimensions, we simply eliminate the longitudinal direction in the spherical L5-branes.
The relevant brane solution would be a transverse 4-brane of fuzzy $S^4$ geometry.
Apparently, this brane solution does not have a time component
in the framework of M(atrix) theory but, as mentioned
earlier, it is possible to express the time component by a matrix
as far as a matrix model building of M-theory in the large $N$ limit is concerned.
Bearing this possibility in mind, we can conjecture an action for
such a fuzzy $S^4$ solution as
\beqar
    \S_4 & = & \frac{r^4 R}{4} \, \Tr \left( [Y_\mu , Y_\nu ]^2 ~+~
    \frac{\bt}{C^{(3)}_{2}} \, {\bf 1}_{N^{(3)}} \right) \, ,
    \label{d8} \\
    \bt &=& \frac{4}{15} \left( \frac{r_*}{r} \right)^4 \, \sim \, 1 \, .
    \label{d9}
\eeqar
There are basically two fundamental parameters, $R$ and $N=N^{(3)} \sim n^3$.
We consider that in the large $N/R$ limit the matrix action (\ref{d8})
describes compactification of M-theory in 4 dimensions.
$R$ is essentially the 11-dimensional Planck length $l_p$; remember that
$R$ is given by $R = gl_s = g^{2/3} l_p$ where $g$ is the
string coupling constant and $l_s$ is the string length scale.
There are no restrictions on the size parameter $r$. This suggests conformal invariance
of the theory of interest. The parameter $\bt$, on the other hand, will be determined
by how we carry out flux compactifications in terms of controlling parameters.
Since the fuzzy $S^4$ solutions are constructed from the L7-branes of $\cp^3 \times S^1$ geometry
on top of the algebraic constraint (\ref{d4}), these solutions are also non-supersymmetric.
Lastly we would like to emphasize that the above action can be used as
a physically interesting 4-dimensional matrix model of M-theory compactification.

\subsection{Purely spherical L5-branes as new solutions in M(atrix) theory}

As we have discussed in (\ref{d6-3}), fuzzy $S^4$ can also be represented by
$N^{(2)}\times N^{(2)}$ block-diagonal matrices $\tilde{Y}_\mu$.
Its matrix dimension is the same as that of fuzzy $\cp^2$.
Thus, as in the case of fuzzy $\cp^2$ solutions,
there are no problems on infinite energy and
we can obtain an L5-brane of $S^4\times S^1$ geometry as a solution to
the original M(atrix) theory without any extra potentials.

The transverse directions of this L5-brane are {\it purely} spherical.
Notice that it is different from the previously proposed spherical L5-brane \cite{CLT}.
The previous solution has been constructed under a condition \cite{CLT}:
\beq
    \ep_{ijklm} X_i X_j X_k X_{l} \sim X_{m}
    \label{d10}
\eeq
where $X_i$'s ($i=1,2,\cdots , 5$) denote matrix coordinates of the
brane solution, four out of five coordinates representing the transverse directions.
Owing to the Levi-Civita tensor, the above condition makes sense when
indices $i, j, \cdots , m$ are distinctive one another.
Strictly speaking, the transverse directions following the condition (\ref{d10})
do not describe $S^4$ geometry but rather part of $\cp^3$ geometry.
In the context of fuzzy $\cp^3$ solutions developed in the present chapter,
this can easily be seen by rewriting the above condition as
\beq
    c_{ij \al} c_{kl \bt}d_{\al \bt \ga} Q_i Q_j Q_k Q_l \sim
    d_{\al \bt \ga} Q_\al Q_\bt \sim Q_\ga
    \label{d11}
\eeq
where we replace $\ep_{ijklm}$ by $c_{ij \al} c_{kl \bt}d_{\al \bt \ga}$
and $X_i$'s by the fuzzy $\cp^3$ coordinates $Q_i$.
As we have seen in (\ref{e9}), $c_{ij \al} c_{kl \bt}d_{\al \bt \ga}$
corresponds to the rank-five invariant tensor of $SU(4)$.
Explicit proportionality in (\ref{d11}) can be read from (\ref{e11}).

As discussed above, in order to obtain purely spherical geometry, we
need to impose an algebraic constraint on $Q_i$.
The resultant solution then becomes an L5-brane of fuzzy $S^4$ geometry
in the transverse directions, Fluctuations of this brane solution can naturally
be described by $Q_i \rightarrow  Q_i + A_i$.
As mentioned in the introduction, there has been a difficulty to include
fluctuations in the previously proposed spherical L5-branes \cite{CLT}.
Our version of a purely spherical L5-brane avoids this difficulty
and provides a new brane solution to M(atrix) theory.

\chapter{Conclusions}

Mathematical and physical aspects of fuzzy spaces have been
explored in this dissertation. As for mathematical part, we
consider construction of fuzzy spaces of certain types. In chapter
2, we review construction of fuzzy complex projective spaces
$\cp^k$ ($k=1,2,\cdots$), following a scheme of geometric
quantization. This construction has particular advantages in
defining  symbols and star products for fuzzy $\cp^k$. Algebraic
construction of fuzzy $\cp^k$ has also been included in this
chapter. In chapter 3, we have presented construction of fuzzy
$S^4$, utilizing the fact that $\cp^3$ is an $S^2$ bundle over
$S^4$. Fuzzy $S^4$ is obtained by imposing an additional
constraint on fuzzy $\cp^3$. We find the constraint is appropriate
by considering commutative limits of functions on fuzzy $S^4$ in
terms of homogeneous coordinates of $\cp^3$. We propose that
coordinates on fuzzy $S^4$ are described by block-diagonal
matrices whose embedding square matrix represents the fuzzy
$\cp^3$. Along the way, we have shown a precise matrix-function
correspondence for fuzzy $S^4$, providing different ways of
counting the number of truncated functions on $S^4$. Because of
its structure, the fuzzy $S^4$ should follow a closed and
associative algebra. Analogously, we also obtain fuzzy $S^8$,
using the fact that $\cp^7$ is a $\cp^3$ bundle over $S^8$.

In the second part of this dissertation, we have considered
physical applications of fuzzy spaces. Fuzzy spaces are
particulary suitable for the studies of matrix models. In chapter
4, we consider matrix models for gravity on fuzzy spaces. Such
models can give a finite mode truncation of ordinary commutative
gravity. We obtain the actions for gravity on fuzzy $S^2$ and on
fuzzy $\cp^2$ in terms of finite dimensional matrices. The
commutative large $N$ limit is also discussed. Lastly, in chapter
5, we have discussed application of fuzzy spaces to M(atrix)
theory. Some of the previously known brane solutions in M(atrix)
theory are reviewed by use of fuzzy $\cp^k$ as ans\"{a}tze. We
show that, with an inclusion of extra potential terms, the
M(atrix) theory also has brane solutions whose transverse
directions are described by fuzzy $S^4$ and fuzzy $\cp^3$. The
extra potentials can be considered as matrix-valued or fuzzy
differential $(2r+1)$-forms or fluxes in M(atrix) theory
($r=1,2,\cdots, k$). Compactification of M(atrix) theory is
discussed by use of these potentials. In particular, we have
conjectured a compactification model of M(atrix) theory in four
dimensions. The resultant action (\ref{d8}) is expressed in
terms of the local coordinates of fuzzy $S^4$ (\ref{d6}) and can
be used as a realistic matrix model of M-theory in four
dimensions.

\chapter*{Acknowledgments}
{\addcontentsline{toc}{chapter}{\protect{Acknowledgments}}}

\noindent I would like to express my sincere gratitude to Professor V.P. Nair
for his invaluable support and guidance. I would also like to
thank members, as well as regular visitors, of the High Energy
theory group at City College of the City University of New York.
Finally I would like to thank my parents and especially my wife
for her constant encouragement.



\end{document}